\documentclass[journal,onecolumn]{IEEEtran}
\IEEEoverridecommandlockouts
\usepackage{cite}
\usepackage{amsmath,amssymb,color,graphicx,caption,subcaption,comment,tikz,url,latexsym} 
\usepackage{algorithmic}
\usepackage{graphicx}
\usepackage{textcomp}
\usepackage{pgfplots}
\usepackage{xcolor}
\usepackage{url}
\usepackage{multirow}
\usepackage{mathtools}
\usepgfplotslibrary{units}
\usetikzlibrary{spy,backgrounds}
\usepackage{pgfplotstable}

\def\BibTeX{{\rm B\kern-.05em{\sc i\kern-.025em b}\kern-.08em
		T\kern-.1667em\lower.7ex\hbox{E}\kern-.125emX}}

\newtheorem{theorem}{Theorem} 

\newtheorem{example}{Example}
\newtheorem{definition}{Definition}
\newtheorem{lemma}{Lemma}
\newtheorem{proposition}[theorem]{Proposition}

\newtheorem{conjecture}[theorem]{Conjecture}
\usepackage{bbm}
\newcommand{\M}{\mathsf{M}}
\newcommand{\m}{\mathsf{m}}
\newcommand{\len}{\textrm{len}}

\newcommand{\fix}{\textnormal{max}}
\newcommand{\Tx}{{{T$_{0}$}}}
\newcommand{\Rel}{{{R$_{1}$}}}
\newcommand{\Rec}{{{R$_{2}$}}}

\newcommand{\E}{\mathbb{E}}
\newcommand{\fxy}{\eta_{1}}
\newcommand{\fyz}{\eta_{2}}
\newcommand{\tR}{{R}}
\newcommand{\mJ}{\mathcal{J}}
\newcommand{\mI}{\mathcal{I}}

\newcommand{\mH}{\mathcal{H}}
\newcommand{\hmH}{\hat{\mH}}

\makeatletter
\newcommand{\s}[1]{%
   \{#1\checknextarg}
\newcommand{\checknextarg}{\@ifnextchar\bgroup{\gobblenextarg}{\}}}
\newcommand{\gobblenextarg}[1]{, #1 \@ifnextchar\bgroup{\gobblenextarg}{\}}}

\newcommand{\D}{\mathcal{D}}

\newcommand{\Ra}[2]{R_{#1,#2}}

\begin{document}
	\interdisplaylinepenalty=0
	\title{Multi-Hop Network with Multiple Decision Centers under Expected-Rate Constraints\\
	}
		\author{Mustapha Hamad, Mich\`ele Wigger, Mireille Sarkiss\thanks{Part of this material was presented   at {IEEE} Inf. Theory Workshop {(ITW)} 2021  \cite{ITW21} and at IEEE Globecom Conference 2021 \cite{GLOBECOM21}.}	
		\thanks{ M. Hamad and  M. Wigger are with LTCI, Telecom Paris, IP Paris, 91120 Palaiseau, France,   \{mustapha.hamad,michele.wigger\}@telecom-paris.fr. }
		\thanks{M. Sarkiss is with SAMOVAR, Telecom SudParis, IP Paris, 91011 Evry, France, 
			mireille.sarkiss@telecom-sudparis.eu}\thanks{The works of M. Hamad and M. Wigger were supported by the European Research Council (ERC) under the European Union’s Horizon 2020 programme, grant agreement number 715111.}
	}
	\allowdisplaybreaks[4]
	\sloppy
	\maketitle

	\begin{abstract}

 We consider a multi-hop distributed hypothesis testing problem with multiple decision centers (DCs) for testing against independence and where the observations obey some Markov chain. For this system, we characterize the fundamental type-II error exponents region, i.e.,  the type-II error exponents that the various DCs can achieve  simultaneously, under expected rate-constraints. Our results show that this fundamental exponents region is boosted compared to the region under maximum-rate constraints, and that it depends on the permissible type-I error probabilities. When all DCs have equal permissible type-I error probabilities, the exponents region is rectangular and all DCs can simultaneously achieve their optimal type-II error exponents. When the DCs have different permissible type-I error probabilities, a tradeoff between the type-II error exponents at the different DCs arises. New achievability and converse proofs are presented. For the achievability, a new multiplexing and rate-sharing strategy is proposed. The converse proof is based on applying different change of measure arguments in parallel and on proving asymptotic Markov chains. For the special cases $K = 2$ and $K = 3$, we provide simplified expressions for the exponents region; a similar simplification is conjectured for arbitrary $K\geq 2$.

	\end{abstract}
	\section{Introduction}
Future wireless systems are driven by the exponential growth of IoT networks and applications with various requirements in terms of rate, reliability, and  energy consumption. 
In applications such as health monitoring, security alerting or automotive car control, the sensing and decision systems aim at accurately detecting hazardous events or anomalies at the decision centers (DCs) by collecting  data about the measurements at  the various sensors. 
The different events can be considered as different hypotheses and are assumed to determine the  joint  probability distribution underlying the data observed at all the terminals. Our focus will be on  binary hypothesis testing, i.e., situations with only two possible events, with one of the two events corresponding to the normal situation, the so called  \emph{null hypothesis}  and the other to an  alert situation  the so called  \emph{alternative hypothesis}. 
There are two types of errors to distinguish here: \emph{type-I error} and \emph{type-II error}. Type-I error corresponds to a false alarm where the decision center decides
on the alternative hypothesis when the true hypothesis is the null hypothesis. Type-II error corresponds
to a missed detection where the decision center decides on the null hypothesis when the true one is the
alternative hypothesis. Since our interest is in alert systems where a missed detection  is more critical, we aim at maximizing the exponential decay of the type-II error probability (called \emph{error exponent}) while only requiring the type-I error probability to stay below a given threshold.

Most of the information theoretic works studied the distributed  binary hypothesis testing problem with
a single sensor that communicates with a single distant DC over  a noise-free link   with a constraint on the maximum allowed communication rate \cite{Ahlswede,Han,Han_Kobayashi,shalaby,Amari,Wagner,Kochman,Watanabe_GHT}.
These results were also extended to setups with noisy communication links \cite{Deniz_noisy_DMCs, Michele_noisy_and_MAC,Gunduz_noisy_strong_converse_TAI}, to setups with privacy and secrecy constraints 
\cite{Deniz_DHT_Privacy,Vincent_Tan_DHT_Privacy}, and to more complicated networks with either interactive communication \cite{Kim1,Kim,Katz1,Katz2}, multiple sensors \cite{Wagner,zhao2016distributed,zhao2018distributed}, multiple decision centers \cite{Michele_BC,Michele3,Michele2,Pierre_BC_collobarative,PierreMichele}, or both of them \cite{Michele,salehkalaibar2020hypothesisv1,Vincent}. The works most closely related to this paper are \cite{salehkalaibar2020hypothesisv1} and \cite{Vincent} which considered a multi-hop setup with $K$ sensors and $K$ DCs. Multi-hop setups are motivated by the stringent energy constraints of IoT devices requiring short-range communication only between neighbouring sensors. 

Specifically,  \cite{salehkalaibar2020hypothesisv1} characterized a  set of  type-II error exponent tuples that are simultaneously achievable at the various DCs  in a multi-hop network with $K$ sensors and $K$ DCs. For the special case of \emph{testing against independence} and when the type-I error probabilities at all the DCs are required to vanish asymptotically,  this set of exponents coincides with the fundamental exponents region, which means that  in this special case no other exponent tuples are achievable. Testing against independence refers to a hypothesis test where under the alternative hypothesis the observations at the various terminals follow   the product of the marginal distributions that they experience under the null hypothesis. The result in \cite{salehkalaibar2020hypothesisv1} further required that the  joint distribution of the various observations under the null hypothesis satisfies certain Markov chains from one relay to the other. Interestingly, in this case, the set of exponent tuples that are simultaneously achievable at the $K$ decision centers is a  $K$-dimensional hypercube, implying that no tradeoff between the exponents arises and each DC can achieve the optimal exponent as if it was the only DC in the system.  When $K=2$, \cite{Vincent} proved the strong converse result that the optimal exponent region does not depend on the permissible type-I error probabilities. 

Above works all focused on  maximum rate-constraints where the length of \emph{any} message sent over the communication link is limited. In this paper, we consider \emph{expected rate-constraints} as in \cite{JSAIT,sadaf_sequential_HT,Watanabe_GHT,Telatar_DHT_Signal_Detection}, where the \emph{expected} length of the message sent over the communication link is constrained. Most closely related are the works in \cite{Sadaf1, JSAIT} which showed that under an expected rate-constraint $R$, the optimal type-II error exponent for testing against independence in the single-sensor and single-DC setup  coincides with the optimal type-II error exponent under a maximum-rate constraint $R/(1-\epsilon)$, for $\epsilon$ denoting the permissible type-I error constraint. In other words, the relaxed expected-rate constraint seems to allow to boost the rate by a factor $(1-\epsilon)^{-1}$ compared to the same setup under a maximum-rate constraint.

In this paper we show that the same conclusion holds for the $K$-hop network with $K$ decision centers considered in \cite{salehkalaibar2020hypothesisv1} when $K=2$ or $K=3$ and when all DCs obey the same type-I error constraint $\epsilon$. In this case, the fundamental exponents region is a $K$-dimensional hypercube where all DCs can simultaneously achieve their optimal type-II error exponents as if they were the only DC in the system, and this exponent coincides with the exponent under a maximum-rate constraint but where the rates of \emph{all  links in the system} are boosted by a factor $(1-\epsilon)^{-1}$. 
In contrast,  when the various DCs have different type-I error probability thresholds, a tradeoff arises between the  type-II error exponents that are simultaneously achievable at the different DCs. 
This tradeoff, which depends on the type-I error thresholds at the different DCs, is the {first of its kind} and we {exactly characterize} it for the studied multi-hop setup.
We notice hence that  under expected rate-constraints a strong converse does not hold, since the optimal type-II error exponents depend on the admissible type-I error probabilities. This result holds for arbitrary $K\geq 2$, for which we derive the fundamental type-II error exponents region.

To prove our achievability results under expected-rate constraints, we propose a new multiplexing and rate-sharing strategy that generalizes the degenerate multiplexing scheme in \cite{JSAIT}. Specifically, we multiplex different coding schemes of different sets of rates on the various links and  with different probabilities, where each multiplexed subscheme is an optimal coding and testing scheme when the maximum rates are limited by the chosen rate-tuple. For $K=2$ and $K=3$, we   explicitly characterize the multiplexing probabilities  in function of the type-I error probability thresholds at the various DCs and we show that one can restrict to only $K+1$ subschemes, instead of $2^K$. We conjecture that a similar simplification holds for arbitrary $K\geq 2$.

Our converse proofs apply several instances of the  change of measure arguments in \cite{tyagi2019strong, GuEffros_1,GuEffros_2} in parallel,  where  we also restrict to jointly typical source sequences as in \cite{GuEffros_1}. In contrast to the related strong converse proofs in \cite{tyagi2019strong,Vincent}, no variational characterizations, or hypercontractivity arguments \cite{liu2017_hyper_conc} are required to prove our desired results. Instead, we rely on arguments showing that certain  Markov chains  hold in an asymptotic regime of infinite blocklengths. 
Notice that our method to circumvent variational characterizations, or hypercontractivity, or blowing-up arguments \cite{MartonBU}, seems to extend also to other converse proofs, see for example the simplified proof of the well-known strong converses  for lossless and lossy compression with side-information at the decoder \cite{wynerziv,oohama2018exponential} presented in \cite{arxiv}.  

\smallskip

We summarize our main contributions for the $K\geq 2$-hop network with  $K$ decision centers that test against independence and when the observations at the terminals obey a specific Markov chain:

\begin{itemize}
	\item We provide an exact characterization of the general fundamental exponents region under expected-rate constraints. This result shows rate-boosts on all the links in the system, and illustrates   a tradeoff between the exponents at all DCs with different type-I error thresholds. 
	\item To prove achievability, we propose a new  coding scheme based on  \emph{multiplexing and rate-sharing strategy}. 
	\item Converses are   proved by several parallel change of measure arguments, by showing certain  Markov chains in the asymptotic regime of infinite blocklengths, and by using the blowing-up lemma. (As we show in the converse for $K$ hops, the blowing-up lemma can be circumvented by extending the change of measure arguments to larger alphabets.)
	\item  We prove that our results simplify for the special cases of $K=2$ or $K=3$ hops, in which case the simplified optimal coding scheme  multiplexes only $K+1$ subschemes (instead of $2^{K}$ subschemes) and the multiplexing probabilities can directly be obtained from the permissible type-I error probabilities at the various DCs.  A similar simplification is conjectured to hold for arbitrary $K\geq 2$ hops.
\end{itemize}

	\textit{Paper organization:}
The remainder of this paper is divided into two main parts, one focusing on the  two-hop network (Sections \ref{sec:Model2}--\ref{sec:converse}) and one considering the general $K$-hop network (Sections \ref{sec:ModelK}--\ref{K_hop_converse}). For the first part, Section~\ref{sec:Model2} describes the two-hop system model, and Section~\ref{sec:previous2} presents the related previous results under maximum-rate constraints.  Section~\ref{sec:coding2} explains and analyses our proposed optimal coding schemes for the setup under expected-rate constraints. Section~\ref{sec:results2} contains our main results, discussion, and numerical analysis for the two-hop network. In Section~\ref{sec:converse}, we provide our converse proof which consists of a main lemma,  a general outer bound, and a simplified one. For the second part, Section~\ref{sec:ModelK} introduces the system model for $K$ hops, presents the related previous results on maximum-rate constraints. It also describes our new optimal coding scheme and the fundamental exponents region under expected-rate constraints, and simplifications on them. The  converse for $K$-Hops is presented in   Section~\ref{K_hop_converse}.

	\textit{Notation:}
	We follow the notation in \cite{ElGamal},\cite{JSAIT}. In particular, we use sans serif font for bit-strings: e.g., $\m$ for a deterministic and $\M$ for a random bit-string. We let  $\mathrm{bin}(m)$ denote the shortest bit-string representation of a positive integer  $m$, and for any bit-string $\m$ we let  $\mathrm{len}(\m)$  and $\mathrm{dec}(\m)$ denote its length and its corresponding positive integer. In addition, $\mathcal{T}_{\mu}^{(n)}$ denotes the strongly typical set given by \cite[Definition 2.8]{Csiszarbook}.

Throughout this manuscript,  $h_{b}(\cdot)$ denotes the binary entropy function, and $D(P\|Q)$  the Kullback-Leibler divergence between two probability mass functions on the same alphabet.

For any positive integer $K$, we denote by $\mathcal{P}(K)$ the power set of all subset of $\{1,\ldots, K\}$ excluding the emptyset.

	
	\section{The Two-Hop System Model}\label{sec:Model2}
	
\begin{figure}[htbp]
		\centerline{\includegraphics[ scale=0.52]{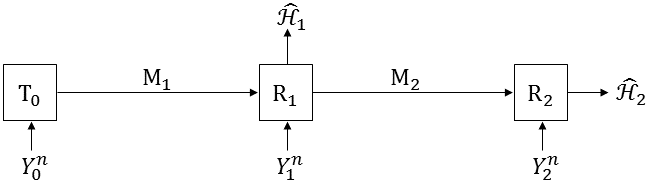}}
		\caption{Cascaded two-hop setup with two decision centers.}
		\label{fig:Cascaded}
	\end{figure}

Consider the distributed hypothesis testing problem in Figure~\ref{fig:Cascaded} with a transmitter \Tx, a relay {\Rel} and a receiver {\Rec} observing sequences $Y_0^n,Y_1^n$ and $Y_2^n$ respectively, forming the Markov chain  
	\begin{equation}\label{eq:Mc}
		Y_0^n \to Y_1^n \to Y_2^n
	\end{equation}
In the special case of testing against independence, i.e., depending on the binary hypothesis $\mathcal{H}\in\{0,1\}$, the tuple $(Y_0^n,Y_1^n,Y_2^n)$ is distributed as:
	\begin{subequations}\label{eq:dist}
		\begin{IEEEeqnarray}{rCl}
			& &\textnormal{under } \mathcal{H} = 0: (Y_0^n,Y_1^n,Y_2^n) \; \textnormal{i.i.d.} \, \sim P_{Y_0Y_1}\cdot P_{Y_2|Y_1} ; \label{eq:H0_dist}\IEEEeqnarraynumspace\\
			& &\textnormal{under } \mathcal{H} = 1: (Y_0^n,Y_1^n,Y_2^n) \; \textnormal{i.i.d.} \,\sim  P_{Y_0}\cdot P_{Y_1}\cdot P_{Y_2}
		\end{IEEEeqnarray} 
	\end{subequations}
	for given probability mass functions (pmfs) $P_{Y_0Y_1}$ and $P_{Y_2|Y_1}$ and where $P_{Y_0}$, $P_{Y_1}$, and $P_{Y_2}$ denote the marginals of the joint pmf $P_{Y_0Y_1Y_2} := P_{Y_0Y_1}P_{Y_2|Y_1}$.
	
	In this two-hop setup, the transmitter {\Tx}  observes the source sequence $Y_0^n$ and sends its bit-string message $\M_1 = \phi_0^{(n)}(Y_0^n)$ to {\Rel}, where the encoding function is of the form $\phi_0^{(n)} : \mathcal{Y}_0^n \to \{0,1\}^{\star}$ and satisfies the \emph{expected} rate constraint
	\begin{equation}\label{eq:Rate1}
		\mathbb{E}\left[\mathrm{len}\left(\M_1\right)\right]\leq nR_1.
	\end{equation} 
	The relay {\Rel} observes the source sequence $Y_1^n$ and with the message $\M_1$ received from {\Tx}, it produces a guess  $\hat{\mathcal{H}}_{1}$ of the hypothesis ${\mathcal{H}}$ using a decision function $g_1^{(n)} : \mathcal{Y}_1^n \times \{0,1\}^{\star} \to \{0,1\}$:
	\begin{equation}
		\hat{\mathcal{H}}_{1} = g_1^{(n)}\left(Y_1^n,\M_1\right) \;   \in\{0,1\}.
	\end{equation}
	Relay {\Rel} also computes a bit-string message $\M_2 = \phi_1^{(n)}\left(Y_1^n,\M_1\right)$ using some encoding function $\phi_1^{(n)}: \mathcal{Y}_1^n\times\{0,1\}^{\star}\to\{0,1\}^{\star}$ that satisfies the expected-rate constraint
	\begin{equation}\label{eq:Rate2}
		\mathbb{E}\left[\mathrm{len}\left(\M_2\right)\right]\leq nR_2.
	\end{equation} Then it sends $\M_2$ to the receiver {\Rec}, which guesses hypothesis $\mathcal{H}$ using   its observation $Y_2^n$ and the received message $\M_2$, i.e.,  using a decision function $g_2^{(n)} : \mathcal{Y}_2^n \times \{0,1\}^{\star} \to \{0,1\}$, it produces the guess:
	\begin{equation}
		\hat{\mathcal{H}}_{2} = g_2^{(n)}\left(Y_2^n,\M_2\right) \;   \in\{0,1\}.
	\end{equation}
	
	The  goal is to design encoding and decision functions such that their type-I error probabilities 
	\begin{IEEEeqnarray}{rCl}
		\alpha_{1,n} &\triangleq& \Pr[\hat{\mathcal{H}}_{1} = 1|\mathcal{H}=0]\\
		\alpha_{2,n} &\triangleq& \Pr[\hat{\mathcal{H}}_{2} = 1|\mathcal{H}=0]
	\end{IEEEeqnarray}
	stay below given thresholds $\epsilon_1 > 0$ and $\epsilon_2 > 0$ and the type-II error probabilities
	\begin{IEEEeqnarray}{rCl}
		\beta_{1,n} &\triangleq& \Pr[\hat{\mathcal{H}}_{1} = 0|\mathcal{H}=1]\\
		\beta_{2,n} &\triangleq& \Pr[\hat{\mathcal{H}}_{2} = 0|\mathcal{H}=1]
	\end{IEEEeqnarray}
	decay to 0 with largest possible exponential decay.  
	
	\bigskip

	\begin{definition} Fix maximum type-I error probabilities $\epsilon_1,\epsilon_2 \in [0,1)$ and rates $R_1,R_2 \geq 0$. The exponent pair $(\theta_1,\theta_2)$ is called \emph{$(\epsilon_1,\epsilon_2)$-achievable} if there exists a sequence of encoding and decision functions $\{\phi_0^{(n)},\phi_1^{(n)},g_1^{(n)},g_2^{(n)}\}_{n\geq 1}$ satisfying $\forall i \in \{1,2\}$:
		\begin{subequations}
			\label{eq:RPconstraints}
		\begin{IEEEeqnarray}{rCl}
			\mathbb{E}[\text{len}(\M_i)] &\leq& nR_i, \label{eq:LEN}\\
			\varlimsup_{n \to \infty}\alpha_{i,n} & \leq& \epsilon_i,\label{type1constraint1}\\ 
			\label{thetaconstraint}
			\varliminf_{n \to \infty}  {1 \over n} \log{1 \over \beta_{i,n}} &\geq& \theta_i.
		\end{IEEEeqnarray}
				\end{subequations}
	\end{definition}
	\bigskip
	
	\begin{definition}\label{eq:Eregion}
		The closure of the set of all $(\epsilon_1,\epsilon_2)$-achievable exponent pairs $(\theta_{1},\theta_{2})$ is called the fundamental \emph{$(\epsilon_1,\epsilon_2)$-exponents region} and is denoted  $\mathcal{E}^*(R_1,R_2,\epsilon_1,\epsilon_2)$. 
			\end{definition}
			
%
%
	\section{Previous Results on Maximum-Rate Constraints for Two Hops} \label{sec:previous2}
\subsection{The Setup}
	The  multi-hop hypothesis testing setup of Figure~\ref{fig:Cascaded} and Equations~\eqref{eq:dist} was also considered in   \cite{Michele} and \cite{Vincent}, but under   \emph{maximum-rate constraints}:
		\begin{equation}\label{eq:FixRates}
			\len(\M_i) \leq nR_i,	\qquad i\in\{1,2\},
		\end{equation} 
		instead of the \emph{expected-rate constraints} \eqref{eq:LEN}. The fundamental exponents region $\mathcal{E}_{\fix}^*(R_1,R_2,\epsilon_1,\epsilon_2)$  for this maximum-rate setup is defined analogously  to Definition~\eqref{eq:Eregion}, but with \eqref{eq:LEN} replaced by \eqref{eq:FixRates}.
		
	In the following  subsection, we report the fundamental exponents region  $\mathcal{E}_{\fix}^*(R_1,R_2,\epsilon_1,\epsilon_2)$ derived in \cite{Vincent}.

\subsection{The Exponents Region}
Define the two functions
			\begin{IEEEeqnarray}{rCl}
				\fxy\left(R_1\right) &:=& \max\limits_{\substack{P_{U_1|Y_0}\colon \\R_1 \geq I\left(U_1;Y_0\right)}} I\left(U_1;Y_1\right)\\
				\fyz\left(R_2\right) &:=& \max\limits_{\substack{P_{U_2|Y_1}\colon \\R_2 \geq I\left(U_2;Y_1\right)}}  I\left(U_2;Y_2\right), \IEEEeqnarraynumspace
			\end{IEEEeqnarray}
			where  the mutual information quantities are calculated with respect to the joint pmfs $P_{U_1Y_0Y_1} := P_{U_1|Y_0}P_{Y_0Y_1}$ and $P_{U_2Y_1Y_2} := P_{U_2|Y_1}P_{Y_1Y_2}$, respectively. As stated in \cite{Ahlswede}, in the above maximization problems it suffices to consider  auxiliary random variables $U_1$ and $U_2$ over alphabets of sizes $|\mathcal{Y}_0|+1$ and $|\mathcal{Y}_1|+1$.

\medskip
\begin{lemma}\label{lem:concavity}
The functions $\eta_1$ and $\eta_2$ are continuous, concave and monotonically non-decreasing on their entire domain $\mathbbm{R}_0^+$.
\end{lemma}
\begin{IEEEproof}
Appendix~\ref{app:concavity} proves the desired properties for $\eta_1$. The proof for $\eta_2$ is analogous and omitted. 
\end{IEEEproof}

\medskip
			
	\begin{theorem}[Theorem 2 in \cite{Vincent}]\label{thm:fixed}	
		Fix $\epsilon_1,\epsilon_2 \in[0,1)$. The fundamental exponents region 	under the maximum-rate constraints \eqref{eq:FixRates} is:
		\begin{IEEEeqnarray}{rCl}
		\lefteqn{\mathcal{E}_{\fix}^*(R_1,R_2,\epsilon_1,\epsilon_2) } \quad\nonumber \\
		&=& \{(\theta_{1},\theta_{2}) \colon \theta_1 \leq \fxy\left(R_1\right),\; 
			\theta_{2} \leq \fxy(R_1)+\fyz(R_2)\}.\IEEEeqnarraynumspace
		\end{IEEEeqnarray}
	\end{theorem}	
	\medskip
	
	We notice that the fundamental exponents region does not depend on the permissible type-I error probabilities $\epsilon_1$ and $\epsilon_2$. We will therefore abbreviate $\mathcal{E}_{\fix}^*(R_1,R_2,\epsilon_1,\epsilon_2)$ by $\mathcal{E}_{\fix}^*(R_1,R_2)$.
	
Notice that  $\fxy(R_1)$ determines the optimal exponent in a point-to-point system where {\Rec} is not present, and $\fyz(R_2)$ determines the optimal exponent in a point-to-point system where {\Tx} is not present \cite{Ahlswede}. In the studied two-hop setup, {\Rec} thus accumulates the optimal exponents achieved over the two links. Since the exponents region is a rectangle, each of the two decision centers, {\Rel} and {\Rec}, can simultaneously achieve their optimal exponents, no tradeoff occurrs between the two exponents. We shall see that this is not always the case under expected-rate constraints.

			\section{Optimal Two-Hop Coding Scheme under  Expected-Rate Constraints}\label{sec:coding2}

The optimal coding scheme under expected-rate constraints depends on whether $\epsilon_1=\epsilon_2$, $\epsilon_1<\epsilon_2$, or $\epsilon_1>\epsilon_2$. The general idea of all the three schemes is that the three terminals {\Tx}, {\Rel}, {\Rec} multiplex  two or three different  subschemes, and the choice of which subscheme to use  depends on the transmitter {\Tx}'s observations $y_0^n$.  To inform all terminals about the choice of the subscheme,  {\Tx} adds one or two flag bits to its message, which the relay {\Rel} forwards to the receiver {\Rec}.  

The main distinguishing feature of the different subschemes is the  choice of the subset of terminals---either only {\Rel} or only {\Rec}, both {\Rel} and {\Rec}, or neither of them---which exploit the information  in the transmitted  messages to produce a guess of  hypothesis $\mathcal{H}$. The other terminals  ignore this communication and simply declare $\hmH=1$. The different subschemes occupy different communication rates, and as we shall see in the following Section~\ref{sec:results2}, the allocation of the rates has to be chosen in function of the desired  tradeoff  between the exponents $\theta_1$ and $\theta_2$. In this section, we formulate the subschemes based on generic hypothesis testing schemes for the two-hop network and the single-hop network with vanishing type-I error probabilities and respecting given rate constraints.  Replacing these generic schemes by the optimal schemes under maximum-rate constraints \cite{Han,Michele} attains the optimal error exponents presented in Theorem~\ref{cor:simplification} ahead. 

		\subsection{The case  $\epsilon_1=\epsilon_2=\epsilon$}\label{sec:scheme_same}
		
We combine two subschemes, where in one subscheme both {\Rel} and {\Rec} attempt to correctly guess the  hypothesis $\mH$ and in the other subscheme both simply declare $\hmH=1$. 
To this end, we partition the set $\mathcal{Y}_0^n$ into subsets $\D_{\emptyset} , \D_{\s{1}{2}}\subseteq \mathcal{Y}_0^n$ so that under $P_{Y_0}^n$ the probability  of subset $\D_{\s{1}{2}}$ is as large as possible but satisfies
\begin{IEEEeqnarray}{rCl}\label{eq:y0eps}
\mathrm{Pr}\left[Y_0^n \in \D_{\s{1}{2}}\right] &\leq& 1- \epsilon.
\end{IEEEeqnarray}
Notice that as $n\to \infty$ the inequality turns into an equality.

Depending on whether $Y_0^n$ lies in $\D_{\emptyset}$ or $\D_{\s{1}{2}}$, the three terminals follow a different subscheme.
			
\underline{If $Y_0^n \in \D_{\emptyset}$:} In this case, none of the terminals attempts to correctly guess the  hypothesis $\mH$. Specifically, \Tx \,and \Rel \,both send
\begin{equation}\label{eq:zerorate}
\M_1=\M_2=[0]
\end{equation}
and {\Rel}and {\Rec} simply declare  
\begin{equation}\label{eq:HyHz1}
\hat{\mathcal{H}}_1=\hat{\mathcal{H}}_2=1.
\end{equation}
			
\underline{If $Y_0^n \in \D_{\s{1}{2}}$:} In this case, both {\Rel} and {\Rec} attempt to correctly  guess $\mH$ based on the transmitted messages. Specifically, 
 {\Tx}, {\Rel}, {\Rec} all apply the encoding/decision functions 
 of a given {two-hop  hypothesis testing scheme with vanishing  type-I error probabilities and respecting maximum-rate constraints   $R_{\s{1}{2},1}$ and   	$R_{\s{1}{2},2}$ on the two links,\footnote{As it will become clear in the subsequent analysis, for the overall scheme to respect rate constraints \eqref{eq:Rate1} and \eqref{eq:Rate2}, it suffices that the two-hop scheme respects the rate constraints   $R_{\s{1}{2},1}$ and   	$R_{\s{1}{2},2}$  on expectation. However, as a consequence of our main result in Theorem~\ref{cor:simplification}, under  vanishing type-I error probabilities, the same type-II error exponents are achievable under both expected- and maximum-rate constraints. There is thus no benefit in considering schemes with expected rates  $R_{\s{1}{2},1}$ and   	$R_{\s{1}{2},2}$, but possibly larger maximum rates.} 
 where these rates are chosen to satisfy
 \begin{subequations}\label{eq:rates_eps}
 \begin{IEEEeqnarray}{rCl}
 	(1-\epsilon)R_{\s{1}{2},1}&\leq &R_1\\
  (1-\epsilon)	R_{\s{1}{2},2}&\leq &	R_2.
  \end{IEEEeqnarray}
\end{subequations} 
  To inform all the terminals about the event $Y_0\in \D_{\s{1}{2}}$ and consequently about the employed scheme, {\Tx} and {\Rel} append the [1]-flag  at the beginning of their messages $\M_1$ and $\M_2$. 
  
\underline{Analysis:}			
By 	\eqref{eq:y0eps} and \eqref{eq:rates_eps}, and because transmission of single bits hardly changes the communication rate for large blocklengths, the overall scheme satisfies the expected-rate constraints $R_1$ and $R_2$ on the two links. 
Appendix~\ref{app1} proves  that when the optimal two-hop hypothesis testing scheme with vanishing type-I error probability \cite{Michele}   is  employed for $Y_0^n \in\D_{\s{1}{2}}$, then the overall scheme meets the permissible type-I error probability  $\epsilon$  and achieves the error exponent given by Equation~\eqref{eq:E1} of Theorem~\ref{cor:simplification}.
				
\subsection{The case $\epsilon_1< \epsilon_2$}\label{sec:scheme_larger}
				
We combine three subschemes, where in each subscheme either no terminal, only  {\Rel}, or both  {\Rel} and  {\Rec} attempt  to correctly guess $\mH$. To this end, we partition the set $\mathcal{Y}_0^n$ into three disjoint subsets $\D_{\emptyset}, \D_{\s{1}},\D_{\s{1}{2}}\subseteq  \mathcal{Y}_0^n$ so that under $P_{Y_0}$ the two sets $\D_{\s{1}}$ and $\D_{\s{1}{2}}$ have largest possible probabilities but limited by
\begin{subequations}\label{eq:inequa}
\begin{IEEEeqnarray}{rCl}
\mathrm{Pr}\left[Y_0^n \in \D_{\s{1}}\right] &\leq & \epsilon_2-\epsilon_1 \label{eq:De0}\\
\mathrm{Pr}\left[Y_0^n \in\D_{\s{1}{2}}\right] &\leq & 1-\epsilon_2. 
\end{IEEEeqnarray}
As a consequence,				
\begin{IEEEeqnarray}{rCl}\label{eq:De}
\mathrm{Pr}\left[Y_0^n \in \D_{\emptyset}\right] &\geq & \epsilon_1.
\end{IEEEeqnarray}
\end{subequations}
Notice that as $n\to \infty$, the three inequalities \eqref{eq:inequa} can hold with equality.  

Choose also nonnegative rates $R_{\s{1},1}$, $R_{\s{1}{2},1}$, 	$R_{\s{1}{2},2}$ satisfying
\begin{subequations}	\label{eq:rate_constrainta}
	\begin{IEEEeqnarray}{rCl}
	(\epsilon_{2}-\epsilon_{1}) R_{\s{1},1} + (1-\epsilon_{2})R_{\s{1}{2},1}&\leq&R_1\\
	(1-\epsilon_2)R_{\s{1}{2},2}&\leq&R_2. 
\end{IEEEeqnarray} 
\end{subequations}

Depending on whether $Y_0^n$ lies in $\D_{\emptyset}$, $\D_{\s{1}}$, or $\D_{\s{1}{2}}$, the three terminals apply a different subscheme satisfying a different pair of maximum-rate constraints, where the subscript $\mI$ of set $\D_{\mI}$ indicates the set of relays that attempt to correctly guess $\mH$ in the event $Y_0^n\in\D_{\mI}$. 
To communicate which of the three subschemes is used,  {\Tx} adds a two-bit flag at the beginning of its message $\M_1$ to  {\Rel}, which  forwards this flag at the beginning of its message $\M_2$ to inform \Rec.
				
\underline{If $Y_0^n \in \D_{\emptyset}$:}  {\Tx} and  {\Rel} send only the flag-bits
\begin{equation}\label{eq:zerorate2}
\M_1=\M_2=[0,0]
\end{equation}
and   {\Rel} and   {\Rec} decide on 
\begin{equation}\label{eq:HyHz122}
\hat{\mathcal{H}}_1=\hat{\mathcal{H}}_2=1.
\end{equation}

\underline{If $Y_0^n \in \D_{\s{1}}$:}	{\Tx} and {\Rel}  apply a given single-hop hypothesis testing scheme with vanishing type-I error probability and expected-rate constraint $R_{\s{1},1}$ for message $\M_1$. Moreover, 
							message $\M_1$ is preceded by flag-bits $[1,0]$, and the relay {\Rel} forwards these flag-bits to {\Rec}:
							\begin{equation}
							\M_2=[1,0]. 
							\end{equation}
							Upon reception of these flag-bits, {\Rec} declares
														\begin{equation}\label{eq:H2trivial}
			\hat{\mathcal{H}}_2=1.
							\end{equation}
							
							We observe that, as indicated by the subscript $\s{1}$ of set $\D_{\s{1}}$, only terminal {\Rel}   attempts to correctly guess $\mH$. Receiver {\Rec}  produces the trivial guess in \eqref{eq:H2trivial} because of its higher admissible type-I error probability $\epsilon_2>\epsilon_1$.  Notice also that no communication rate is required for message $\M_2$ in the limit as $n\to \infty$.

	\underline{If $Y_0^n \in \D_{\s{1}{2}}$:}	
							{\Tx, \Rel, \Rec} apply a given  two-hop hypothesis testing scheme  with vanishing type-I error probabilities and satisfying the expected-rate constraints $R_{\s{1}{2},1}$ and 	$R_{\s{1}{2},2}$.

	\underline{Analysis:}	By \eqref{eq:inequa} and \eqref{eq:rate_constrainta}, and because transmission of two bits hardly changes the rate for sufficiently large blocklengths, the proposed overall scheme respects the expected-rate constraints $R_1$ and $R_2$ for large values of $n$. 
								 Appendix~\ref{app2} proves that  when the optimal single-hop and two-hop hypothesis testing schemes under maximum-rate constraints $R_{\s{1},1}$ and $(R_{\s{1}{2},1}, R_{\s{1}{2},2})$ with vanishing type-I error probability \cite{Han,Michele} are used, then the overall scheme satisfies the type-I error constraints $\epsilon_1$ and $\epsilon_2$ and achieves the error exponents in Equation~\eqref{eq:E2} of Theorem~\ref{cor:simplification}.
								
%
%
								\subsection{The case $\epsilon_1 > \epsilon_2$}
								\label{sec:scheme1_larger}
								
							We combine three subschemes, where in each subscheme either no terminal, only  {\Rec}, or both  {\Rel} and  {\Rec} attempt  to correctly guess $\mH$. To this end, we partition the set $\mathcal{Y}_0^n$ into three disjoint subsets $\D_{\emptyset}, \D_{\s{2}},\D_{\s{1}{2}}\subseteq  \mathcal{Y}_0^n$ so that under $P_{Y_0^n}$ the two sets $\D_{\s{2}}$ and $\D_{\s{1}{2}}$ have largest possible probabilities but limited by
\begin{subequations}\label{eq:inequa2}
\begin{IEEEeqnarray}{rCl}
\mathrm{Pr}\left[Y_0^n \in \D_{\s{2}}\right] &\leq & \epsilon_1-\epsilon_2 \\
\mathrm{Pr}\left[Y_0^n \in\D_{\s{1}{2}}\right] &\leq & 1-\epsilon_1. 
\end{IEEEeqnarray}
As a consequence,				
\begin{IEEEeqnarray}{rCl}\label{eq:De2}
\mathrm{Pr}\left[Y_0^n \in \D_{\emptyset}\right] &\geq & \epsilon_2.
\end{IEEEeqnarray}
\end{subequations}
Notice that as $n\to \infty$, the three inequalities \eqref{eq:inequa2}  hold with equality.  

Choose also nonnegative rates $R_{\s{2},1}$, $R_{\s{1}{2},1}$, $R_{\s{2},2}$, and $R_{\s{1}{2},2}$ satisfying
\begin{IEEEeqnarray}{rCl}\label{eq:rate_constraint}
(\epsilon_{1}-\epsilon_{2}) R_{\s{1},1} + (1-\epsilon_{1})R_{\s{1}{2},1} & \leq & R_1\\
(\epsilon_{1}-\epsilon_{2}) R_{\s{2},2} + (1-\epsilon_{1})R_{\s{1}{2},2} &\leq & R_2.
\end{IEEEeqnarray}

Depending on whether $Y_0^n$ lies in $\D_{\emptyset}$, $\D_{\s{2}}$, or $\D_{\s{1}{2}}$,  the three terminals apply a different subscheme. The subscript $\mI$ of set $\D_{\mI}$ again indicates the set of terminals that attempt to correctly guess $\mH$ in the event $Y_0^n\in\D_{\mI}$, and $R_{\mI,1}, R_{\mI,2}$ indicate the maximum rates of the subscheme employed under $Y_0^n\in\D_{\mI}$. (An exception is the event $Y_0^n\in \D_{\emptyset}$, where both rates are 0.) Flag-bits are used at the beginning of the messages $\M_1$ and $\M_2$ to inform {\Rel} and {\Rec} about which of the subschemes is employed. 
			
\underline{If $Y_0^n \in \D_{\emptyset}$:}  All three terminals, {\Tx},  {\Rel}, and {\Rec} apply the degenerate scheme in \eqref{eq:zerorate2}--\eqref{eq:HyHz122}.

	\underline{If $Y_0^n \in \D_{\s{2}}$:}  	As indicated by the subscript of set $\D_{\s{2}}$, only {\Rec} makes a serious attempt to correctly guess $\mH$, while {\Rel} always declares 
	\begin{equation}\label{eq:dec1}
	\hmH_1=1,
	\end{equation}
	irrespective of the received message and its observations. This implies that under this subscheme, $\alpha_{1,n}=1$ and $\beta_{1,n}=0$. Besides this decision, 
	\Tx, \Rel, and \Rec \,  apply a given two-hop distributed hypothesis testing scheme  with vanishing type-I error probabilities and respecting the maximum-rate constraints $R_{\s{2},1}$ and $R_{\s{2},2}$ for messages $\M_1$ and $\M_2$. Moreover, both {\Tx} and {\Rel} append the two-bit flag [0,1] at the beginning of these two messages to inform all the terminals about the employed scheme.

				Notice that in the optimal two-hop hypothesis testing  scheme \cite{Michele}, 									
														  the relay {\Rel}  computes a tentative decision based on $\M_1$ and $Y_1^n$, which influences the message $\M_2$  sent to {\Rec} and allows the latter to improve its  type-I error probability.
														  Here we propose that {\Rel} itself ignores its tentative decision, because the naive decision \eqref{eq:dec1} is sufficient to satisfy the constraint $\epsilon_1$ on its type-I error probability and is also the most-favorable decision to maximize the type-II error exponent.

\underline{If $Y_0^n \in \D_{\s{1}{2}}$:}  	Both decision centers {\Rel} and {\Rec} attempt to correctly guess $\mH$. Specifically, 		\Tx, \Rel, and {\Rec} apply a given  two-hop hypothesis testing scheme with vanishing type-I error probabilities and respecting the maximum-rate constraints $R_{\s{1}{2},1}$ and $R_{\s{1}{2},2}$ for messages $\M_1$ and $\M_2$. Moreover, both {\Tx} and {\Rel} append the two-bit flag [1,1] at the beginning of these two messages to inform all the terminals about the employed scheme.

								\underline{Analysis:}  Similarly to the case $\epsilon_1<\epsilon_2$, it can be shown that the described scheme respects the expected-rate constraints \eqref{eq:Rate1} and \eqref{eq:Rate2} on both links, and that when  the optimal two-hop scheme \cite{Michele} is employed, then the described scheme  achieves the error exponents in Equation~\eqref{eq:E3_simplified} of Theorem~\ref{cor:simplification}.

			\section{Exponents Region for  the Two-Hop Network under Expected-Rate Constraints}\label{sec:results2}
			
	The fundamental exponents region $\mathcal{E}^*(R_1,R_2,\epsilon_1,\epsilon_2)$ has a different form, depending on the three cases $\epsilon_1=\epsilon_2$, $\epsilon_1 <\epsilon_2$, or $\epsilon_1>\epsilon_2$. 
	
	\begin{theorem}\label{cor:simplification}Given $\epsilon_1,\epsilon_2, R_1, R_2\geq 0$. 
		
		If $\epsilon_1=\epsilon_2=\epsilon$,   then $\mathcal{E}^*(R_1,R_2,\epsilon,\epsilon)$ {is the set of} all nonnegative ($\theta_{1},\theta_{2}$) pairs  satisfying
		\begin{subequations}\label{eq:E1}
			\begin{IEEEeqnarray}{rCl}
				\theta_{1} &\leq&\fxy(R_1/(1-\epsilon))\\
				\theta_{2} &\leq& \fxy(R_1/(1-\epsilon)) + \fyz(R_2/(1-\epsilon)).
			\end{IEEEeqnarray}
		\end{subequations}
		
		If $\epsilon_1< \epsilon_2$, then  $\mathcal{E}^*(R_1,R_2,\epsilon_1,\epsilon_2)$ {is the set of} all nonnegative ($\theta_{1},\theta_{2}$) pairs satisfying
		\begin{subequations}\label{eq:E2}
			\begin{IEEEeqnarray}{rCl}
				\theta_{1} &\leq& \min\left\{\fxy\left({R}_{\{1\},1}\right), \fxy\left(\tR_{\{1,2\},1}\right)\right\}\label{eq:E2theta1} \\
				\theta_{2} &\leq& \fxy\left(\tR_{\{1,2\},1}\right)+\fyz\left(R_{2}/(1-\epsilon_2)\right),\label{eq:E2theta2}
			\end{IEEEeqnarray}
			for some  rates $\tR_{\{1\},1}, \tR_{\{1,2\},1}\geq 0$ so that
			\begin{IEEEeqnarray}{rCl}\label{eq:E2R1}
				R_1 &\geq& (\epsilon_2-\epsilon_1) \tR_{\{1\},1}+ (1-\epsilon_{2}) \tR_{\{1,2\},1}. 
			\end{IEEEeqnarray}
		\end{subequations}
		
		If $\epsilon_1> \epsilon_2$, then $\mathcal{E}^*(R_1,R_2,\epsilon_1,\epsilon_2)$ {is the set of} all nonnegative ($\theta_{1},\theta_{2}$) pairs satisfying
		\begin{subequations}\label{eq:E3_simplified}
			\begin{IEEEeqnarray}{rCl}
				\theta_{1} &\leq& \fxy(\tR_{\{1,2\},1})\label{eq:E3_simplified_theta1} \\
				\theta_{2} &\leq& \min \big\{ \fxy(\tR_{\{1,2\},1})+\fyz\big(\tR_{\{1,2\},2}\big), \; \nonumber \\
				&& \hspace{1.5cm} \fxy\big(\tR_{\{2\},1}\big)+\fyz\big(\tR_{\{2\},2}\big) \big\} \label{eq:E3_simplified_theta2}
			\end{IEEEeqnarray}
			for some  rates $\tR_{\{1,2\},1}$, $\tR_{\{2\},1}$, $\tR_{\{1,2\},2}$, $\tR_{\{2\},2} \geq 0$, so that
			\begin{IEEEeqnarray}{rCl}
				R_1 &\geq& (\epsilon_{1}-\epsilon_{2}) \tR_{\{2\},1} + (1-\epsilon_1) \tR_{\{1,2\},1} \label{eq:E3_simplified_R1}\\
				R_2 &\geq&(\epsilon_{1}-\epsilon_{2}) \tR_{\{2\},2}+  (1-\epsilon_1) \tR_{\{1,2\},2}. \label{eq:E3_simplified_R2}
			\end{IEEEeqnarray}
		\end{subequations}
	\end{theorem}
	\begin{IEEEproof} 
		Achievability is based on the schemes in Section~\ref{sec:coding2}, see Appendices \ref{app1}, \ref{app2} for their analyses. The converse is proved in Section~\ref{sec:converse}. 
	\end{IEEEproof}

		\bigskip
		
We observe from above theorem, that for $\epsilon_1=\epsilon_2=\epsilon$, the fundamental exponents region $\mathcal{E}^*(R_1,R_2,\epsilon,\epsilon)$ is a rectangle. Also, compared to the fundamental exponents region under maximum-rate constraints, here the rates are boosted by a factor $(1-\epsilon)^{-1}$:
\begin{equation}
\mathcal{E}^*(R_1,R_2,\epsilon,\epsilon) = \mathcal{E}^*_{\fix}\left(\frac{R_1}{(1-\epsilon)},\frac{R_2}{(1-\epsilon)}\right).
\end{equation}
In particular,  for $\epsilon_1=\epsilon_2=0$ the fundamental exponents regions under maximum- and expected-rates coincide:
 \begin{equation}
 \mathcal{E}^*(R_1,R_2,0,0)=  \mathcal{E}_{\fix}^*(R_1,R_2).
 \end{equation}

For $\epsilon_1 \neq \epsilon_2$, the fundamental exponents region $\mathcal{E}^*(R_1,R_2,\epsilon_1,\epsilon_2)$ is not a rectangle, as can be verified by the numerical results in Figures~\ref{fig:DSBS_Proposed_scheme_vs_FL2}, \ref{fig:DSBS_Proposed_scheme_vs_FL}, and \ref{fig:DSBS_Proposed_scheme_vs_FL_comparing_rate_distribution} in the next subsection. In fact, one  observes a tradeoff between the two exponents $\theta_1$ and $\theta_2$, which is driven by the choice of the rates $R_{\mI,1}, R_{\mI,2}$ for $\mI \in \mathcal{P}(2)$, where $\mathcal{P}(2)$ is the power set of all subsets of $\{1,2\}$ excluding the emptyset, i.e. $\mathcal{P}(2) = \{\{1\},\{2\},\{1,2\}\}$. More specifically, for $\epsilon_1<\epsilon_2$ the choice 
\begin{subequations}\label{eq:opt2}
\begin{IEEEeqnarray}{rCl}
	R_{\s{1}{2},1}&=&R_1/(1-\epsilon_2)\\
	R_{\s{1},1}&=&0
	\end{IEEEeqnarray} 
	\end{subequations}
	 maximizes exponent $\theta_2$, which then evaluates to 
\begin{equation}\label{eq:maxtheta2}
\theta_2=\theta_{2,\max}:= \eta_1\left(R_1/(1-\epsilon_2)\right) +  \eta_2\left(R_2/(1-\epsilon_2)\right),
\end{equation}
but completely degrades $\theta_1$ to $\theta_1=0$.  
(Notice that for large $R_1/(1-\epsilon_2)$ above choice \eqref{eq:opt2} might not be the unique optimizer and other optimizers will still allow to attain a positive $\eta_1$.) 

On the other hand, the choice
\begin{equation}\label{eq:opt1}
R_{\s{1},1}=R_{\s{1}{2},1}= R_1 /(1-\epsilon_1)
\end{equation}  maximizes exponent $\theta_1$, which then evaluates to
\begin{equation}
\theta_1=\theta_{1,\max}:= \eta_1\left(R_1/(1-\epsilon_1)\right), 
\end{equation}
 but it  degrades $\theta_2$ to 
 \begin{equation}
 \theta_2=\theta_{2,\textnormal{deg}}:=  \eta_1\left(R_1/(1-\epsilon_1)\right) +  \eta_2\left(R_2/(1-\epsilon_2)\right)< \theta_{2,\max}. 
 \end{equation}
Varying the rate $R_{\s{1}{2},1}$ between the choices in  \eqref{eq:opt2} and \eqref{eq:opt1}, (and accordingly varying also rate $R_{\s{1},1}$ to meet \eqref{eq:E2R1}) achieves the entire Pareto-optimal boundary of the fundamental exponents region $\mathcal{E}^*(R_1,R_2,\epsilon_1,\epsilon_2)$. \\

For $\epsilon_1>\epsilon_2$ the choice 
\begin{subequations}\label{eq:opt1b}
	\begin{IEEEeqnarray}{rCl}
		R_{\s{1}{2},1}&=&R_1/(1-\epsilon_1)\\
		R_{\s{2},1}&=&0
	\end{IEEEeqnarray} 
\end{subequations}
maximizes exponent $\theta_1$, which then evaluates to 
\begin{equation}\label{eq:maxtheta1}
\theta_1=\theta_{1,\max}
\end{equation}
and degrades $\theta_2$ to 
\begin{IEEEeqnarray}{rCl}
\theta_2=\theta_{2,\textnormal{deg}}'&:= & \min\big\{\eta_1\left(R_1/(1-\epsilon_1)\right) +  \eta_2(R_{\s{1}{2},2}), \; \nonumber \\
&& \hspace{4.2cm} \eta_2(R_{\s{2},2})\big\}, \IEEEeqnarraynumspace \label{eq:theta_2_deg'_def}
\end{IEEEeqnarray} for $R_{\s{2},2}$ and $R_{\s{1}{2},2}$ satisfying \eqref{eq:E3_simplified_R2}.
(Notice again that for large values of $R_1/(1-\epsilon_1)$ the optimizer in \eqref{eq:opt1b} might not be unique and other optimizers might lead to a larger value of $\theta_2$.)

On the other hand, the choice
\begin{subequations}\label{eq:opt2b}
\begin{IEEEeqnarray}{rCl}
R_{\s{2},1}=R_{\s{1}{2},1}&=&R_1 /(1-\epsilon_2)\\
R_{\s{2},2}=R_{\s{1}{2},2}&= &R_2 /(1-\epsilon_2)
\end{IEEEeqnarray}
\end{subequations}  maximizes exponent $\theta_2$, which then evaluates to
$\theta_2=\theta_{2,\max}$, 
but it  degrades $\theta_1$ to  
\begin{equation}
\theta_1=\theta_{1,\textnormal{deg}}:=  \eta_1\left(R_1/(1-\epsilon_2)\right) 
\end{equation}
Varying the rate $R_{\s{1}{2},1}$ between the choices in  \eqref{eq:opt1b} and \eqref{eq:opt2b} (and  varying the rates $R_{\s{1},1}, R_{\s{1}{2},2}, R_{\s{1},2}$ accordingly), achieves  the  entire Pareto-optimal boundary of the fundamental exponents region $\mathcal{E}^*(R_1,R_2,\epsilon_1,\epsilon_2)$.

 Notice  that in our two-hop system with expected-rate constraints, exponents $\theta_{1,\max}$ and $\theta_{2,\max}$ defined in \eqref{eq:maxtheta1} and \eqref{eq:maxtheta2}, are the largest possible exponents achievable at  the two decision centers,  irrespective of the ordering of $\epsilon_1$  and $\epsilon_2$. By Theorem~\ref{thm1}, they  coincide with the optimal exponents under \emph{maximum-rate constraints} $R_1/(1-\epsilon_1)$ and $R_2/(1-\epsilon_1)$ for the two links in case of \eqref{eq:maxtheta1}, and maximum-rate constraints $R_1/(1-\epsilon_2)$ and $R_2/(1-\epsilon_2)$ in case of \eqref{eq:maxtheta2}. We thus observe that whenever $\epsilon_1\neq \epsilon_2$,  the rate-boosts that \emph{expected}-rate constraints allow to obtain over \emph{maximum}-rate constraints  depend on the permissible type-I error probabilities and also on the tradeoff between the two exponents $\theta_1$ and $\theta_2$. In this view, notice that when the focus is on maximizing $\theta_2$, then for $\epsilon_1 < \epsilon_2$ one has to entirely sacrifice $\theta_1$, whereas for $\epsilon_1>\epsilon_2$ positive $\theta_1$-exponents are possible but  the rate-boost experienced by $\theta_1$  is  reduced from  $(1-\epsilon_1)^{-1}$, which is the boost experienced for its maximum $\theta_{1,\max}$, to the smaller factor $(1-\epsilon_2)^{-1}$.

								\subsection{Numerical Simulations}\label{sec:numerical}
				
				In this section, we illustrate the benefits of exploiting the relaxed expected-rate constraints in \eqref{eq:Rate1} and \eqref{eq:Rate2} compared to the more stringent maximum-rate constraints \eqref{eq:FixRates} at hand of some examples. We also show for $\epsilon_1 < \epsilon_2$ the benefits of ``Rate-sharing'' on the first link and the corresponding tradeoff, where the rate $R_1$ is split into $(\epsilon_2-\epsilon_{1})\tR_{\{1\},1}$ and $(1-\epsilon_{2})\tR_{\{1,2\},1}$ as in \eqref{eq:E2}, instead of restricting to a single rate choice for the communication on the first link $\tR_{\{1,2\},1} = R_1/(1-\epsilon_{1})$. For the case $\epsilon_1 < \epsilon_2$, ``Rate-sharing'' on the second link does not have any added value. However, for the case $\epsilon_1 > \epsilon_2$, we illustrate the benefits of ``Rate-sharing'' on both links and the resulting tradeoff from varying the choices of the rates $\tR_{\{1,2\},1}$, $\tR_{\{2\},1}$, $\tR_{\{1,2\},2}$ and $\tR_{\{2\},2}$ that satisfy \eqref{eq:E3_simplified}. This tradeoff stems from multiplexing three coding subschemes among which we have two full versions of the basic two-hop scheme and one degraded subscheme as explained in Subsection~\ref{sec:scheme1_larger}.
				
				Throughout this section we consider the following example. 
				\begin{example}\label{ex:two_hop}
				Let $Y_0,S,T$ be  independent Bernoulli random variables of parameters $p_{Y_0}=0.4,p_S=0.8,p_T=0.8$ and set $Y_1=Y_0 \oplus T$ and $Y_2=Y_1 \oplus S$.

				We first consider the case $\epsilon_1=0.05< \epsilon_2=0.15$, and plot the optimal exponents region $\mathcal{E}^*(R_1,R_2,\epsilon_1,\epsilon_2)$ in Figure~\ref{fig:DSBS_Proposed_scheme_vs_FL2} for symmetric rates $R_1=R_2=0.5$. We note a tradeoff between the type-II error exponents $\theta_1$ and $\theta_2$, which is not present neither for the case $\epsilon_1=\epsilon_2$, nor for the same setup under maximum-rate constraints. (This tradeoff occurs because both exponents have to be optimized over the same choices of rates $\tR_{\{1\},1},\tR_{\{1,2\},1}$.) The figure also shows a sub-optimal version of the exponents region in Theorem~\ref{cor:simplification}, where we set $\tR_{\{1\},1}=\tR_{\{1,2\},1}=R_1/(1-\epsilon_1)$ and thus obtain $\mathcal{E}^*_{\max}(R_1/(1-\epsilon_1), R_2/(1-\epsilon_2))$. Comparing these two regions, we observe that using two different rates $\tR_{\{1\},1}$ and $\tR_{\{1,2\},1}$ (i.e., two different versions of the basic two-hop scheme) allows to obtain a better tradeoff between the two exponents. 
				For futher comparison,  Figure~\ref{fig:DSBS_Proposed_scheme_vs_FL2}  also shows the exponents region $\mathcal{E}^*_{\fix}(R_1, R_2)$ under maximum-rate constraints, so as to illustrate the gain provided by having the relaxed expected-rate constraints instead of maximum-rate constraints. 

				\begin{figure}[htbp]
												\begin{center}
					\begin{tikzpicture} [every pin/.style={fill=white},scale=.875]
						\begin{axis}[scale=1,
							width=.47\columnwidth,
							scale only axis,
							xmin=0,
							xmax=0.18,
							xmajorgrids,
							xlabel={$\theta_{1}$},
							x tick label style={
								/pgf/number format/.cd,
								fixed,
								precision=2,
								/tikz/.cd
							},
							ymin=0.3,
							ymax=0.38,
							ymajorgrids,
							ylabel={$\theta_2$},
							axis x line*=bottom,
							axis y line*=left,
							legend pos=south west,
							legend style={draw=none,fill=none,legend cell align=left, font=\normalsize}
							]

							\addplot[color=teal,solid,line width=2pt]
							table[row sep=crcr]{											
								0					0.375149407228070\\
								9.65894031423886e-14	0.375149407228070\\
								0.00711536871724117	0.374517176038806\\
								0.0141862538100548	0.373883773904402\\
								0.0212118435729523	0.373249205160872\\
								0.0281913021116575	0.372613474108246\\
								0.0351237681785439	0.371976585011117\\
								0.0420083539204619	0.371338542099168\\
								0.0488441435296696	0.370699349567698\\
								0.0556301917873212	0.370059011578125\\
								0.0623655224874962	0.369417532258488\\
								0.0690491267280540	0.368774915703936\\
								0.0756799610525716	0.368131165977200\\
								0.0822569454252817	0.367486287109063\\
								0.0887789610181113	0.366840283098822\\
								0.0952448477856276	0.366193157914727\\
								0.101653401799731	0.365544915494427\\
								0.108003372311205	0.364895559745400\\
								0.114293458499529	0.364245094545369\\
								0.120522305865461	0.363593523742720\\
								0.126688502212481	0.362940851156903\\
								0.132790573152924	0.362287080578831\\
								0.138826977061905	0.361632215771266\\
								0.144796099386432	0.360976260469201\\
								0.150696246197410	0.360319218380234\\
								0.156525636847444	0.359661093184933\\
								0.162282395565877	0.359001888537195\\
								0.167964541750087	0.358341608032599\\
								0.169743069578874    0.358132875286663\\
								0.169743069578874	0\\};
							\addlegendentry{$\mathcal{E}^*(R_1,R_2,\epsilon_{1},\epsilon_{2}) $.}
							
							\addplot[color=red,dashed,line width=2pt]
							table[row sep=crcr]{
								0  	 0.358132875286663\\
								0.169743069578874    0.358132875286663\\
								0.169743069578874    0\\};
							\addlegendentry{$\mathcal{E}^*_{\fix}\left(\frac{R_1}{(1-\epsilon_1)},\frac{R_2}{(1-\epsilon_2)}\right)$}
							
							\addplot[color=teal,dashdotted,line width=2pt]
							table[row sep=crcr]{
								0  	 0.325872480392762\\
								0.162282395565877   	0.325872480392762\\
								0.162282395565877    0\\};
							\addlegendentry{$\mathcal{E}^*_{\fix}(R_1,R_2)$}
							
						\end{axis}				
					\end{tikzpicture}
					\caption{Exponents regions for Example~\ref{ex:two_hop} when  $\epsilon_1=0.05<\epsilon_2=0.15$ and $R_1=R_2=0.5$.}
					\label{fig:DSBS_Proposed_scheme_vs_FL2} 
					\end{center}
				\end{figure}
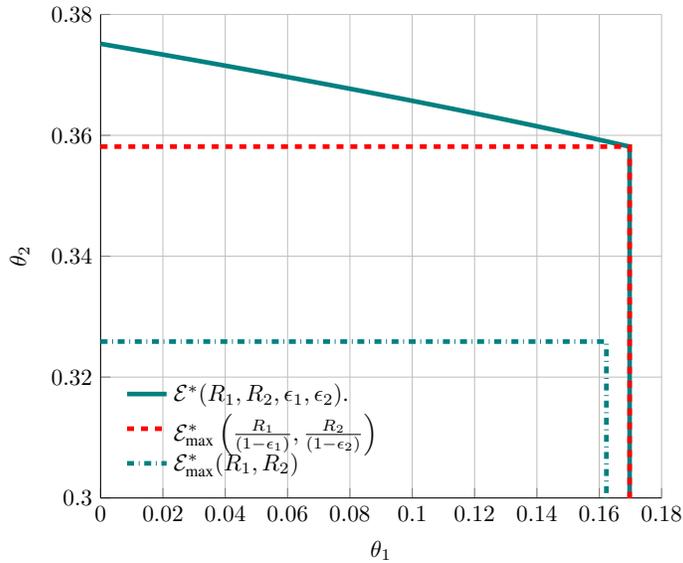

				We then consider the case $\epsilon_1=0.15>\epsilon_2=0.05$. Here we consider three sub-cases for the rates:  symmetric rates   $R_1=R_2=0.5$ or asymmetric rates  $R_1=0.75  > R_2=0.25$ or $R_1=0.25  > R_2=0.75$.  
				
				In Figure~\ref{fig:DSBS_Proposed_scheme_vs_FL} we plot the optimal exponents region $\mathcal{E}^*(R_1,R_2,\epsilon_1,\epsilon_2)$ in Theorem~\ref{cor:simplification} for the first sub-case $R_1=R_2=0.5$, and we compare it  with the exponents region under maximum-rate constraints $\mathcal{E}^*_{\fix}(R_1,R_2)$ and  with sub-optimal versions of Theorem~\ref{thm1}  where we either set  $\tR_{\{1,2\},1}=\tR_{\{2\},1}$, for which we obtain $\mathcal{E}^*_{\fix}\left(\frac{R_1}{(1-\epsilon_{2})},\frac{R_2}{(1-\epsilon_2)}\right)$, or we set $\tR_{\{1,2\},2}=\tR_{\{2\},2}$, for which we have a tradeoff between the type-II error exponents due to rate-sharing on the first link. 
				Comparing all these regions, we see that  rate-sharing on the first link allows to obtain a smooth tradeoff between the exponents, while rate-sharing on both links (i.e., having two full versions of the basic two-hop scheme) yields an even improved  tradeoff.				
				
				Figure~\ref{fig:DSBS_Proposed_scheme_vs_FL_comparing_rate_distribution} compares the  exponents regions under expected rate-constraints for all three sub-cases. Clearly, $\theta_1$ is increasing in $R_1$, but $\theta_2$ is not necessarily increasing in $R_2$, since it also depends on $R_1$. In fact, exponents region $\mathcal{E}^*(0.25,0.75,\epsilon_1,\epsilon_2)$ is completely included in exponents region $\mathcal{E}^*(0.5,0.5,\epsilon_1,\epsilon_2)$.   
				To understand this phenomena, notice that the maximum achievable exponents on each communication link are $\eta_1^*(R_1) = I(Y_0;Y_1)  = 0.26766$ and $\eta_2^*(R_2) = I(Y_1;Y_2)  = 0.27433$. Recall also that the $\theta_2$-error exponent is an accumulation of the error exponents given by both functions $\eta_1(\cdot) + \eta_2(\cdot)$. The similar behaviours of the  two functions $\eta_1(r) \approx \eta_2(r)$ ($r\in[0,1]$), together with the concavity and monotonicity  of these functions, induce that to obtain the largest $\theta_2$ values in this example, the total rate  should be distributed almost equally between both links. In contrast, since the $\theta_1$-error exponent depends only  on rate $R_1$, the largest value is achieved by putting all available rate to $R_1$.  All of the above explains the superiority of the error exponent region obtained when $R_1=R_2=0.5$ over the one obtained when $R_1=0.25, R_2=0.75$, and the tradeoff between the exponents regions for the sub-cases $R_1=R_2=0.5$ and $R_1=0.75, R_2=0.25$.

				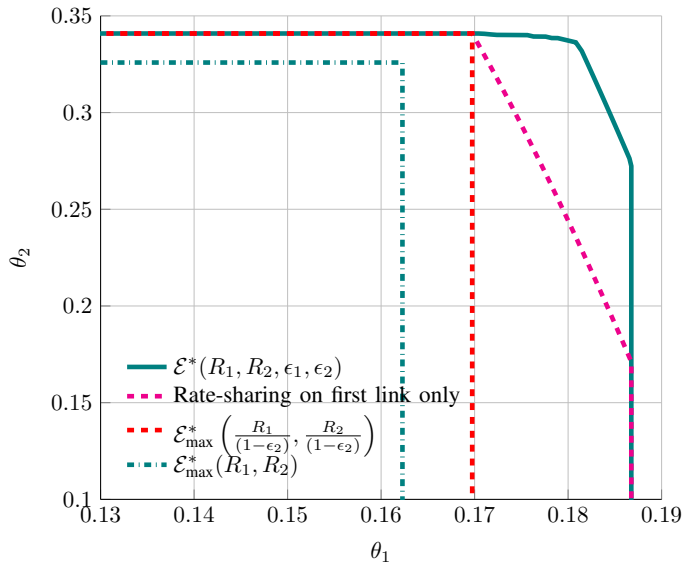
\begin{figure}[htbp]
				\begin{center}
					\begin{tikzpicture} [every pin/.style={fill=white},scale=.875]
						\begin{axis}[scale=1,
							width=.47\columnwidth,
							scale only axis,
							xmin=0.13,
							xmax=0.19,
							xmajorgrids,
							xlabel={$\theta_{1}$},
							ymin=0.1,
							ymax=0.35,
							ymajorgrids,
							ylabel={$\theta_2$},
							axis x line*=bottom,
							axis y line*=left,
							legend pos=south west,
							legend style={draw=none,fill=none,legend cell align=left, font=\normalsize}
							]
							
							\addplot[color=teal,solid,line width=2pt]
							table[row sep=crcr]{												
								0					0.340885797698502\\
								0.169743069706874	0.340885797698502\\
								0.170403689164158	0.340876366858099\\
								0.171063233936723	0.340722577355750\\
								0.171721700405229	0.340411490391450\\
								0.172379084924128	0.340161193583547\\
								0.173035383821312	0.340142274864155\\
								0.173690593397760	0.340121243244965\\
								0.174344709927168	0.340098098243934\\
								0.174997729655584	0.340072839330183\\
								0.175649648801028	0.340045465923868\\
								0.176300463553102	0.339334666549005\\
								0.176950170072602	0.339302057795125\\
								0.177598764491114	0.339267335634734\\
								0.178246242910601	0.338545189364651\\
								0.178892601402988	0.338505245890200\\
								0.179537836009735	0.337774912535338\\
								0.180181942741397	0.337039519337324\\
								0.180824917577184	0.336299086100763\\
								0.181466756464502	0.331800007038230\\
								0.182107455318493	0.325029867874832\\
								0.182747010021555	0.318209682510148\\
								0.183385416422864	0.311340405772141\\
								0.184022670337875	0.304422962593512\\
								0.184658767547818	0.297458249504565\\
								0.185293703799182	0.290447136014997\\
								0.185927474803185	0.283390465895617\\
								0.186560076235237	0.276289058369637\\
								0.186759601321881	0.272352262830495\\
								0.186759601321881	0\\};
							\addlegendentry{$\mathcal{E}^*(R_1,R_2,\epsilon_1,\epsilon_2)$}

							\addplot[color=magenta,dashed,line width=2pt]
							table[row sep=crcr]{													
								0					0.340885797570502\\
								0.169743069706874	0.340885797570502\\
								0.170403689164158	0.335227642938375\\
								0.171063233936723	0.329494226837738\\
								0.171721700405229	0.323687565181861\\
								0.172379084924128	0.317809565723856\\
								0.173035383821312	0.311862038186362\\
								0.173690593397760	0.305846703043338\\
								0.174344709927168	0.299765199177939\\
								0.174997729655584	0.293619090596745\\
								0.175649648801028	0.287409872346545\\
								0.176300463553102	0.281138975753178\\
								0.176950170072602	0.274807773080806\\
								0.177598764491114	0.268417581693086\\
								0.178246242910601	0.261969667784148\\
								0.178892601402988	0.255465249736315\\
								0.179537836009735	0.248905501152524\\
								0.180181942741397	0.242291553604088\\
								0.180824917577184	0.235624499128371\\
								0.181466756464502	0.228905392505933\\
								0.182107455318493	0.222135253342534\\
								0.182747010021555	0.215315067977851\\
								0.183385416422864	0.208445791239844\\
								0.184022670337875	0.201528348061215\\
								0.184658767547818	0.194563634972267\\
								0.185293703799182	0.187552521482699\\
								0.185927474803185	0.180495851363320\\
								0.186560076235237	0.173394443837340\\
								0.186759600912282    0.171142727863628\\
								0.186759600912282	0\\};
							\addlegendentry{Rate-sharing on first link only}
							
							\addplot[color=red,dashed,line width=2pt]
							table[row sep=crcr]{
								0  	 0.340885797570502\\
								0.169743069706874   	0.340885797570502\\
								0.169743069706874   	0.171142727863628\\
								0.169743069706874   	0\\};
							\addlegendentry{$\mathcal{E}^*_{\fix}\left(\frac{R_1}{(1-\epsilon_2)},\frac{R_2}{(1-\epsilon_2)}\right)$}
							
							\addplot[color=teal,dashdotted,line width=2pt]
							table[row sep=crcr]{
								0  	 0.325872480392762\\
								0.162282395565877   	0.325872480392762\\
								0.162282395565877    0\\};
							\addlegendentry{$\mathcal{E}^*_{\fix}(R_1,R_2)$}
							
						\end{axis}				
					\end{tikzpicture}
					\caption{Exponents regions under expected- and maximum-rate constraints  for Example~\ref{ex:two_hop} when  $\epsilon_1=0.15> \epsilon_2=0.05$ and $R_1=R_2=0.5$.}
					\label{fig:DSBS_Proposed_scheme_vs_FL} 
					\end{center}
				\end{figure}

				\begin{figure}[htbp]
				\begin{center}
					\begin{tikzpicture} [every pin/.style={fill=white},scale=.875]
						\begin{axis}[scale=1,
							width=.47\columnwidth,
							scale only axis,
							xmin=0,
							xmax=0.26,
							xmajorgrids,
							xlabel={$\theta_{1}$},
							ymin=0.05,
							ymax=0.35,
							ymajorgrids,
							ylabel={$\theta_2$},
							axis x line*=bottom,
							axis y line*=left,
							legend pos=south west,
							legend style={draw=none,fill=none,legend cell align=left, font=\normalsize}
							]
							
							\addplot[color=teal,solid,line width=2pt]
							table[row sep=crcr]{												
								0					0.340885797698502\\
								0.169743069706874	0.340885797698502\\
								0.170403689164158	0.340876366858099\\
								0.171063233936723	0.340722577355750\\
								0.171721700405229	0.340411490391450\\
								0.172379084924128	0.340161193583547\\
								0.173035383821312	0.340142274864155\\
								0.173690593397760	0.340121243244965\\
								0.174344709927168	0.340098098243934\\
								0.174997729655584	0.340072839330183\\
								0.175649648801028	0.340045465923868\\
								0.176300463553102	0.339334666549005\\
								0.176950170072602	0.339302057795125\\
								0.177598764491114	0.339267335634734\\
								0.178246242910601	0.338545189364651\\
								0.178892601402988	0.338505245890200\\
								0.179537836009735	0.337774912535338\\
								0.180181942741397	0.337039519337324\\
								0.180824917577184	0.336299086100763\\
								0.181466756464502	0.331800007038230\\
								0.182107455318493	0.325029867874832\\
								0.182747010021555	0.318209682510148\\
								0.183385416422864	0.311340405772141\\
								0.184022670337875	0.304422962593512\\
								0.184658767547818	0.297458249504565\\
								0.185293703799182	0.290447136014997\\
								0.185927474803185	0.283390465895617\\
								0.186560076235237	0.276289058369637\\
								0.186759601321881	0.272352262830495\\
								0.186759601321881	0\\};
							\addlegendentry{$\mathcal{E}^*(0.5R,0.5R,\epsilon_1,\epsilon_2)$}
							
							\addplot[color=blue,solid,line width=2pt]
							table[row sep=crcr]{												
								0					0.3259\\
								0.2389	0.3259\\
								0.2399	0.3257\\
								0.2437	0.3241\\
								0.2456	0.3229\\
								0.2474	0.3212\\
								0.2492	0.3192\\
								0.2501	0.3179\\
								0.251	0.3163\\
								0.2519	0.3143\\
								0.2527	0.3021\\
								0.2536	0.2882\\
								0.2544	0.2733\\
								0.2544	0\\};
							\addlegendentry{$\mathcal{E}^*(0.75R,0.25R,\epsilon_1,\epsilon_2)$}
							
							\addplot[color=cyan,dashed,line width=2pt]
							table[row sep=crcr]{												
								0					0.328235408699108\\
								0.089803652263579	0.328235408699108\\
								0.090806888855447	0.328235408699108\\
								0.090806888855447	0.328050006857630\\
								0.092309203906715	0.327947618905668\\
								0.093808457844206	0.326984271621701\\
								0.094307526380863	0.324812771846084\\
								0.098287700674230	0.293917606335432\\
								0.099774557210598	0.272387001204354\\
								0.099774557210598	0\\};
							\addlegendentry{$\mathcal{E}^*(0.25R,0.75R,\epsilon_1,\epsilon_2)$}

						\end{axis}				
					\end{tikzpicture}
					\caption{Exponents regions under  symmetric and asymmetric expected-rate constraints for Example~\ref{ex:two_hop} and when $\epsilon_1=0.15> \epsilon_2=0.05$ and $R=1$.}
					\label{fig:DSBS_Proposed_scheme_vs_FL_comparing_rate_distribution} 
					\end{center}
				\end{figure}
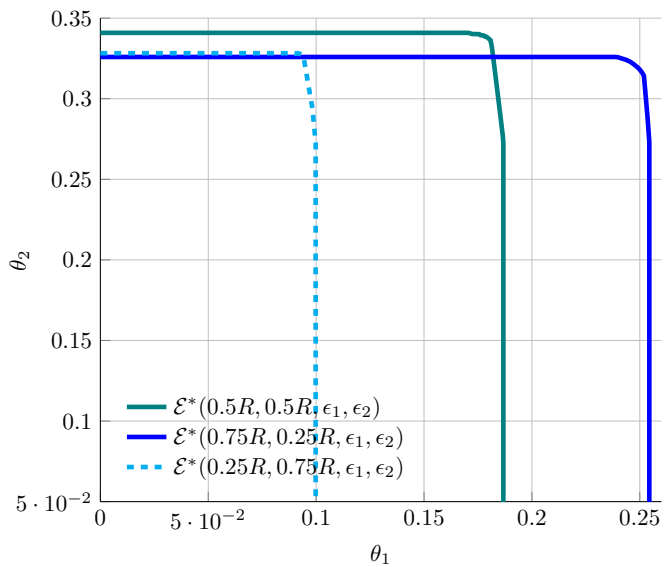

	\end{example}

\section{Converse Proof to Theorem~\ref{cor:simplification}}\label{sec:converse}

The converse is outlined  as follows. Subsection~\ref{general_converse}  proposes the  auxiliary Lemma~\ref{lem:receiverconverse} (proved in  Appendix~\ref{app_lemma1}), and  applies this lemma to derive a general outer bound on the exponents region that is valid for all values of $\epsilon_1, \epsilon_2 \in [0,1)$.
Subsection~\ref{app:cor1_cor2}  simplifies the general outer bound depending on the three cases $\epsilon_1=\epsilon_2$, $\epsilon_1<\epsilon_2$, or $\epsilon_1>\epsilon_2$.

\subsection{An Auxiliary Lemma and a General Outer Bound}\label{general_converse}
Consider 
	a sequence (in $n$) of encoding and decision functions $\{(\phi_1^{(n)}, \phi_2^{(n)}, g_1^{(n)}, g_2^{(n)})\}$ satisfying the constraints on the rates and  error probabilities in \eqref{eq:RPconstraints}.

\begin{lemma}\label{lem:receiverconverse}
	Fix a small number $\eta>0$, a blocklength $n$, and a set $\D\subseteq \mathcal{Y}_0^n\times\mathcal{Y}_1^n$ of  probability exceeding $\eta$. Let  the tuple ($\tilde{\M}_1,\tilde{\M}_2,\tilde{Y}_0^n,\tilde{Y}_1^n,\tilde{Y}_2^n$)  follow the pmf 
	\begin{IEEEeqnarray}{rCl}
		\lefteqn{P_{{\tilde{\M}_1}{\tilde{\M}_2}\tilde{Y}_0^n\tilde{Y}_1^n\tilde{Y}_2^n}(\m_1,\m_2,y_0^n,y_1^n,y_2^n) \triangleq} \quad \nonumber \\
		&& P_{Y_0^nY_1^nY_2^n}(y_0^n,y_1^n,y_2^n)\cdot{\mathbbm{1} \{(y_0^n,y_1^n)\in \D\} \over P_{Y_0^nY_1^n}(\D)}\nonumber\\ && \quad \cdot{\mathbbm{1}\{\phi_1^{(n)}(y_0^n)=\m_1\}}
		\cdot{\mathbbm{1}\{\phi_2^{(n)}(y_1^n,\phi_1(y_0^n))=\m_2\}}. \IEEEeqnarraynumspace \label{pmftildedoubleprime2_lemma}
	\end{IEEEeqnarray}
	Further, define 
\begin{IEEEeqnarray}{rCl}	
	{U_1}& \triangleq &(\tilde{\M}_1,\tilde{Y}_0^{T-1},\tilde{Y}_1^{T-1},T)\\
	{U_2}& \triangleq & (\tilde{\M}_2,\tilde{Y}_0^{T-1},\tilde{Y}_1^{T-1},T)\\
	\tilde{Y}_i & \triangleq &\tilde{Y}_{i,T}, \quad i\in\{0,1,2\},
		\end{IEEEeqnarray}
	 where $T$ is uniform over $\{1,\ldots,n\}$ and independent of all previously defined random variables. Notice the Markov chain $U_2 \to \tilde{Y}_1\to \tilde{Y}_2$.
	The following (in)equalities hold:
	\begin{IEEEeqnarray}{rCl}
		H(\tilde{M}_1) &\geq& nI(U_1;\tilde{Y}_0) + \log P_{Y_0^nY_1^n}(\D),\\
		H(\tilde{M}_2) &\geq& nI(U_2;\tilde{Y}_1) + \log P_{Y_0^nY_1^n}(\D),\\
		I(U_1;\tilde{Y}_1|\tilde{Y}_0) &= &\o_1(n), 
	\end{IEEEeqnarray}
	where $\o_1(n)$ is a function that tends to 0 as $n\to \infty$.

If
	\begin{equation}\label{eq:lemma1_cond1}
	\Pr[\hat{\mathcal{H}}_2=0|\mathcal{H}=0,Y_0^n=y_0^n,Y_1^n=y_1^n]\geq \eta, \;\; \forall (y_0^n,y_1^n) \in \D,
	\end{equation}
	then
	\begin{IEEEeqnarray}{rCl}
		-{1\over n}\log\beta_{2,n}
		&\leq& I(U_1;\tilde{Y}_1) + I(U_2;\tilde{Y}_2) + \o_2(n),
	\end{IEEEeqnarray}
	and if
	\begin{equation}\label{lem:cond2}
	\Pr[\hat{\mathcal{H}}_1=0|\mathcal{H}=0,Y_0^n=y_0^n,Y_1^n=y_1^n] \geq \eta, \;\; \forall (y_0^n,y_1^n) \in \D,
	\end{equation}
	then
	\begin{equation}
	-{1\over n}\log\beta_{1,n} \leq I(U_1;\tilde{Y}_1) + \o_3(n),
	\end{equation}
	where $\o_2(n),\o_3(n)$ are functions that tend to $0$ as $n \to \infty$.
\end{lemma}
\begin{IEEEproof}
	See Appendix~\ref{app_lemma1}.
\end{IEEEproof}	
\medskip

With this lemma, we can prove the desired general outer bound on the exponents region.

\medskip 

\begin{proposition}\label{thm1} Given $\epsilon_1,\epsilon_2, R_1, R_2\geq 0$.
	The fundamental exponents region $\mathcal{E}^*(R_1,R_2,\epsilon_1,\epsilon_2)$ is included in the  set of all ($\theta_{1},\theta_{2}$) pairs  satisfying 
	\begin{subequations}\label{eq:E3}
		\begin{IEEEeqnarray}{rCl}
			\theta_{1} &\leq& \min\{ \fxy(R_{\{1\},1}),\; \fxy(R_{\{1,2\},1})\}, \label{eq:theta1E3}\\
			\theta_{2} &\leq& \min\big\{  \fxy(R_{\{1,2\},1})+ \fyz(R_{\{1,2\},2}),\; \nonumber \\
			&& \hspace{1.5cm}  \fxy(R_{\{2\},1}) + \fyz(R_{\{2\},2}) \big\}, \label{eq:theta2E3}\IEEEeqnarraynumspace
		\end{IEEEeqnarray}
		for    rates $R_{\{1\},1}, R_{\{1,2\},1},  R_{\{1,2\},2}, R_{\{2\},1}, R_{\{2\},2}\geq 0$ and  numbers $\sigma_{\{1\}}, \sigma_{\{2\}}, \sigma_{\{1,2\}}\geq 0$ so that $\sigma_{\{1\}}+ \sigma_{\{2\}}+\sigma_{\{1,2\}}\leq 1$ and 
				\begin{IEEEeqnarray}{rCl}
					\sigma_{\{1\}}+  \sigma_{\{1,2\}} & \geq & 1- \epsilon_1\\
					\sigma_{\{2\}}+  \sigma_{\{1,2\}} & \geq & 1- \epsilon_2\\
					\sigma_{\{1,2\}} & \geq & \max\{1- \epsilon_1-\epsilon_2, 0\}, 
				\end{IEEEeqnarray}
				and so that the following rate constraints are satisfied:
		\begin{IEEEeqnarray}{rCl}
			R_1& \geq & \sigma_{\{1\}} R_{\{1\},1}+  \sigma_{\{1,2\}}  R_{\{1,2\},1} +\sigma_{\{2\}}R_{\{2\},1} , \label{eq:R1une} \\
			R_2& \geq & \sigma_{\{1,2\}}R_{\{1,2\},2} +  \sigma_{\{2\}} R_{\{2\},2}. \label{eq:R2une}
		\end{IEEEeqnarray}
	\end{subequations}
\end{proposition}
It can be shown that the outer bound on the fundamental exponents region given in this proposition is tight. We however only need and prove the converse result here.
\begin{IEEEproof} 
	Fix a positive $\eta>0$. 
	 Set $\mu_n=n^{-1/3}$, and define the sets 
	\begin{IEEEeqnarray}{rCl}\label{Bn}
		\mathcal{B}_{1}(\eta) &\triangleq& \{(y_0^n,y_1^n) \in \mathcal{T}_{\mu_n}^{(n)}(P_{Y_0Y_1}) \colon \nonumber \\ && \;\; \mathrm{Pr}[\hat{\mathcal{H}}_1=0 | Y_0^n = y_0^n, Y_1^n=y_1^n, \mathcal{H}=0] \geq \eta\}, \nonumber\\
		\IEEEeqnarraynumspace\label{eq:Bn1def}\\
		\mathcal{B}_{2}(\eta) &\triangleq& \{(y_0^n,y_1^n) \in \mathcal{T}_{\mu_n}^{(n)}(P_{Y_0Y_1}) \colon \nonumber \\ && \;\; \mathrm{Pr}[\hat{\mathcal{H}}_2=0 | Y_0^n = y_0^n, Y_1^n=y_1^n, \mathcal{H}=0] \geq \eta\},\nonumber \\
		\label{eq:Bn2def}\\
		\D_{\{1,2\}}(\eta) &\triangleq& 	\mathcal{B}_{1}(\eta) \cap 	\mathcal{B}_{2}(\eta),\\
		\D_{\{1\}}(\eta) &\triangleq& \mathcal{B}_{1}(\eta) \backslash \D_{\{1,2\}}(\eta), \label{D1n}\\
		\D_{\{2\}}(\eta) &\triangleq& \mathcal{B}_{2}(\eta) \backslash \D_{\{1,2\}}(\eta). \label{D2n} 
	\end{IEEEeqnarray}
	Further define for each $n$ the probabilities
	\begin{IEEEeqnarray}{rCl}
		\Delta_{\mI} &\triangleq& P_{Y_0^nY_1^n}(\D_{\mI}(\eta)), \quad\mI \in \mathcal{P}(2),
	\end{IEEEeqnarray}
	and notice that by the laws of probability
	\begin{IEEEeqnarray}{rCl}\label{eq:sum}
		\Delta_{\{1,2\}} +\Delta_{\{1\}} &=& P_{Y_0^nY_1^n}(\mathcal{B}_{1}(\eta)) \\
		\Delta_{\{1,2\}} +\Delta_{\{2\}}& =& P_{Y_0^nY_1^n}(\mathcal{B}_{2}(\eta))\\
		\Delta_{\{1,2\}} & \geq & P_{Y_0^nY_1^n}(\mathcal{B}_{1}(\eta)) + P_{Y_0^nY_1^n}(\mathcal{B}_{2}(\eta))-1.\IEEEeqnarraynumspace
	\end{IEEEeqnarray}

Now by the type-I error probability constraints (\ref{type1constraint1}), we have for $j \in \{1,2\}$:
\begin{IEEEeqnarray}{rCl}
	\lefteqn{1 - \epsilon_j} \quad \nonumber\\
	  &\leq&  \sum_{y_0^{n},y_1^n}\Pr[\hat{\mathcal{H}}_j=0|Y_0^n=y_0^n,Y_1^n=y_1^n,\mathcal{H}=0]\nonumber\\
	&&\hfill\cdot P_{Y_0^nY_1^n}(y_0^n,y_1^n)\IEEEeqnarraynumspace
	\\
	&\leq&\sum_{(y_0^n,y_1^n) \in \mathcal{T}_{\mu_n}^{(n)}}\Pr[\hat{\mathcal{H}}_j=0|Y_0^n=y_0^n,Y_1^n=y_1^n,\mathcal{H}=0]\nonumber \\
	&& \hfill \cdot P_{Y_0^nY_1^n}(y_0^n,y_1^n)\nonumber\IEEEeqnarraynumspace\\
	&&+\hspace{-1mm}\sum_{(y_0^n,y_1^n) \notin \mathcal{T}_{\mu_n}^{(n)}}\Pr[\hat{\mathcal{H}}_j=0|Y_0^n=y_0^n,Y_1^n=y_1^n,\mathcal{H}=0]\nonumber\\
	&&\hfill \cdot P_{Y_0^nY_1^n}(y_0^n,y_1^n)
	\IEEEeqnarraynumspace\\
	&\leq&\sum_{\substack{(y_0^n,y_1^n) \in \\ \mathcal{T}_{\mu_n}^{(n)} \cap \bar{\mathcal{B}}_j(\eta)}} \hspace{-3mm}\Pr[\hat{\mathcal{H}}_j=0|Y_0^n=y_0^n,Y_1^n=y_1^n,\mathcal{H}=0] \nonumber \\
	&& \hfill \cdot P_{Y_0^nY_1^n}(y_0^n,y_1^n)\nonumber \IEEEeqnarraynumspace\\
	&&+\hspace{-2mm}\sum_{(y_0^n,y_1^n) \in \mathcal{B}_j(\eta)}\hspace{-3mm} P_{Y_0^nY_1^n}(y_0^n,y_1^n) + \left(1- P_{Y_0Y_1}^{n}(\mathcal{T}_{\mu_n}^{(n)})\right) \IEEEeqnarraynumspace
	\\
	&\leq& \eta(1- P_{Y_0^nY_1^n}(\mathcal{B}_j(\eta))) + P_{Y_0^nY_1^n}(\mathcal{B}_j(\eta)) \nonumber \\
	&& \hfill  +  P_{Y_0Y_1}^{n}(\overline{\mathcal{T}}_{\mu_n}^{(n)}). \IEEEeqnarraynumspace \label{eq:Bi_derivation} 
\end{IEEEeqnarray}
Moreover, by \cite[Remark to Lemma~2.12]{Csiszarbook}, the probability that the pair ($Y_0^n,Y_1^n$) lies in the jointly strong typical set $\mathcal{T}_{\mu_n}^{(n)}(P_{Y_0Y_1})$ satisfies
\begin{equation}\label{Txy}
	P_{Y_0Y_1}^{n}\left(\mathcal{T}_{\mu_n}^{(n)}(P_{Y_0Y_1})\right) \geq 1 - {\vert{\mathcal{Y}_0}\vert\ \vert{\mathcal{Y}_1}\vert\ \over{4 \mu_n^2 n}},
\end{equation}
Thus, by \eqref{eq:Bi_derivation} and \eqref{Txy}:
\begin{IEEEeqnarray}{rCl}\label{eq:PB}
	P_{Y_0^nY_1^n}(\mathcal{B}_{j}(\eta)) &\geq& {1 - \epsilon_j - \eta \over{1 - \eta}} - {\vert{\mathcal{Y}_0}\vert\vert{\mathcal{Y}_1}\vert \over{(1-\eta)4 \mu_n^2 n}}, \quad j\in\{1,2\}, \IEEEeqnarraynumspace
\end{IEEEeqnarray}
	and we thus conclude that in the limit $n\to \infty$ and $\eta \downarrow 0$:
	\begin{subequations} \label{eq:Delta_conditions}
		\begin{IEEEeqnarray}{rCl}
			\varliminf_{\eta \downarrow 0} \varliminf_{n\to \infty}( \Delta_{\{1,2\}}+ \Delta_{\{1\}} )& \geq & 1- \epsilon_1\\
				\varliminf_{\eta \downarrow 0} \varliminf_{n\to \infty}(\Delta_{\{1,2\}}+ \Delta_{\{2\}}) & \geq & 1- \epsilon_2 \\
			\varliminf_{\eta \downarrow 0} \varliminf_{n\to \infty}\Delta_{\{1,2\}} & \geq & \max\{1- \epsilon_1-\epsilon_2, 0\}\\
			\varlimsup_{\eta \downarrow 0} \varliminf_{n\to \infty}\sum_{\mI \in \mathcal{P}(2)}\Delta_{\mI} & \leq &  1.
		\end{IEEEeqnarray}
	\end{subequations}

	We proceed by applying Lemma~\ref{lem:receiverconverse} to  subset $\D_{\mI}$ for all $\mI\in\mathcal{P}(2)$  with  $\Delta_{\mI} \geq \eta$. This allows to conclude that for any $\mI\in \mathcal{P}(2)$ with  $\Delta_{\mI} \geq \eta$ there exists a pair $(U_{\mI,1},U_{\mI,2})$ satisfying the Markov chain $U_{\mI,2} \to \tilde{Y}_{\mI,1} \to \tilde{Y}_{\mI,2}$ and the (in)equalities 
	\begin{IEEEeqnarray}{rCl}
		H(\tilde{M}_{\mI,1}) &\geq& n {I(U_{\mI,1};\tilde{Y}_{\mI,0})} + \log P_{Y_0^nY_1^n}(\D_{\mI}),\quad \mI\in \mathcal{P}(2), \nonumber \\
		\label{eq:M1i} \IEEEeqnarraynumspace\\
		H(\tilde{M}_{\mI,2}) &\geq& nI(U_{\mI,2};\tilde{Y}_{\mI,1}) + \log P_{Y_0^nY_1^n}(\D_{\mI}), \nonumber \\
		&& \qquad \qquad \qquad \qquad \qquad  \; \mI \in \{\{1,2\},\{2\}\}, \IEEEeqnarraynumspace\label{eq:M2i}\\
		\o_{\mI,1}(n) & = & I(U_{\mI,1}; \tilde{Y}_{\mI,1}|\tilde{Y}_{\mI,{0}})  , \qquad \mI \in \mathcal{P}(2),
	\end{IEEEeqnarray}
	and 
	\begin{IEEEeqnarray}{rCl}
		-\frac{1}{n} \log \beta_{1,n}
		&& \leq I(U_{\mI,1};\tilde{Y}_{\mI,1}) + \o_{\mI,2}(n),  \quad \; \mI\in\{\{1\},\{1,2\}\}, \nonumber \\ 
		\IEEEeqnarraynumspace 
	\end{IEEEeqnarray}
	\begin{IEEEeqnarray}{rCl}
		-\frac{1}{n} \log \beta_{2,n}
		&& \leq I(U_{\mI,1};\tilde{Y}_{\mI,1}) + I(U_{\mI,2};\tilde{Y}_{\mI,2}) + \o_{\mI,3}(n), \nonumber\\
		&& \qquad \qquad \qquad \qquad \qquad \; \mI\in\{\{2\},\{1,2\}\}, \label{eq:sumI} \IEEEeqnarraynumspace
	\end{IEEEeqnarray}
	where for each $\mI$  the functions $\o_{\mI,1}(n)$, $\o_{\mI,2}(n)$, $\o_{\mI,3}(n) \to 0$ as $n\to \infty$ and the random variables $\tilde{Y}_{\mI,0}$, $\tilde{Y}_{\mI,1}$, $\tilde{Y}_{\mI,2}$, $\tilde{\M}_{\mI,1}$, $\tilde{\M}_{\mI,2}$ are defined as in the lemma, when applied to the subset $\D_{\mI}$. 
	
	To simplify exposition, we assume  $\eta$  very small and $\Delta_{\mI}\geq \eta$ for all sets $\mI \in \mathcal{P}(2)$. Otherwise the proof is similar but omitted here.
	
	To summarize:
	\begin{IEEEeqnarray}{rCl}
		\lefteqn{-\frac{1}{n} \log \beta_{1,n}}  \nonumber \\  &\leq& \min\left\{I\left(U_{\{1\},1};\tilde{Y}_{\{1\},1}\right);I\left(U_{\{1,2\},1};\tilde{Y}_{\{1,2\},1}\right)\right\} + \o_2(n), \nonumber \\ \label{eq:R1thetaslasteq}\\
		\lefteqn{-\frac{1}{n} \log \beta_{2,n}} \quad \nonumber\\
		 &\leq& \min\left\{I(U_{\{1,2\},1};\tilde{Y}_{\{1,2\},1}) + I(U_{\{1,2\},2};\tilde{Y}_{\{1,2\},2}); \right.\nonumber \\ 
		&& \left. \qquad \;\; I(U_{\{2\},1};\tilde{Y}_{\{2\},1}) + I(U_{\{2\},2};\tilde{Y}_{\{2\},2})\right\} +\o_3(n), \nonumber \\ \label{eq:R1R2thetaslasteq}\IEEEeqnarraynumspace
	\end{IEEEeqnarray}
	where $\o_2(n)$ and $\o_3(n)$ are functions tending to 0 as $n\to \infty$.
	
	Further, define the following random variables
	\begin{equation}
	{\tilde{L}_{\mI,j}} \triangleq \mathrm{len}(\tilde{\M}_{\mI,j}), \quad  j\in\{1,2\},.\; \mI \in \mathcal{P}(2).
	\end{equation}
	By the rate constraints  \eqref{eq:Rate1} and \eqref{eq:Rate2}, and the definition of the random variables ${\tilde{L}_{\mI,j}}$, we obtain by the total law of expectations:
	\begin{IEEEeqnarray}{rCl}
		nR_1 &\geq& \mathbb{E}[L_1]\\
		&\geq&\sum_{\mI\in{\mathcal{P}(2)}}\mathbb{E}[{\tilde{L}_{\mI,1}}]\Delta_{\mI}. \label{ELi}
	\end{IEEEeqnarray} 
	Moreover,  
	\begin{IEEEeqnarray}{rCl}
		H({\tilde{\M}_{\mI,1}}) &=& H({\tilde{\M}_{\mI,1}},{\tilde{L}_{\mI,1}})\\
		&=& \sum_{l_i} \Pr[{\tilde{L}_{\mI,1}} = l_\mI]H({\tilde{\M}_{\mI,1}}|{\tilde{L}_{\mI,1}}=l_\mI) \nonumber \\
		&& \quad + \; H({\tilde{L}_{\mI,1}})\IEEEeqnarraynumspace\\
		&\leq& \sum_{l_\mI} \Pr[{\tilde{L}_{\mI,1}} = l_\mI]l_\mI + H({\tilde{L}_{\mI,1}}) \label{HMi_ineq1}\\
		&=& \mathbb{E}[{\tilde{L}_{\mI,1}}] + H({\tilde{L}_{\mI,1}}),
	\end{IEEEeqnarray}
	which combined with \eqref{ELi} establishes
	\begin{IEEEeqnarray}{rCl}
		\hspace{-2mm}	\sum_{\mI\in\mathcal{P}(2)} \Delta_{\mI} H({\tilde{\M}_{\mI,1}}) &\leq& \sum_{\mI\in\mathcal{P}(2)} \Delta_{\mI}\mathbb{E}[{\tilde{L}_{\mI,1}}] +  \Delta_{\mI} H({\tilde{L}_{\mI,1}}) \IEEEeqnarraynumspace\\
		&\leq & {nR_1} \left(1 +\sum_{\mI\in\mathcal{P}(2)} h_b\left({\Delta_{\mI} \over nR_1}\right)\right), \label{M1ub'}\IEEEeqnarraynumspace
	\end{IEEEeqnarray}
	where \eqref{M1ub'} holds by \eqref{ELi} and because the entropy of the discrete and positive random variable ${\tilde{L}_{\mI,1}}$ of  mean   $\E{[{\tilde{L}_{\mI,1}}]} \leq {nR_1\over \Delta_{\mI}}$ is bounded by $ \frac{n R_1}{\Delta_{\mI}} \cdot h_b\left({\Delta_{\mI} \over nR_1}\right)$, see \cite[Theorem 12.1.1]{cover}. 
	
	In a similar way, we obtain 
	\begin{IEEEeqnarray}{rCl}
		\sum_{\substack{\mI \in \\ \{\{1,2\},\{2\}\}}}\hspace{-2mm}\Delta_\mI H({\tilde{\M}_{\mI,2}}) \leq {nR_2} \left(1 +\sum_{\substack{\mI\in \\ \{\{1,2\},\{2\}\}}}\hspace{-2mm} h_b\left({\Delta_\mI \over nR_2}\right)\right). \nonumber \\ \IEEEeqnarraynumspace \label{M2ub'}
	\end{IEEEeqnarray}
	Then by combining  \eqref{M1ub'} and \eqref{M2ub'} with \eqref{eq:M1i} and \eqref{eq:M2i}, noting \eqref{eq:sum} and \eqref{eq:PB}, and considering also \eqref{eq:R1thetaslasteq} and \eqref{eq:R1R2thetaslasteq}, we have proved so far that for all $n\geq 1$ there exist joint pmfs $P_{U_{\mI,1}\tilde{Y}_{\mI,0}\tilde{Y}_{\mI,1}}=P_{U_{\mI,1}|\tilde{Y}_{\mI,0}}P_{\tilde{Y}_{\mI,0}\tilde{Y}_{\mI,1}}$ (abbreviated as $P_{\mI,1}^{(n)}$) for $\mI\in\mathcal{P}(2)$, and $P_{U_{\mI,2}\tilde{Y}_{\mI,1}\tilde{Y}_{\mI,2}}=P_{U_{\mI,2}|\tilde{Y}_{\mI,1}}P_{\tilde{Y}_{\mI,1}\tilde{Y}_{\mI,2}} $ (abbreviated as $P_{\mI,2}^{(n)}$) for $\mI\in\{\{1,2\},\{2\}\}$ so that the following conditions hold (where $I_{P}$ indicates that the mutual information should be calculated according to a pmf $P$):
	\begin{subequations}\label{eq:conditions}
		\begin{IEEEeqnarray}{rCl}			
			R_1 &\geq  &\sum_{\mI \in \mathcal{P}(2)} \big(I_{P_{\mI,1}^{(n)}}({U}_{\mI,1};\tilde{Y}_{\mI,0}) + g_{\mI,1}(n)\big)\cdot g_{\mI,2}(n,\eta), \nonumber \\ \label{eq:R111} \IEEEeqnarraynumspace \\
			R_2 &\geq  & \hspace{-2mm} \sum_{\substack{\mI\in\\\{\{1,2\},\{2\}\}}}\hspace{-2mm}\big(I_{P_{\mI,2}^{(n)}}({U}_{\mI,2};\tilde{Y}_{\mI,1}) + g_{\mI,1}(n)\big)\cdot g_{\mI,2}(n,\eta), \nonumber \\ \label{eq:R222}\\
			\theta_1 &\leq& \min \left\{I_{P_{\{1\},1}^{(n)}}({U}_{\{1\},1};\tilde{Y}_{\{1\},1}), \right. \nonumber \\
			&& \qquad \hfill \left.   I_{P_{\{1,2\},2}^{(n)}}({U}_{\{1,2\},1};\tilde{Y}_{\{1,2\},1})\right\} + g_{1,3}(n), \label{eq:theta111}\IEEEeqnarraynumspace\\
			\theta_2 &\leq& \min \left\{I_{P_{\{1,2\},1}}^{(n)}({U}_{\{1,2\},1};\tilde{Y}_{\{1,2\},1}) \right. \nonumber \\ 
			&& \hfill \quad \left. + \; I_{P_{\{1,2\},2}}^{(n)}(U_{\{1,2\},2};\tilde{Y}_{\{1,2\},2}),\qquad \IEEEeqnarraynumspace\right.\nonumber\\
			&&\left. \qquad \quad I_{P_{\{2\},1}}^{(n)}({U}_{\{1,2\},1};\tilde{Y}_{\{2\},1}) \right. \nonumber \\
			&& \hfill \left. + \; I_{P_{\{2\},2}^{(n)}}(U_{\{2\},2};\tilde{Y}_{\{2\},2})\right\} + g_{2,3}(n) , \nonumber \\ \label{eq:theta222}\IEEEeqnarraynumspace\\
			g_{\mI,4}(n) &=& I_{P_{\mI,1}^{(n)}}(\tilde{Y}_{\mI,1};{U}_{\mI,1}|\tilde{Y}_{\mI,0}) , \hfill \mI \in \mathcal{P}(2) , \quad \IEEEeqnarraynumspace \label{eq:last_cond}
		\end{IEEEeqnarray}
	\end{subequations}
	for some nonnegative functions $g_{\mI,1}(n)$, $g_{\mI,2}(n,\eta)$, $g_{k,3}(n)$, $g_{\mI,4}(n)$ with the following asymptotic behaviors:
	\begin{IEEEeqnarray}{rCl}
		\lim_{n\to \infty} g_{\mI,1}(n) &=& 0, \quad  \forall \, \mI \in \mathcal{P}(2),\IEEEeqnarraynumspace\\ 
		\lim_{n\to \infty} g_{k,3}(n) &=& 0, \quad  \forall \, k \in \{1,2\},\IEEEeqnarraynumspace\\
		\lim_{n\to \infty} g_{\mI,4}(n)& = & 0, \quad  \forall \, \mI \in \mathcal{P}(2),\IEEEeqnarraynumspace\\
		\varliminf_{n\to \infty} \left(g_{\{1\},2}(n,\eta)+g_{\{1,2\},2}(n,\eta)\right) & \geq  & \frac{1-\epsilon_1-\eta}{1-\eta},\\
		\varliminf_{n\to \infty} \left(g_{\{1,2\},2}(n,\eta)+g_{\{2\},2}(n,\eta)\right) & \geq  & \frac{1-\epsilon_2-\eta}{1-\eta}, \IEEEeqnarraynumspace\\
\varliminf_{\eta \downarrow 0}\varliminf_{n\to \infty} g_{\{1,2\},2}(n,\eta) & \geq  & \max\{1- \epsilon_1-\epsilon_2, 0\}. \nonumber \\
	\end{IEEEeqnarray}

	We next observe that by Carath\'eodory's theorem \cite[Appendix C]{ElGamal} for each $n$ there must exist random variables ${U}_{\{1\},1},U_{\{1,2\},1},U_{\{2\},1},U_{\{1,2\},2},U_{\{2\},2}$ satisfying \eqref{eq:conditions} over alphabets of sizes
	\begin{align}
	\vert {\mathcal{U}}_{\mI,1}\vert &\leq \vert \mathcal{Y}_0\vert\cdot\vert \mathcal{Y}_1\vert + 2, \qquad \quad \mI \in \mathcal{P}(2),\\
	\vert {\mathcal{U}}_{\mI,2} \vert &\leq \vert {\mathcal{U}}_{\mI,1} \vert\cdot|\mathcal{Y}_0|\cdot\vert \mathcal{Y}_1\vert + 1, \quad \mI \in \{\{1,2\},\{2\}\}.
	\end{align}
	Then we invoke the Bolzano-Weierstrass theorem and  consider an increasing  subsequence of positive numbers $\{n_k\}_{k=1}^\infty$ such that the following  subsequences converge: 
		\begin{IEEEeqnarray}{rCl}
			\lim_{k\to \infty} P_{\tilde{Y}_{\mI,0}\tilde{Y}_{\mI,1}{U}_{\mI,1}}^{(n_k)}&=& P_{Y_{\mI,0}Y_{\mI,1}U_{\mI,1}}^{*}, \quad \mI \in \mathcal{P}(2), \\
			\lim_{k\to \infty} P_{\tilde{Y}_{\mI,1}\tilde{Y}_{\mI,2}{U}_{\mI,2}}^{(n_k)}&=& P_{Y_{\mI,1}Y_{\mI,2}U_{\mI,2}}^{*}, \quad \mI \in \{\{1,2\},\{2\}\}. \IEEEeqnarraynumspace
				\end{IEEEeqnarray}
		Considering  further an appropriate sequence of diminishing $\eta$-values, we conclude  	
by \eqref{eq:R111}--\eqref{eq:theta222} and \eqref{eq:Delta_conditions} that:
	\begin{IEEEeqnarray}{rCl}			
		R_1& \geq & \sigma_{\{1\}}\cdot  I_{P_{\{1\},1}^{*}}({U}_{\{1\},1};{Y}_{\{1\},0}) \nonumber \\
		&& \; + \; \sigma_{\{1,2\}} \cdot I_{P_{\{1,2\},1}^{*}}({U}_{\{1,2\},1};{Y}_{\{1,2\},0})\nonumber\\
		&& \; +\;  \sigma_{\{2\}} \cdot I_{P_{\{2\},1}^{*}}({U}_{\{2\},1};{Y}_{\{2\},0}),\label{eq:R_1_f} \\
		R_2 &\geq  &\sigma_{\{1,2\}} \cdot I_{P_{\{1,2\},2}^{*}}({U}_{\{1,2\},2};{Y}_{\{1,2\},1})\nonumber\\
		&& \; +\;  \sigma_{\{2\}} \cdot I_{P_{\{2\},2}^{*}}({U}_{\{2\},2};{Y}_{\{2\},1}),\\
		\theta_1 &\leq & \min \left\{I_{P_{\{1\},1}^{*}}({U}_{\{1\},1};{Y}_{\{1\},1}), \right. \nonumber \\
		&& \hfill \left. I_{P_{\{1,2\},1}^{*}}({U}_{\{1,2\},1};{Y}_{\{1,2\},1})\right\},\label{theta_1_f} \IEEEeqnarraynumspace\\
		\theta_2 &\leq &\min \left\{I_{P_{\{1,2\},2}^{*}}({U}_{\{1,2\},1};{Y}_{\{1,2\},1}) \right. \nonumber \\
		&& \hfill \left. \,+\, I_{P_{\{1,2\},2}^{*}}({U}_{\{1,2\},2};{Y}_{\{1,2\},2}), \right. \nonumber \\ 
		&& \qquad \quad I_{P_{\{2\},1}^{*}}({U}_{\{2\},1};{Y}_{\{2\},1}) \nonumber \\ 
		&& \hfill \left. \,+\, I_{P_{\{2\},2}^{*}}({U}_{\{2\},2};{Y}_{\{2\},2}) \right\} \label{theta_2_f}\end{IEEEeqnarray}
for some numbers $\sigma_{\{1\}}, \sigma_{\{2\}}, \sigma_{\{1,2\}} >0$ satisfying $\sigma_{\{1\}}+\sigma_{\{2\}}+ \sigma_{\{1,2\}}  \leq 1$
	and 
	\begin{subequations}\label{eq:cond_epsilon}
		\begin{IEEEeqnarray}{rCl}\label{eq:sigma_constraints_KHop}
			\sigma_{\{1\}}+\sigma_{\{1,2\}}  &{\geq}  &   1-\epsilon_1, \\
			\sigma_{\{1,2\}}& \geq & \max\{1- \epsilon_1-\epsilon_2, 0\},\\
			\sigma_{\{2\}}+\sigma_{\{1,2\}}& {\geq} & 1- \epsilon_2. 
		\end{IEEEeqnarray}
	\end{subequations}
	Notice further that since for any $\mI \in \mathcal{P}(2)$ and any $k$ the pair $(\tilde{Y}_{\mI,0}^{(n_k)},\tilde{Y}_{\mI,1}^{(n_k)})$ lies in the jointly typical set $\mathcal{T}^{(n_k)}_{\mu_{n_k}}(P_{Y_0Y_1})$, we have  $\vert P_{\tilde{Y}_{\mI,0}\tilde{Y}_{\mI,1}} - P_{Y_0Y_1}\vert \leq \mu_{n_k}$ and thus the limiting pmfs satisfy $P^*_{Y_{\mI,0}Y_{\mI,1}}=P_{Y_0Y_1}$. Moreover, since for each $n_k$ the random variable $\tilde{Y}_{\mI,2}$ is drawn according to $P_{Y_2|Y_1}$ given $\tilde{Y}_{\mI,1}$, irrespective of $\tilde{Y}_{\mI,0}$, the limiting pmfs also satisfy $P_{Y_{\mI,2}|Y_{\mI,0}Y_{\mI,1}}^*=P_{Y_2|Y_1}$. 
	We also notice for all $\mI \in \{\{1,2\},\{2\}\}$ that  under $P_{Y_{\mI,1}Y_{\mI,2}U_{\mI,2}}^*$ the Markov chain
	\begin{IEEEeqnarray}{rCl}\label{eq:MC2_1_genconv}
		U_{\mI,2}\to Y_{\mI,1} \to Y_{\mI,2}, 
	\end{IEEEeqnarray}	
	holds because  $U_{\mI,2}\to \tilde{Y}_{\mI,1} \to \tilde{Y}_{\mI,2}$  forms a Markov chain for any $n_k$. Finally, by continuity considerations and by \eqref{eq:last_cond}, the following Markov chain must hold under $P_{Y_{\mI,0}Y_{\mI,1}U_{\mI,1}}^*$ for all $\mI \in \mathcal{P}(2)$:
	\begin{IEEEeqnarray}{rCl}\label{eq:MC2_2_genconv}
		U_{\mI,1}\to Y_{\mI,0} \to Y_{\mI,1}.
	\end{IEEEeqnarray}
	Using the definitions of the functions $\fxy(\cdot)$ and $\fyz(\cdot)$, we thus  proved that for any pair of achievable exponents $(\theta_1, \theta_2)$ there exist rates $R_{\{1\},1},R_{\{1,2\},1}, R_{\{2\},1}, R_{\{1,2\},2}, R_{\{2\},2}>0$  satisfying 
	\begin{subequations}\label{eq:E3_generalconverse}
		\begin{IEEEeqnarray}{rCl}
			\theta_{1} &\leq& \min\left\{\fxy(R_{\{1\},1}), \fxy(R_{\{1,2\},1})\right\},\IEEEeqnarraynumspace \\
			\theta_{2} &\leq&  \min\left\{\fxy(R_{\{1,2\},1})+\fyz(R_{\{1,2\},2}),\right. \nonumber \\
			&& \left.\hspace{1.5cm}\; \fxy(R_{\{2\},1})+\fyz(R_{\{2\},2})\right\} ,
		\end{IEEEeqnarray}
		and numbers $\sigma_{\{1\}}, \sigma_{\{2\}}, \sigma_{\{1,2\}} >0$ satisfying $\sigma_{\{1\}}+\sigma_{\{2\}}+ \sigma_{\{1,2\}}  \leq 1$, Inequalities \eqref{eq:cond_epsilon}, and the following two rate constraints:
		\begin{IEEEeqnarray}{rCl}
			R_1& \geq & \sigma_{\{1\}} \cdot  R_{\{1\},1}+  \sigma_{\{1,2\}} \cdot R_{\{1,2\},1}+ \sigma_{\{2\}} \cdot R_{\{2\},1} ,\IEEEeqnarraynumspace		\\
			R_2& \geq & \sigma_{\{1,2\}} \cdot  R_{\{1,2\},2}+ \sigma_{\{2\}} \cdot  R_{\{2\},2}.
		\end{IEEEeqnarray}
	\end{subequations}
\end{IEEEproof}

\subsection{Simplification of the Outer Bound in Proposition~\ref{thm1}}\label{app:cor1_cor2}

We proceed to simplify the outer bound in Proposition~\ref{thm1} depending  on the cases $\epsilon_1=\epsilon_2$, $\epsilon_1 < \epsilon_2$, or $\epsilon_1 >\epsilon_2$. To this end, fix an exponent pair $(\theta_1,\theta_2)$ in $\mathcal{E}^\star(R_1,R_2,\epsilon_1,\epsilon_2)$,  rates $R_{\{1\},1}, R_{\{1,2\},1}, R_{\{1,2\},2}, R_{\{2\},1}, R_{\{2\},2}\geq 0$ and numbers $\sigma_{\{1\}},\sigma_{\{2\}},\sigma_{\{1,2\}}\geq 0$ summing to less than 1 and satisfying constraints \eqref{eq:E3}.

\subsubsection{The case $\epsilon_1=\epsilon_2$}

By \eqref{eq:E3}: 
\begin{IEEEeqnarray}{rCl}
	\theta_1 & \leq & \min\{\fxy(R_{\{1\},1}), \fxy({R}_{\{1,2\},1})\} \\
	&\stackrel{(a)}{\leq}& \frac{\sigma_{\{1\}} \fxy(R_{\{1\},1}) + \sigma_{\{1,2\}} \fxy({R}_{\{1,2\},1})}{\sigma_{\{1\}}+\sigma_{\{1,2\}}} \\
	&\stackrel{(b)}{\leq}& \fxy\left (\frac{\sigma_{\{1\}} R_{\{1\},1} + \sigma_{\{1,2\}} {R}_{\{1,2\},1}}{\sigma_{\{1\}}+\sigma_{\{1,2\}}}\right)\\
	&\stackrel{(c)}{\leq}& \fxy\left(R_1/(1-\epsilon)\right),
\end{IEEEeqnarray}
where $(a)$ holds because the minimum is never larger than any linear combination; $(b)$ holds by the concavity of the function $\fxy(\cdot)$; and $(c)$ holds by the monotonicity of the function $\fxy(\cdot)$ and  because  by \eqref{eq:E3} we have $\sigma_{\{1\}} R_{\{1\},1} + \sigma_{\{1,2\}} {R}_{\{1,2\},1} \leq R_1$ and $\sigma_{\{1\}}+\sigma_{\{1,2\}} \geq 1-\epsilon$. 

Following similar steps, one can prove that 
\begin{IEEEeqnarray}{rCl}
	\theta_2 & \leq &\min\Big\{\fxy\left(R_{\{1,2\},1}\right)+\fyz\left(R_{\{1,2\},2}\right), \nonumber \\
	&& \hspace{15mm}\fxy\left(R_{\{2\},1}\right)+\fyz\left(R_{\{2\},2}\right)\Big\} \\
	&\stackrel{(d)}{\leq}& \frac{\sigma_{\{2\}}  \fxy\left(R_{\{2\},1}\right) +\sigma_{\{2\}}  \fyz\left({R}_{\{2\},2}\right)}{\sigma_{\{2\}} +\sigma_{\{1,2\}} } \nonumber
	 \\
	&& +\frac{\sigma_{\{1,2\}}  \fxy\left(R_{\{1,2\},1}\right) +\sigma_{\{1,2\}}  \fyz\left({R}_{\{1,2\},2}\right)}{\sigma_{\{2\}} +\sigma_{\{1,2\}}}\\
	&\stackrel{(e)}{\leq}& \fxy\left(\frac{\sigma_{\{2\}}  R_{\{2\},1} + \sigma_{\{1,2\}} {R}_{\{1,2\},1}}{\sigma_{\{2\}} +\sigma_{\{1,2\}}}\right) \nonumber \\ 
	&& + \fyz\left(\frac{\sigma_{\{2\}} R_{\{2\},2} +\sigma_{\{1,2\}}R_{\{1,2\},2}}{\sigma_{\{2\}} +\sigma_{\{1,2\}}}\right)\\
	&\stackrel{(f)}{\leq}& \fxy\left({R}_{1}/(1-\epsilon)\right) + \fyz\left({R}_2/(1-\epsilon)\right),
\end{IEEEeqnarray}
where $(d)$ holds again because the minimum is never larger than any linear combination; $(e)$ holds by the concavity of the functions $\fxy(\cdot)$ and $\fyz(\cdot)$; and $(f)$ holds because by \eqref{eq:E3} we have 
$\sigma_{\{2\}} R_{\{2\},i} + \sigma_{\{1,2\}} {R}_{\{1,2\},i} \leq R_i$, for $i\in\{1,2\}$, and  $\sigma_{\{2\}}+\sigma_{\{1,2\}} \geq 1-\epsilon$. 

This concludes the converse  proof to \eqref{eq:E1}.

\subsubsection{The case $\epsilon_1< \epsilon_2$}

Choose nonnegative numbers $a_1, a_{1,2}, b_1, b_{1,2}, c_{1,2}$ satisfying
\begin{subequations}\label{eq:0}
	\begin{IEEEeqnarray}{rCl}
		a_{1} + a_{1,2} &\leq& \sigma_{\{1\}} \\
		b_{1} + b_{1,2} &\leq& \sigma_{\{1,2\}} \\
		c_{1,2} &\leq& \sigma_{\{2\}} \\
		a_{1,2} +b_{1,2}= b_{1,2} + c_{1,2}  &=& 1-\epsilon_2\\
		a_{1} + b_{1} &=& \epsilon_2-\epsilon_1.
	\end{IEEEeqnarray}
\end{subequations}
Notice that this set of (in)equalities is equivalent to the two equalities  $a_{1,2}=c_{1,2} = 1- \epsilon_2-b_{1,2}$ and  $a_1= \epsilon_2-\epsilon_1-b_1$ and the three inequalities: 
\begin{subequations}\label{eq:i}
	\begin{IEEEeqnarray}{rCl}
		1- \epsilon_1 - b_{1,2}-b_1 &\leq& \sigma_{\{1\}} \\
		b_{1} + b_{1,2} &\leq& \sigma_{\{1,2\}} \\
		1- \epsilon_2- b_{1,2}  &\leq& \sigma_{\{2\}}.
	\end{IEEEeqnarray}
\end{subequations}
Through the Fourier-Motzkin Elimination (FME) Algorithm, it can be verified that above three inequalities \eqref{eq:i} have a nonnegative solution pair $(b_{1}, b_{1,2})$, with corresponding nonnegative values for $a_{1,2},c_{1,2}, a_1$, whenever
\begin{subequations}\label{eq:two}
\begin{IEEEeqnarray}{rCl}
0 & \leq & \sigma_{\mathcal{I}}, \,\, \quad \qquad \qquad  \mathcal{I} \in \mathcal{P}(2),\\
		1- \epsilon_i &\leq& \sigma_{\{i\}} +\sigma_{\{1,2\}}, \quad i\in\{1,2\},\\
	0 & \leq &	\epsilon_2- \epsilon_1,
	\end{IEEEeqnarray}
\end{subequations}
which hold by assumption, see \eqref{eq:E3}. The existence of the desired nonnegative numbers $a_1, a_{1,2}, b_1, b_{1,2}, c_{1,2}$ satisfying \eqref{eq:0}  is thus established. 

With the chosen numbers, we form 
\begin{subequations}
	\begin{IEEEeqnarray}{rCl}
		\tilde{R}_{\{1,2\},1} & := &\max \bigg\{\frac{a_{1,2}R_{\{1\},1} +b_{1,2}R_{\{1,2\},1}}{1-\epsilon_2}, \nonumber \\
		&& \hspace{12mm}\frac{b_{1,2}R_{\{1,2\},1} +c_{1,2}R_{\{2\},1}}{1-\epsilon_2} \bigg\}, \IEEEeqnarraynumspace\\
		\tilde{R}_{\{1,2\},2} & := & \frac{b_{1,2}R_{\{1,2\},2} +c_{1,2}R_{\{2\},2}}{1-\epsilon_2} , \IEEEeqnarraynumspace\\
		\tilde{R}_{\{1\},1} & := & \frac{a_{1}R_{\{1\},1} +b_{1}R_{\{1,2\},1}}{\epsilon_2-\epsilon_1}. 
	\end{IEEEeqnarray}
\end{subequations}
We show that  exponents $(\theta_1, \theta_2)$ and rates $\tilde{R}_{\{1\},1}$, $\tilde{R}_{\{1,2\},1}$ and $\tilde{R}_{\{1,2\},2}$  satisfy constraints \eqref{eq:E2}. To this end, notice that 
\begin{IEEEeqnarray}{rCl}
	\theta_1 & \leq & \min\{\fxy(R_{\{1\},1}), \fxy({R}_{\{1,2\},1})\} \\
	&\stackrel{(a)}{\leq}& \frac{a_1 \fxy(R_{\{1\},1}) + b_1 \fxy({R}_{\{1,2\},1})}{\epsilon_{2}-\epsilon_{1}} \\
	&\stackrel{(b)}{\leq}& \fxy\left (\frac{a_1 R_{\{1\},1} + b_1 {R}_{\{1,2\},1}}{\epsilon_{2}-\epsilon_{1}}\right)\\
	&\stackrel{(c)}{\leq}& \fxy\left(\tilde{R}_{\{1\},1}\right),\label{eq:b1}
\end{IEEEeqnarray}
where $(a)$ holds because the minimum is smaller than any linear combination and because $a_1+b_1=\epsilon_2-\epsilon_1$; $(b)$ holds by the concavity of the function $\fxy(\cdot)$; and $(c)$ holds by the definition of rate $R_{\{1\},1}$. In a similar way we have:
\begin{IEEEeqnarray}{rCl}
	\theta_1 & \leq & \min\left\{\fxy\left(R_{\{1\},1}\right), \fxy\left({R}_{\{1,2\},1}\right)\right\} \\
	&{\leq}& \frac{a_{1,2} \fxy\left(R_{\{1\},1}\right) + b_{1,2} \fxy\left({R}_{\{1,2\},1}\right)}{1-\epsilon_{2}} \\
	&{\leq}& \fxy\left(\frac{a_{1,2} R_{\{1\},1} + b_{1,2} {R}_{\{1,2\},1}}{1-\epsilon_{2}}\right)\\
	&{\leq}& \fxy\left(\tilde{R}_{\{1,2\},1}\right),\label{eq:b2}
\end{IEEEeqnarray}
where the last step holds by the monotonicity of the function $\fxy(\cdot)$ and because by definition $\tilde{R}_{\{1,2\},1}\geq \frac{a_{1,2} R_{\{1\},1} + b_{1,2} {R}_{\{1,2\},1}}{1-\epsilon_{2}}$. Thus, by \eqref{eq:b1} and \eqref{eq:b2}:
\begin{IEEEeqnarray}{rCl}\label{eq:wtp1}
	\theta_1 & \leq & \min\left\{\fxy\left(\tilde{R}_{\{1\},1}\right), \fxy\left(\tilde{R}_{\{1,2\},1}\right)\right\}.
\end{IEEEeqnarray}

We continue to notice
\begin{IEEEeqnarray}{rCl}
	\theta_2 & \leq & \min\Big\{\fxy\left(R_{\{1,2\},1}\right)+\fyz\left(R_{\{1,2\},2}\right), \nonumber \\
	&& \hspace{12mm}\fxy\left(R_{\{2\},1}\right)+\fyz\left(R_{\{2\},2}\right)\Big\} \\
	&\stackrel{(d)}{\leq}& \frac{b_{1,2} \fxy\left(R_{\{1,2\},1}\right) + b_{1,2} \fyz\left({R}_{\{1,2\},2}\right)}{1-\epsilon_{2}} \nonumber \\
	&& + \frac{c_{1,2} \fxy\left(R_{\{2\},1}\right) + c_{1,2} \fyz\left({R}_{\{2\},2}\right)}{1-\epsilon_{2}} \\
	&\stackrel{(e)}{\leq}& \fxy\left(\frac{b_{1,2} R_{\{1,2\},1} + c_{1,2} {R}_{\{2\},1}}{1-\epsilon_{2}}\right) \nonumber \\ 
	&& + \fyz\left(\frac{b_{1,2}R_{\{1,2\},2} +c_{1,2}R_{\{2\},2}}{1-\epsilon_{2}}\right)\\
	&\stackrel{(f)}{\leq}& \fxy\left(\tilde{R}_{\{1,2\},1}\right) + \fyz\left(\tilde{R}_{\{1,2\},2}\right),\label{eq:wtp2}
\end{IEEEeqnarray}
where $(d)$ holds because the minimum is smaller than any linear combination and because $b_{1,2}+c_{1,2}=1-\epsilon_2$; $(e)$ holds concavity of the functions $\fxy(\cdot)$ and $\fyz(\cdot)$; and $(f)$ holds by the definitions of rates $R_{\{1,2\},1}$ and $R_{\{1,2\},2}$ and by the monotonicity of the function $\fxy(\cdot)$. 

From the rate constraints in \eqref{eq:E3}, we further obtain
\begin{IEEEeqnarray}{rCl}
	R_1 &\geq& \sigma_{\{1\}} R_{\{1\},1}+ \sigma_{\{2\}}  R_{\{2\},1}   + \sigma_{\{1,2\}}  R_{\{1,2\},1}   \\
	& \stackrel{(g)}{\geq}  & (a_{1} + a_{1,2}) R_{\{1\},1}+ c_{1,2}R_{\{2\},1}  +(b_{1}+b_{1,2}) R_{\{1,2\},1}\nonumber \\\\
	&=  & (\epsilon_2-\epsilon_1) \left(\frac{a_{1}R_{\{1\},1} +b_{1}R_{\{1,2\},1}}{\epsilon_2-\epsilon_1}\right) \nonumber\\
	&& + (1-\epsilon_2)  \left( \frac{a_{1,2}R_{\{1\},1} + c_{1,2}R_{\{2\},1} +b_{1,2}R_{\{1,2\},1} }{1-\epsilon_2}\right)\nonumber \\\\
	& \stackrel{(h)}\geq& (\epsilon_2-\epsilon_1) \tilde{R}_{\{1\},1}  + (1-\epsilon_2)   \tilde{R}_{\{1,2\},1} \IEEEeqnarraynumspace \label{eq:wtp3}
\end{IEEEeqnarray}
and 
\begin{IEEEeqnarray}{rCl}
	R_2 &\geq& \sigma_{\{1,2\}} R_{\{1,2\},2} + \sigma_{\{2\}} R_{\{2\},2}     \\
	& \stackrel{(g)}\geq& b_{1,2} R_{\{1,2\},2} + c_{1,2} R_{\{2\},2}     \\
	& =  & (1-\epsilon_2)  \left( \frac{b_{1,2}R_{\{1,2\},2} + c_{1,2}R_{\{2\},2}}{1-\epsilon_2}\right)\\
	& \stackrel{(h)}=&  (1-\epsilon_2)   \tilde{R}_{\{1,2\},2},\IEEEeqnarraynumspace\label{eq:wtp4}
\end{IEEEeqnarray}
where inequalities $(g)$ hold because $a_{1}+a_{1,2}  \leq \sigma_{\{1\}}$, $c_{1,2} \leq \sigma_{\{2\}}$, and $b_1+b_{1,2} \leq \sigma_{\{1,2\}}$, see \eqref{eq:0}; and $(h)$ holds by the definitions of rates $\tilde{R}_{\{1\},1}$, $\tilde{R}_{\{1,2\},1}$, and $\tilde{R}_{\{1,2\},2}$ and because 
\begin{IEEEeqnarray}{rCl}
	\lefteqn{a_{1,2}R_{\{1\},1} + c_{1,2}R_{\{2\},1} +b_{1,2}R_{\{1,2\},1} } \qquad \quad \nonumber\\
	&\geq& \max \{ a_{1,2}R_{\{1\},1} +b_{1,2}R_{\{1,2\}, 1}; \nonumber \\ 
	&& \qquad \;\; b_{1,2} R_{\{1,2\},1}+ c_{1,2}R_{\{2\},1}\}.
	\IEEEeqnarraynumspace
\end{IEEEeqnarray}

The desired converse result to \eqref{eq:E2} then follows by combining \eqref{eq:wtp1}, \eqref{eq:wtp2}, \eqref{eq:wtp3}, and \eqref{eq:wtp4}, and by noticing that by the monotonicity of the function $\fyz(\cdot)$ there is no loss in optimality to restrict to rates $\tilde{R}_{\{1,2\},2} = R_2/(1-\epsilon_2)$. 

\subsubsection{The case $\epsilon_1 > \epsilon_2$}
The proof is similar to the case $\epsilon_1<\epsilon_2$. We present it here for completeness.

Choose nonnegative numbers  $a_{1,2},b_2, b_{1,2}, c_2, c_{1,2}$ satisfying
\begin{subequations}\label{eq:fff}
	\begin{IEEEeqnarray}{rCl}
		a_{1,2} &\leq& \sigma_{\{1\}} \\ 
		b_{2} + b_{1,2} &\leq& \sigma_{\{1,2\}} \\
		c_2 + c_{1,2} &\leq& \sigma_{\{2\}} \\
		a_{1,2} +b_{1,2} = b_{1,2} + c_{1,2} &= & 1-\epsilon_1\\
		b_{2} + c_{2} &=& \epsilon_1-\epsilon_2, 
	\end{IEEEeqnarray}
\end{subequations}
which is equivalent to the three equalities $a_{1,2}=c_{1,2}=1-\epsilon_1 - b_{1,2}$ and $c_2=\epsilon_1-\epsilon_2- b_2$ and the three inequalities
\begin{subequations}\label{eq:ii}
	\begin{IEEEeqnarray}{rCl}
		1-\epsilon_1 - b_{1,2} &\leq& \sigma_{\{1\}} \\ 
		b_{2} + b_{1,2} &\leq& \sigma_{\{1,2\}} \\
		1-\epsilon_2 	-b_{2} - b_{1,2}  &\leq& \sigma_{\{2\}}.
	\end{IEEEeqnarray}
\end{subequations}
Through FME it can be shown that a nonnegative pair $(b_{2}, b_{1,2})$ satisfying \eqref{eq:ii} exists and the corresponding  values for $a_{1,2},c_{1,2},c_2$ are non-negative whenever
\begin{subequations}
\begin{IEEEeqnarray}{rCl}
0 & \leq & \sigma_{\mathcal{I}}, \,\, \quad \qquad \qquad  \mathcal{I} \in \mathcal{P}(2),\\
		1- \epsilon_i &\leq& \sigma_{\{i\}} +\sigma_{\{1,2\}}, \quad i\in\{1,2\},\\
	0 & \leq &	\epsilon_1- \epsilon_2,
	\end{IEEEeqnarray} 
\end{subequations}
which hold by assumption, see \eqref{eq:E3}. 

Define the new rates
\begin{IEEEeqnarray}{rCl}
	\tilde{R}_{\{1,2\},1} & := &\max \bigg\{\frac{a_{1,2}R_{\{1\},1} +b_{1,2}R_{\{1,2\},1}}{1-\epsilon_1}, \nonumber \\
	&& \hspace{12mm}\frac{b_{1,2}R_{\{1,2\},1} +c_{1,2}R_{\{2\},1}}{1-\epsilon_1} \bigg\}, \IEEEeqnarraynumspace\\
	\tilde{R}_{\{1,2\},2} & := & \frac{b_{1,2}R_{\{1,2\},2} +c_{1,2}R_{\{2\},2}}{1-\epsilon_1} , \IEEEeqnarraynumspace\\
	\tilde{R}_{\{2\},i} & := & \frac{b_{2}R_{\{1,2\},i}+c_{2}R_{\{2\},i}}{\epsilon_1-\epsilon_2}, \qquad i\in\{1,2\} \IEEEeqnarraynumspace
\end{IEEEeqnarray}
We show that the exponents $\theta_1, \theta_2$ and the rates $\tilde{R}_{\{2\},1}$,$\tilde{R}_{\{2\},2}$, $\tilde{R}_{\{1,2\},1}$ and $\tilde{R}_{\{1,2\},2}$  satisfy constraints \eqref{eq:E3_simplified}.
To this end, notice  that by similar arguments as in the preceding subsections:  
\begin{IEEEeqnarray}{rCl}
	\theta_1 & \leq & \min\big\{\fxy\big(R_{\{1\},1}\big), \fxy\big({R}_{\{1,2\},1}\big)\big\} \\
	&\leq& \frac{a_{1,2} \fxy\big(R_{\{1\},1}\big) + b_{1,2} \fxy\big({R}_{\{1,2\},1}\big)}{1-\epsilon_{1}} \\
	&\leq& \fxy\left(\frac{a_{1,2} R_{\{1\},1} + b_{1,2} {R}_{\{1,2\},1}}{1-\epsilon_{1}}\right)\\
	&\leq& \fxy\left(\tilde{R}_{\{1,2\},1}\right). \label{eq:wtp1b}
\end{IEEEeqnarray}
Moreover, 
\begin{IEEEeqnarray}{rCl}
	\theta_2 & \leq & \min\big\{\fxy\big(R_{\{1,2\},1}\big)+\fyz\big(R_{\{1,2\},2}\big), \nonumber \\
	&& \hspace{12mm}\fxy\big(R_{\{2\},1})+\fyz(R_{\{2\},2})\big\} \\
	&\leq& \frac{b_{1,2} \fxy\big(R_{\{1,2\},1}\big) + b_{1,2} \fyz\big({R}_{\{1,2\},2}\big)}{1-\epsilon_{1}} \nonumber \\
	&& + \frac{c_{1,2} \fxy\big(R_{\{2\},1}\big) + c_{1,2} \fyz\big({R}_{\{2\},2}\big)}{1-\epsilon_{1}} \\
	&\leq& \fxy\left(\frac{b_{1,2} R_{\{1,2\},1} + c_{1,2} {R}_{\{2\},1}}{1-\epsilon_{1}}\right) \nonumber \\ 
	&& + \fyz\left(\frac{b_{1,2}R_{\{1,2\},2} +c_{1,2}R_{\{2\},2}}{1-\epsilon_{1}}\right)\\
	&\leq& \fxy\left(\tilde{R}_{\{1,2\},1}\right) + \fyz\left(\tilde{R}_{\{1,2\},2}\right)\label{eq:d1}
\end{IEEEeqnarray}
and
\begin{IEEEeqnarray}{rCl}
	\theta_2 &\leq& \frac{b_{2} \fxy\big(R_{\{1,2\},1}\big) + b_{2} \fyz\big({R}_{\{1,2\},2}\big)}{\epsilon_1-\epsilon_{2}} \nonumber \\
	&& + \frac{c_{2} \fxy\big(R_{\{2\},1}\big) + c_{2} \fyz\big({R}_{\{2\},2}\big)}{\epsilon_1-\epsilon_{2}} \\
	&\leq& \fxy\left(\frac{b_{2} R_{\{1,2\},1} + c_{2} {R}_{\{2\},1}}{\epsilon_1-\epsilon_{2}}\right) \nonumber \\ 
	&& + \fyz\left(\frac{b_{2}R_{\{1,2\},2} +c_{2}R_{\{2\},2}}{\epsilon_1-\epsilon_{2}}\right)\\
	&\leq& \fxy\left(\tilde{R}_{\{2\},1}\right) + \fyz\left(\tilde{R}_{\{2\},2}\right).\label{eq:d2}
\end{IEEEeqnarray}
Combining \eqref{eq:d1} and \eqref{eq:d2} we obtain:
\begin{IEEEeqnarray}{rCl}
	\theta_2 &\leq& \min \bigg\{\fxy\left(\tilde{R}_{\{1,2\},1}\right) + \fxy\left(\tilde{R}_{\{1,2\},2}\right), \nonumber \\
	&& \hspace{1.4cm}  \fxy\left(\tilde{R}_{\{2\},1}\right) + \fxy\left(\tilde{R}_{\{2\},2}\right)\bigg\}. \label{eq:wtp2b}
\end{IEEEeqnarray}

From the rate constraints in \eqref{eq:E3}, inequalities \eqref{eq:fff}, and the definitions of the rates $\tilde{R}_{\{2\},1}, 	\tilde{R}_{\{2\},2} , 	\tilde{R}_{\{1,2\},1} , 	\tilde{R}_{\{1,2\},2}$,  we obtain:
\begin{IEEEeqnarray}{rCl}
	R_1 &\geq& \sigma_{\{1\}} R_{\{1\},1} + \sigma_{\{1,2\}} R_{\{1,2\},1} + \sigma_{\{2\}} R_{\{2\},1}    \\
	& \geq  &  a_{1,2}R_{\{1\},1}+(c_{2}+c_{1,2}) R_{\{2\},1}  +(b_{2} + b_{1,2}) R_{\{1,2\},1} \nonumber\\\\
	& =  & (\epsilon_1-\epsilon_2) \left(\frac{b_{2}R_{\{1,2\},1}+c_{2}R_{\{2\},1}}{\epsilon_1-\epsilon_2}\right) \nonumber\\
	&& + (1-\epsilon_1)  \left( \frac{a_{1,2}R_{\{1\},1} + c_{1,2}R_{\{2\},1} +b_{1,2}R_{\{1,2\},1} }{1-\epsilon_1}\right)\nonumber\\\\
	&\geq& (\epsilon_1-\epsilon_2) \tilde{R}_{\{2\},1}  + (1-\epsilon_1)   \tilde{R}_{\{1,2\},1}\IEEEeqnarraynumspace \label{eq:wtp3b}
\end{IEEEeqnarray}
and 
\begin{IEEEeqnarray}{rCl}
	R_2 &\geq& \sigma_{\s{1}{2}} R_{\{1,2\},2} + \sigma_{\s{2}} R_{\{2\},2}     \\
	& \geq  & (b_2+b_{1,2}) R_{\{1,2\},2} +(c_2+c_{1,2}) R_{\{2\},2}     \\ 
	& = & (1-\epsilon_1)  \left( \frac{b_{1,2}R_{\{1,2\},2} + c_{1,2}R_{\{2\},2}}{1-\epsilon_1}\right) \nonumber \\
	&& + (\epsilon_1-\epsilon_2)  \left( \frac{b_{2}R_{\{1,2\},2} + c_{2}R_{\{2\},2}}{1-\epsilon_1}\right)\\
	&=&  (1-\epsilon_1)   \tilde{R}_{\{1,2\},2} + (\epsilon_{1}-\epsilon_2)   \tilde{R}_{\{2\},2}. \IEEEeqnarraynumspace \label{eq:wtp4b}
\end{IEEEeqnarray}
Combining \eqref{eq:wtp1b}, \eqref{eq:wtp2b}, \eqref{eq:wtp3b}, and \eqref{eq:wtp4b} establishes the desired converse result in \eqref{eq:E3_simplified}.

	\section{A system with $K$-hops }\label{sec:ModelK}
We generalize our setup and results to $K$ hops, i.e., to $K-1$ relays. 
\subsection{System Model}

Consider a system with a transmitter T$_0$ observing the source sequence $Y_0^n$, $K-1$ relays labelled $\text{R}_1,\ldots,\text{R}_{K-1} $ and observing sequences $Y_{1}^n, \ldots, Y_{K-1}^n$, respectively, and a receiver R$_{K}$ observing sequence $Y_{K}^n$.  

The source sequences $(Y_0^n,Y_1^n,\ldots,Y_{K}^n)$ are distributed according to one of two distributions depending on a binary hypothesis $\mathcal{H}\in\{0,1\}$:
\begin{subequations}\label{eq:dist_Khop}
	\begin{IEEEeqnarray}{rCl}
		& &\textnormal{if } \mathcal{H} = 0: (Y_0^n,Y_1^n,\ldots,Y_{K}^n)  \textnormal{ i.i.d. } \sim \, P_{Y_0Y_1\cdots Y_{K}}; \label{eq:H0_dist_Khop}\\
		& &\textnormal{if } \mathcal{H} = 1: (Y_0^n,Y_1^n,\ldots,Y_{K}^n)  \textnormal{ i.i.d. } \sim\, P_{Y_0}\cdot P_{Y_1}\cdots P_{Y_K}.\nonumber\\
	\end{IEEEeqnarray} 
\end{subequations}

\begin{figure}[htbp]
	\centerline{\includegraphics[ scale=0.52]{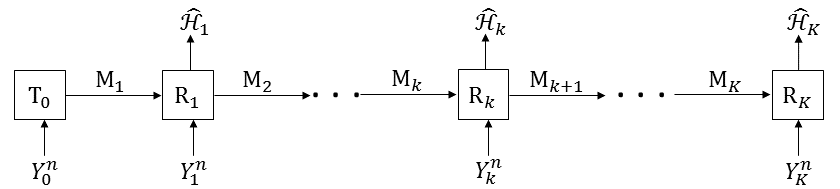}}
	\caption{Cascaded $K$-hop setup with $K$ decision centers.}
	\label{fig:Khop}
\end{figure}

Communication  takes place over $K$  hops as illustrated in Figure~\ref{fig:Khop}.  
The transmitter T$_{0}$  sends a message $\M_1 = \phi_0^{(n)}(Y_0^n)$ to the first relay R$_1$, which  sends a message $\M_2=\phi_1^{(n)}(Y_1^n,\M_1)$ to the second relay and so on. The communication is thus described by encoding functions 
\begin{IEEEeqnarray}{rCl}
	\phi_0^{(n)} &\colon& \mathcal{Y}_0^n \to \{0,1\}^\star \\
	\phi_k^{(n)} & \colon & \mathcal{Y}_k^n \times \{0,1\}^\star \to \{0,1\}^\star, \quad k\in\{1,\ldots, K-1\},\IEEEeqnarraynumspace 
\end{IEEEeqnarray} 
so that the produced message strings
\begin{IEEEeqnarray}{rCl}
	\M_1&=& \phi_0^{(n)}({Y}_0^n) \\
	\M_{k+1}& = & 	\phi_{k}^{(n)} ({Y}_k^n, \M_{k}) , \quad  k\in\{1,\ldots, K-1\},
\end{IEEEeqnarray} 
satisfy the expected-rate constraints 
\begin{equation}\label{eq:Ratek}
	\mathbb{E}\left[\mathrm{len}\left(\M_{k}\right)\right]\leq nR_{k}, \quad k \in \{1,\ldots,K\}.
\end{equation}

Each relay R$_1$, \ldots, R$_{K-1}$ as well as the receiver R$_K$, produces a guess of the hypothesis $\mathcal{H}$. 
These guesses are described by guessing functions 
\begin{equation}
	g_k^{(n)} \colon \mathcal{Y}_{k}^n \times \{0,1\}^{\star} \to \{0,1\}, \quad k\in\{1,\ldots, K\}, 
\end{equation}
where we request that the guesses
\begin{IEEEeqnarray}{rCl}
	\hat{\mathcal{H}}_{k,n} = g_k^{(n)}( Y_k^n, \M_{k}), 	\quad k\in\{1,\ldots, K\},
\end{IEEEeqnarray}  have type-I error probabilities 
\begin{IEEEeqnarray}{rCl}
	\alpha_{k,n} &\triangleq& \Pr[\hat{\mathcal{H}}_{k} = 1|\mathcal{H}=0], \quad k\in\{1,\ldots,K\},
\end{IEEEeqnarray}
not exceeding given thresholds $\epsilon_1,\epsilon_2,\ldots,\epsilon_K> 0$, and  type-II error probabilities
\begin{IEEEeqnarray}{rCl}
	\beta_{k,n} &\triangleq& \Pr[\hat{\mathcal{H}}_{k} = 0|\mathcal{H}=1],\quad  k\in\{1,\ldots,K\},
\end{IEEEeqnarray}
decaying to 0 exponentially fast with largest possible exponents.  

\medskip

\begin{definition} Given maximum type-I error probabilities $\epsilon_1,\epsilon_2,\ldots,\epsilon_K \in [0,1)$ and rates $R_1,R_2, \ldots, R_K \geq 0$. The exponent tuple $(\theta_1,\theta_2, \ldots, \theta_{K})$ is called \emph{$(\epsilon_1,\epsilon_2, \ldots, \epsilon_K)$-achievable} if there exists a sequence of encoding and decision functions $\big\{\phi_0^{(n)},\phi_1^{(n)}, \ldots, \phi_{K-1}^{(n)},g_1^{(n)},g_2^{(n)}, \cdots g_K^{(n)}\big\}_{n\geq 1}$ satisfying for each $k \in \{1,\ldots,K\}$:
	\begin{subequations}\label{eq:Kconditions}
		\begin{IEEEeqnarray}{rCl}
			\mathbb{E}[\text{len}(\M_k)] &\leq& nR_k,\label{rate_constraint_Khop} \\
			\varlimsup_{n \to \infty}\alpha_{k,n} & \leq& \epsilon_k,\label{type1constraint1_Khop}\\ 
			\label{thetaconstraint_Khop}
			\varliminf_{n \to \infty}  {1 \over n} \log{1 \over \beta_{k,n}} &\geq& \theta_k.
		\end{IEEEeqnarray}
	\end{subequations}
\end{definition}
\begin{definition}\label{def:EregionK} 
	The fundamental exponents region $\mathcal{E}^*(R_1,R_2,\ldots,R_K,\epsilon_1,\epsilon_2,\ldots,\epsilon_K)$ is defined as the closure of the set of all $(\epsilon_1,\epsilon_2,\ldots,\epsilon_K)$-achievable exponent pairs $(\theta_{1},\theta_{2},\ldots,\theta_{K})$ for given  rates $R_1,\dots, R_K\geq 0$.
\end{definition}

\subsection{Previous Results under Maximum-Rate Constraints}\label{sec:maxresults_K}
	The  $K$-hop hypothesis testing setup of Figure~\ref{fig:Khop} and Equations~\eqref{eq:dist_Khop}} was also considered in   \cite{salehkalaibar2020hypothesisv1}, but under   \emph{maximum-rate constraints}:
		\begin{equation}\label{eq:FixRatesK}
			\len(\M_i) \leq nR_i,	\qquad i\in\{1,\ldots, K\},
		\end{equation} 
		instead of the \emph{expected-rate constraints} \eqref{eq:Ratek}. The fundamental exponents region $\mathcal{E}_{\fix}^*(R_1,\ldots, R_K,\epsilon_1,\ldots,\epsilon_K)$  for this maximum-rate setup is defined analogously  to Definition~\ref{def:EregionK}, but with \eqref{eq:Ratek} replaced by \eqref{eq:FixRatesK}.
	
	The fundamental exponents region of this setup was only established for vanishing type-I error probabilities, i.e., when $\epsilon_1=\ldots =\epsilon_K=0$. 
	\begin{definition}
	For any $\ell\in\{1,\ldots, K\}$, define the function
	\begin{IEEEeqnarray}{rCl}
		\eta_\ell \colon  \mathbb{R}_0^+ & \to & \mathbb{R}_0^+ \\
		R & \mapsto&\max_{\substack{P_{U|Y_{\ell-1}}\colon \\R \geq I\left(U;Y_{\ell-1}\right)}} I\left(U;Y_\ell\right).
	\end{IEEEeqnarray}
\end{definition}
\medskip

The functions $\eta_1,\ldots, \eta_K$ are  concave and monotonically non-decreasing. The proof is analogous to the proof of Lemma~\ref{lem:concavity} presented in  Appendix~\ref{app:concavity}, and  omitted for brevity. Notice further that in the maximization determining $\eta_\ell(R)$ it suffices to consider distributions $P_{U|Y_{\ell-1}}$ on alphabets of sizes $|\mathcal{Y}_{\ell-1}|+1$, see \cite{Ahlswede}.

\medskip
			
	\begin{theorem}[Proposition 5 in \cite{salehkalaibar2020hypothesisv1}]\label{thm:fixedK}	
		The fundamental exponents region 	under the maximum-rate constraints \eqref{eq:FixRatesK} and vanishing type-I error constraints satisfies
		\begin{IEEEeqnarray}{rCl}
		\lefteqn{\mathcal{E}_{\fix}^*(R_1,\ldots,R_K,0,\ldots,0) } \nonumber \\
		&=& \left\{(\theta_{1},\ldots, \theta_{K}) \colon \theta_k \leq \sum_{\ell=1}^{k}  \eta_\ell(R_\ell), \; k \in \{1,\ldots, K\}\right\} \IEEEeqnarraynumspace
		\end{IEEEeqnarray}
	\end{theorem}	
	\medskip
Notice that in this $K$-hop setup, each decision center accumulates all the error exponents on the various links from the transmitter to this decision center. The fundamental exponents region is thus given by a $K$-dimensional hyperrectangle.  That means, each decision center can simultaneously achieve its optimal error exponent as if the other decision centers were not present in the system. 

We abbreviate $\mathcal{E}_{\fix}^*(R_1,\ldots,R_K,0,\ldots,0)$ by $\mathcal{E}_{\fix}^*(R_1,\ldots,R_K)$.

\subsection{Optimal Coding Scheme for $K$ Hops under Expected-Rate Constraints}\label{sec:schemeK}
Similarly to the two-hop scheme,   the terminals multiplex  different subschemes depending on the sequence $Y_0^n$ observed at the transmitter {\Tx}.  To this end, partition  the set $\mathcal{Y}_0^n$ into disjoint subsets $\D_{\emptyset}$ and $\{\D_{\mI}\}_{\mI \in \mathcal{P}(K)}$ so that the probabilities 
\begin{equation}\label{eq:sigmaI}
\sigma_{\mI} := \Pr[ Y_0^n \in \D_{\mI}]
\end{equation}
satisfy 
		\begin{subequations}\label{eq:sigma_eps}
\begin{IEEEeqnarray}{rCl}
1- \sum_{k \in \mathcal{S}} \epsilon_k &\leq&  \sum_{\substack{\mI \in \mathcal{P}(K)\colon \\  \mathcal{S} \subseteq  \mI }} \sigma_{\mI}, \quad \mathcal{S} \subseteq \{1,\ldots, K\},\\
\sum_{\mI \in \mathcal{P}(K)} \sigma_{\mI} &\leq& 1. 
\end{IEEEeqnarray}
		\end{subequations}


In our multiplexed schemes, the index $\mI$ of $\D_{\mI}$ indicates that if 
 {\Tx}'s observation $Y_0^n$ lies in  $\D_{\mI}$, then all decision centers R$_k$, for  $k\in\mI$, attempt to correctly guess hypothesis $\mathcal{H}$, while all decision centers R$_k$, for  $k\notin\mI$, simply declare $\hat{\mathcal{H}}_k=1$.
%
If $Y_0^n\in  \D_\emptyset$, then   \emph{all} decision centers R$_{1},\ldots,$ R$_{K}$ simply declare $\hmH=1$. 

The transmitter {\Tx} adds $K$ flag-bits to its message $\M_1$ to inform {\Rel} about the set $\D_{\mI}$  containing its observation $Y_0^n$, and thus about  the choice of the employed coding scheme. These flag-bits are forwarded by all relays R$_{1},\ldots,$ R$_{K-1}$  at the beginning of their messages $\M_2, \ldots, \M_K$ so as to pass the information to all terminals in the network.

We describe the different multiplexed coding schemes in more detail. Let $\ell_{\mI}^*$ be the largest index in set $\mI$:
\begin{equation}
\ell_\mI^*:= \max_{k\in \mI} k,
\end{equation}
and chooses a set of rates 
\begin{equation}
\{R_{\mI,\ell}\colon \quad \mI \in \mathcal{P}(K), \ \ell \in \{1,\ldots, \ell_{\mI}^*\} \} 
\end{equation} 
satisfying 
		\begin{IEEEeqnarray}{rCl}\label{eq:rate_constrainta3}
			R_{\ell} &>& \sum_{\substack{\mI \in \mathcal{P}(K) \colon \\ \ell_{\mI}^* \geq \ell}} \sigma_{\mI} \cdot R_{\mI,\ell} , \quad \ell \in \{1,\ldots,K\}.
		\end{IEEEeqnarray}	
	We will see that the choice of the various rates  determines the tradeoff between the different exponents $\theta_1,\ldots, \theta_K$. 
Rates $\{R_{{\mI}_\ell}\colon \ell \in \{1,\ldots, \ell_{\mI}^*\} \}$ are used in the subscheme employed when $Y_0^n \in \D_{\mI}$, where under this event only the messages on  the first $\ell_{\mI}^*$ links have positive  rates, while messages on the last $K-{\mI}_\ell^*$ links are of zero rate. The reason is that decision center R$_{{\mI}_\ell ^*+1},\ldots,$ R$_{K}$ simply declare $\hmH=1$ and thus  messages $\M_{\ell_{\mI}^*+1},\ldots, \M_K$  only have to convey the zero-rate information that $Y_0^n \in \D_{\mI}$.

\noindent\underline{\textit{Subscheme for $Y_0^n \in \D_{\emptyset}$:}}  
All terminals T$_0$ and R$_1$, \ldots, R$_{K-1}$ send the length-$K$ all-zero bit string  over the respective communication links: 
\begin{equation}
\M_1=\cdots = \M_K=[0,0,\ldots,0].
\end{equation} Upon receiving this all-zero flag, relays {R$_1$, \ldots, R$_{K-1}$} and  receiver {R$_K$} all declare 
\begin{equation}
\hat{\mathcal{H}}_{1} =\cdots= \hat{\mathcal{H}}_{K} = 1. 
\end{equation}
Communication is thus  only used to inform  the relays and the receiver about the scheme to employ, or equivalently the event $Y_0^n\in\D_{\emptyset}$, without providing any further information about the correct hypothesis.\\

\noindent\underline{\textit{Subscheme for $Y_0^n \in \D_{\mI}$, for $\mI \in \mathcal{P}(K)$:}} In this case, only decision centers R$_{k}$, for $k \in \mI$, attempt to correctly guess  hypothesis $\mH$; all other decision centers R$_{k}$, for $k\notin \mI$, directly declare $\hat{\mathcal{H}}_k=1$.

Terminals {\Tx}, {\Rel}, \ldots, R$_{\ell_{\mI}^*}$ apply a given $\ell_{\mI}^*$-hop hypothesis testing scheme with vanishing type-I error probabilities and respecting the maximum-rate constraints $R_{\mI,1}, \ldots, R_{\mI,\ell_{\mI}^*}$ on the first $\ell_{\mI}^*$ links. 
To inform all relays and the receiver about the scheme to use, terminals T$_0$, R$_1,\ldots,  $R$_{K-1}$ append a   $K$-length flag sequence describing set $\mI$ at the beginning of their messages. We propose that this flag sequence shows bit 1 at all positions $k\in\mI$ and bit 0 at  all positions $k\notin \mI$. Notice that Messages $\M_{\ell_{\mI}^*+1},\ldots, \M_{K}$ consist of  only the flag sequence. 

All decision centers R$_k$ with $k\in \mI$ declare the hypothesis indicated by the employed multi-hop hypothesis testing scheme. The remaining  decision centers R$_k$ with $k\notin \mI$ simply declare
\begin{equation}
\hmH_{k}=1, \quad k \notin \mI.
\end{equation}

	\underline{Analysis:}	By \eqref{eq:sigmaI} and \eqref{eq:rate_constrainta3}, and because transmission of $K$ bits hardly changes the rate for sufficiently large blocklengths, the proposed overall scheme respects the expected-rate constraints $R_1, \ldots, R_K$ on the $K$ links for large values of $n$. 
	Appendix~\ref{app:schemeK} proves that  when the optimal multi-hop hypothesis testing schemes with vanishing type-I error probability \cite{salehkalaibar2020hypothesisv1} are used as the various subschemes, then the overall scheme satisfies the type-I error constraints $\epsilon_1, \ldots, \epsilon_K$  and achieves the error exponents in the following Theorem~\ref{thm3}.

\subsection{Results on the Exponents Region}
%
%

\begin{theorem}\label{thm3}
	The fundamental exponents region $\mathcal{E}^*(R_1,\ldots, R_K,\epsilon_1,\ldots,\epsilon_K)$ is 
	equal to the set of all nonnegative tuples ($\theta_{1},\ldots,\theta_{K}$)  satisfying  
	\begin{subequations}\label{eq:E1_Khop}
		\begin{IEEEeqnarray}{rCl}\label{eq:56a}
			\theta_{k} &\leq& \min_{\substack{\mI\in\mathcal{P}(K) \colon \\ k\in\mI}}
			\sum_{\ell =1}^{k}
			\eta_{\ell}(R_{\mI,\ell}), \IEEEeqnarraynumspace
		\end{IEEEeqnarray}
		for some  nonnegative rates $\{R_{\mI,1}, \ldots, R_{\mI, \ell^*_{\mI}}\}_{\mI\in\mathcal{P}(K)}$ and nonnegative numbers $\{\sigma_{\mI}\}_{\mI \in \mathcal{P}(K)}$  satisfying
		\begin{IEEEeqnarray}{rCl}
			R_k &\geq& \sum_{\substack{\mI \in \mathcal{P}(K)\colon\\ k  \leq \ell^*_{\mI}}}  \sigma_{\mI}  \cdot R_{\mI, k} , \quad k\in\{1,\ldots,K\},\label{eq:Rkkk}\IEEEeqnarraynumspace\\
			\max\left\{0, 1- \sum_{k \in \mathcal{S}} \epsilon_k\right\}  &\leq&  \sum_{\substack{\mI \in \mathcal{P}(K)\colon \\  \mathcal{S} \subseteq  \mI }} \sigma_{\mI}, \quad \mathcal{S} \subseteq \{1,\ldots, K\},\label{eq:sigma_sum_constraint_lb}\\
			\sum_{\mI \in \mathcal{P}(K)} \sigma_{\mI} &\leq& 1. \label{eq:sigma_sum_constraint_ub}
		\end{IEEEeqnarray}
	\end{subequations}
\end{theorem}
\begin{IEEEproof} 
	Achievability is based on the coding scheme presented in the previous subsection and analyzed in Appendix~\ref{app:schemeK}. The converse is proved in the next Section~\ref{K_hop_converse}. 
\end{IEEEproof}

Similar observations apply to  the general Theorem~\ref{thm3} as  for  $K=2$. In particular, irrespective of the ordering of the permissible type-I error probabilities,  the largest exponent achievable  at a decision center $k$ is given by 
\begin{equation}\label{eq:theta_max}
\theta_{k,\max}:=  \sum_{\ell=1}^{k}  \eta_{\ell}\left(\frac{R_\ell}{1-\epsilon_k}\right). 
\end{equation}
It coincides with the optimal exponent under maximum-rate constraint  and  vanishing type-I error probabilities, see  Theorem~\ref{thm:fixedK}, but where the rates are boosted by the factor $(1-\epsilon_k)^{-1}$. 
In fact, $\theta_k=\theta_{k,\max}$ is achieved by choosing  the first $k$ rates as:\footnote{This choice assumes that the ordering \eqref{eq:orderingK} is strict, i.e., no two $\epsilon$-values coincide. Moreover, when some of the available rates $R_1,\ldots, R_{k}$ are sufficiently large so as to saturate the functions $\eta_\ell(R_\ell)$, then other choices are possible.}
\begin{IEEEeqnarray}{rCl}\label{eq:FixK}
R_{\mI,\ell}&= & \frac{R_{\ell}}{1-\epsilon_k}, \quad k\in\mI, \; \ell \in\{1,\ldots, k\}.  \IEEEeqnarraynumspace\label{eq:rate_choice_maximizing_theta_k}
\end{IEEEeqnarray}
%
This choice imposes that $\sigma_{\mI} R_{\mI,\ell}=0$ for all $\mI$ not containing $k$ and all $\ell \in\{1,\ldots, k\}$. As a consequence, the optimal performance for a decision center R$_{k'}$, for $k'<k$, is 
\begin{IEEEeqnarray}{rCl}
\theta_{k'} &=&  \sum_{\ell=1}^{k'}  \eta_{\ell}\left(\frac{R_\ell}{1-\epsilon_k}\right), \quad \textnormal{ if } \epsilon_{k'} > \epsilon_{k} \label{eq:Rk'}\\
\theta_{k'} &=& 0, \quad \textnormal{ if } \epsilon_{k'} < \epsilon_{k},\label{eq:Rk''}
\end{IEEEeqnarray}
where the performance  in \eqref{eq:Rk'} is obtained by setting $\sigma_{\mI}=0$ for all $\mI$ containing  an index $k'<k$  with $\epsilon_k'>\epsilon_k$ and by setting the corresponding  rates to infinity. Notice that $\sigma_{\mI}$ cannot be chosen equal to 0 for all sets $\mI$ containing index $k'<k$ when $\epsilon_{k'}<\epsilon_k$ because  Constraint \eqref{eq:sigma_sum_constraint_lb} implies that at least one of these $\sigma$-values is positive, which by $\sigma_{\mI} R_{\mI,\ell}=0$ implies that the corresponding rates $R_{\mI,\ell}=0$, for all $\ell=1,\ldots,k$,  causing $\theta_{k'}$ to degrade to 0. 
We conclude that under \eqref{eq:theta_max}, for  any $k'<k$, when $\epsilon_k' \geq \epsilon_k$ then exponent  $\theta_{k'}$ is degraded from its maximum value because all rates are only boosted by the factor $(1-\epsilon_k)^{-1}$ and not by the larger factor $(1-\epsilon_{k'})^{-1}$, and when $\epsilon_k' < \epsilon_k$   the exponent $\theta_{k'}$ 
 completely degrades to 0.  

With appropriate choices for the rates on the last $(K-k)$ links, different  tradeoffs between the exponents $\theta_{k+1},\ldots,\theta_{K}$ can be achieved. In particular, it is possible that an exponent $\theta_{k'}$, for $k'>k$, experiences its maximum rate-boost $(1-\epsilon_{k'})^{-1}$ on some of these links. On the first $k$ links, any exponent $\theta_{k+1},\ldots, \theta_K$  experiences a rate-boost of $(1-\epsilon_k)^{-1}$ if the corresponding $\epsilon_{k'}>\epsilon_k$, whereas the contributions of the first $k$ links  completely degrade to 0 if  $\epsilon_{k'}<\epsilon_k$.

\medskip

Further notice the following property of the region in Theorem~\ref{thm3}.
\begin{lemma}\label{lem:moving}
Consider a set of  nonnegative numbers $\{R_{\mI,1}, \ldots, R_{\mI, \ell^*_{\mI}}\}_{\mI\in\mathcal{P}(K)}$ and  $\{\sigma_{\mI}\}_{\mI \in \mathcal{P}(K)}$  satisfying \eqref{eq:E1_Khop} for exponents $(\theta_1,\ldots, \theta_K)$. Let $\mI', \mI'' \in \mathcal{P}(K)$ and $\Gamma\in[0,\sigma_{\mI''}]$ be so that 
\begin{equation}\label{eq:set_inc}
\mI' \subseteq \mI''
\end{equation}
 and 
 \begin{equation}\label{eq:De}
\max \left\{0, 1- \sum_{k \in \mathcal{S}} \epsilon_k \right\}  + \Gamma \leq  \sum_{\substack{\mI \in \mathcal{P}(K)\colon  \\  \mathcal{S} \subseteq  \mI }} \sigma_{\mI}, \quad \mathcal{S} \subseteq \mI'', \mathcal{S} \nsubseteq \mI'.
 \end{equation}
 
 Then, the new nonnegative numbers
\begin{IEEEeqnarray}{rCl}
\tilde \sigma_{\mI'} & = & \sigma_{\mI'} + \Gamma  \label{eq:sigma1}\\
\tilde \sigma_{\mI''} & = & \sigma_{\mI''} - \Gamma  \label{eq:sigma2}\\
\tilde \sigma_{\mI} & = & \sigma_{\mI}, \quad \mI \in \mathcal{P}(K) \backslash \{ \mI', \mI''\}, \label{eq:new_sigma3}
\end{IEEEeqnarray}
and rates, for  $\ell\in\{1,\ldots, K\}$,
\begin{IEEEeqnarray}{rCl}
\tilde  R_{\mI',\ell} & = & \frac{ \sigma_{\mI'} \cdot R_{\mI',\ell} +  \Gamma \cdot R_{\mI'',\ell}}{ \tilde \sigma_{\mI'} } ,\\
\tilde  R_{\mI,\ell} & = & R_{\mI,\ell} , \quad \mI \in \mathcal{P}(K) \backslash \{\mI'\}. \label{eq:Rn}
\end{IEEEeqnarray}
also   satisfy \eqref{eq:E1_Khop} for exponents $(\theta_1,\ldots, \theta_K)$.
\end{lemma}
\begin{IEEEproof}
Above rate-definitions essentially only shift the term $\Gamma \cdot R_{\mI'',\ell}$ from $\sigma_{\mI''} R_{\mI'', \ell}$ to $\tilde{\sigma}_{\mI'}R_{\mI',\ell}$, and therefore the rate constraints \eqref{eq:Rkkk} remain valid also for the new numbers. Similarly, constraint \eqref{eq:sigma_sum_constraint_ub} remains valid since the sum of all $\sigma$-values is preserved. Notice further that the $\sigma$-values included in Constraint \eqref{eq:sigma_sum_constraint_lb}  for $\mathcal{S} \nsubseteq \mI''$ remain unchanged by \eqref{eq:new_sigma3} and for $\mathcal{S} \subseteq \mI'$ their sum is preserved  by \eqref{eq:sigma1} and \eqref{eq:sigma2}. For $\mathcal{S} \nsubseteq \mI'$ but $\mathcal{S} \subseteq \mI''$, Constraint  \eqref{eq:sigma_sum_constraint_lb} is satisfied by Assumption \eqref{eq:De}. It remains to check the validity of \eqref{eq:56a} for the new rate-values. By \eqref{eq:Rn} the constraint remains unchanged for all $k \notin \mI'$. For $k\in \mI'$, we notice that by \eqref{eq:set_inc} the minimum in \eqref{eq:56a} includes both sets $\mI'$ and $\mI''$ and this minimum cannot be smaller for the new rates because: 
 \begin{IEEEeqnarray}{rCl}
 \lefteqn{\min \left\{ \sum_{\ell=1}^{k} \eta_\ell\left( {R}_{\mI', \ell}\right) , \; \sum_{\ell=1}^{k} \eta_\ell\left( {R}_{\mI'', \ell}\right) \right\}} \quad \nonumber\\
& \leq	& \min \left\{ \sum_{\ell=1}^{k}\left( \frac{ \sigma_{\mI'} }{\tilde \sigma_{\mI'} }  \eta_\ell\left( {R}_{\mI', \ell}\right)+ \frac{ \Gamma}{\tilde \sigma_{\mI'} }   \eta_\ell\left( {R}_{\mI'', \ell}\right) \right), \right.  \nonumber \\
& & \hspace{1cm} \; \left. \sum_{\ell=1}^{k} \eta_\ell\left( {R}_{\mI'', \ell}\right) \right\}\\
&\leq & \min \left\{ \sum_{\ell=1}^{k} \eta_\ell\left( \tilde{R}_{\mI', \ell}\right) , \; \sum_{\ell=1}^{k} \eta_\ell\left(\tilde{R}_{\mI'', \ell}\right) \right\},
 \end{IEEEeqnarray}
 where the first inequality holds because the minimum of two numbers cannot exceed any convex combination of the numbers, and the second inequality holds  by the concavity and monotonicity of the functions $\{\eta_\ell(\cdot)\}_{\ell}$. 
\end{IEEEproof}

Above lemma indicates that when evaluating the fundamental exponents region $\mathcal{E}^*(R_1,\ldots, R_K,\epsilon_1,\ldots,\epsilon_K)$ in Theorem~\ref{thm3} one can restrict to sets of parameters $\{\sigma_{\mI}\}$ that satisfy some of the constraints \eqref{eq:sigma_sum_constraint_lb} with equality and set certain $\sigma$-values to $0$. In fact, we conjecture that the simplified expression for the exponents region  $\mathcal{E}^*(R_1,\ldots, R_K,\epsilon_1,\ldots,\epsilon_K)$  in Conjecture~\ref{cor:simplification_3hop_generalcase} ahead holds, where we define a permutation $\pi\colon \{1,\ldots, K\}\to \{1,\ldots, K\}$ that  orders the $\epsilon$-values in decreasing order:
\begin{equation} \label{eq:orderingK}
\epsilon_{\pi(1)} \geq \epsilon_{\pi(2) } \geq  \cdots \geq  \epsilon_{\pi(K)},
\end{equation} 
and sets $\epsilon_{\pi(0)}:=1$. We observe that the expression in Conjecture~\ref{cor:simplification_3hop_generalcase} is obtained from Theorem~\ref{thm3} by setting
\begin{IEEEeqnarray}{rCl}\label{eq:simp_par}
\sigma_{\s{\pi(i)}{\cdots}{\pi(K)}}  = \epsilon_{\pi(i-1)} - \epsilon_{\pi(i)}, \quad i\in\{1,\ldots, K\}, \IEEEeqnarraynumspace
\end{IEEEeqnarray}
and all other $\sigma$-values to $0$, and by renaming rates $R_{\s{\pi(i)}{\cdots}{\pi(K)},\ell}$ to $R_{i,\ell}$ and $\ell_{\s{\pi(i)}{\cdots}{\pi(K)}}^*$ to $\ell_i^*$. The region in Conjecture~\ref{cor:simplification_3hop_generalcase} is thus achievable.

\medskip
\begin{conjecture}\label{cor:simplification_3hop_generalcase}
	The fundamental exponents region $\mathcal{E}^*(R_1,\ldots, R_K,\epsilon_1,\ldots, \epsilon_K)$ is the set of all exponent tuples ($\theta_{1},\ldots,,\theta_K$)  that satisfy 
	\begin{subequations}\label{eq:E3_simplified_3hop_generalcase}
		\begin{IEEEeqnarray}{rCl}
			\theta_{k} &\leq& \min_{i\in\{1,\ldots, \pi(k)\}} \left[ \sum_{\ell=1}^{k}  \eta_{\ell}\left(\Ra{i}{\ell}\right)\right] , \quad k \in \{1,\ldots,K\}, \label{eq:E3_simplified_theta1_3hop_generalcase} \IEEEeqnarraynumspace
		\end{IEEEeqnarray}
		for some nonnegative rates $\{\Ra{i}{\ell}\}$ satisfying
		\begin{IEEEeqnarray}{rCl}
			R_{\ell} &\geq& \sum_{\substack{i\in\{1,\ldots, K\}  \colon \\ \ell_i^* \geq \ell}}  \left(\epsilon_{\pi(i-1)}-\epsilon_{\pi(i)}\right) \Ra{i}{\ell} , \quad \ell \in \{1,\ldots,K\}, \label{eq:E3_simplified_R_3Hop_generalcase} \nonumber\\
		\end{IEEEeqnarray}
		where 
		\begin{equation}
		\ell_i^*:=\max_{\ell} \{ \ell \colon \ell \in \{ \pi(i), \ldots, \pi(K)\}\}. 
		\end{equation}
	\end{subequations}
\end{conjecture}
Linking this conjecture to the coding scheme in the previous Subsection~\ref{sec:schemeK}, we observe that if it holds, then the optimal coding scheme only multiplexes $K+1$ coding schemes (instead of $2^K$ schemes as implied by Theorem~\ref{thm3}), where the $i$-th scheme is applied with probability $\epsilon_{\pi(i-1)} - \epsilon_{\pi(i)}$ and is intended only for the  decision centers with $(K-i+1)$-th smallest type-I error constraints.

\begin{proposition}\label{prop:K3}
Conjecture~\ref{cor:simplification_3hop_generalcase} holds for $K=3$.
\end{proposition}
\begin{IEEEproof}
As already mentioned, achievability of the region in \eqref{eq:E3_simplified_3hop_generalcase} for any value of $K$ follows by specializing the region in Theorem~\ref{thm3} to the parameter choice in \eqref{eq:simp_par} and setting all other $\sigma$-values to 0, and by renaming rates $R_{\s{\pi(i)}{\cdots}{\pi(K)},\ell}$ as $R_{i,\ell}$. The converse for $K=3$ is proved in Section~\ref{sec:Simp}.
\end{IEEEproof}
\medskip

\section{Converse Proof to Theorem~\ref{thm3}}\label{K_hop_converse}

%

Fix an exponent-tuple $(\theta_1,\ldots, \theta_K)$ in the exponents region $\mathcal{E}^*(R_1,\ldots, R_K, \epsilon_1,\ldots, \epsilon_K)$, and a sequence (in $n$) of encoding and decision functions $\{(\phi_0^{(n)}, \phi_1^{(n)}, \ldots, \phi_K^{(n)}, g_1^{(n)}, \ldots, g_K^{(n)})\}_{n\geq 1}$ achieving this tuple, i.e., satisfying constraints \eqref{eq:Kconditions}. 

Our proof relies on the following lemma:\vspace{2mm}

\begin{lemma}\label{lem:receiverconverseKhop}
 Fix a positive $\delta>0$ and  blocklength $n$ and a set $\D\subseteq \mathcal{Y}_0^n\times\mathcal{Y}_1^n\times\cdots\times\mathcal{Y}_{K}^n$ of  probability exceeding $\delta$, and let  the tuple ($\tilde{\M}_1,\tilde{\M}_2,\ldots,\tilde{\M}_K,\tilde{Y}_0^n,\tilde{Y}_1^n,\ldots,\tilde{Y}_K^n$)  follow the pmf 
	\begin{IEEEeqnarray}{rCl}
		\lefteqn{P_{{\tilde{\M}_1}{\tilde{\M}_2}\cdots {\tilde{\M}_K}\tilde{Y}_0^n\tilde{Y}_1^n\cdots\tilde{Y}_K^n}(\m_1,\m_2,\ldots,\m_K,y_0^n,y_1^n,\ldots,y_K^n) \triangleq} \quad \nonumber \\
		&& P_{Y_0^nY_1^n\cdots Y_K^n}(y_0^n,y_1^n,\ldots,y_K^n)\cdot{\mathbbm{1} \{(y_0^n,y_1^n,\ldots,y_{K}^n)\in \D\} \over P_{Y_0^nY_1^n\ldots Y_{K}^n}(\D)}\nonumber\\ && \quad \cdot{\mathbbm{1}\{\phi_1(y_0^n)=\m_1\}}
		\cdot{\mathbbm{1}\{\phi_2(y_1^n,\phi_1(y_0^n))=\m_2\}}\cdot\cdots\nonumber\\
		&& \quad \cdot{\mathbbm{1}\{\phi_K(y_{K-1}^n,\phi_{K-1}(y_{K-2}^n,\phi_{K-2}(\cdots,\phi_1(y_0^n)))=\m_K\}}. \nonumber\\ \label{pmftildedoubleprime2_lemma_Khop}
	\end{IEEEeqnarray}
	Further, define  the  auxiliary random variables 
\begin{IEEEeqnarray}{rCl}
	{U}_k &\triangleq& (\tilde{\M}_k,\tilde{Y}_0^{T-1}\tilde{Y}_1^{T-1},\ldots,\tilde{Y}_{K}^{T-1},T), \quad k \in \{1,\ldots,K\},\nonumber\\ \\
	\tilde{Y}_k& \triangleq&\tilde{Y}_{k,T}, \quad k \in \{0,1,\ldots,K\},
	\end{IEEEeqnarray}
	where $T$ is uniform over $\{1,\ldots,n\}$ and independent of the tuple ($\tilde{\M}_1,\tilde{\M}_2,\ldots,\tilde{\M}_K,\tilde{Y}_0^n,\tilde{Y}_1^n,\ldots,\tilde{Y}_K^n$).
	
	For any $k\in\{1,\ldots, K\}$
	the following (in)equalities hold:
	\begin{IEEEeqnarray}{rCl}
		H(\tilde{M}_k) &\geq& nI(U_{k};\tilde{Y}_{k-1}) + \log P_{Y_0^nY_1^n\ldots Y_{K}^n}(\D), \label{eq:r}\nonumber\\\\
		I(U_k ;\tilde{Y}_{k}|\tilde{Y}_{k-1}) &= &\o_{1,k}(n), \label{eq:s}
	\end{IEEEeqnarray}
	where $\o_{1,k}(n)$ is a function that tends to 0 as $n\to \infty$.
	
If further for some  $k\in\{1,\ldots, K\}$, decision center   R$_k$ decides on the null hypothesis for all tuples $(y_0^n,\ldots,y_{K}^n) \in \D$\footnote{Notice that once we fix the realizations of all observed sequences $Y_0^n,\ldots,Y_{K}^n$, the decision $\hat{\mathcal{H}}_k$ is either determinstically 0 or 1.},
	\begin{IEEEeqnarray}{rCl}\label{eq:lemma_cond1Khop}
	\hat{\mathcal{H}}_{k}=0, 
	\end{IEEEeqnarray}
	then 
	\begin{IEEEeqnarray}{rCl}
		\lefteqn{-{1\over n}\log\Pr[\hat{\mathcal{H}}_{k} = 0| \mathcal{H}=1,(Y_0^n,\ldots,Y_{K}^n) \in \D]} \qquad \qquad \qquad \qquad  \nonumber\\
		&& \qquad \leq \sum_{\ell=1}^k
		I(U_\ell;\tilde{Y}_\ell) + \o_{2,k}(n),\label{eq:t} \IEEEeqnarraynumspace
	\end{IEEEeqnarray}
	where $\o_{2,k}(n)$ are functions that tend to $0$ as $n \to \infty$.
\end{lemma}
\begin{IEEEproof}
	See Appendix~\ref{app_lemmaK}.
\end{IEEEproof}	

%
%

We continue to prove Theorem~\ref{thm3}.
 Set $\mu_n=n^{-1/(K+1)}$.
Define for each index $k\in\{1,\ldots,K\}$ the set 
\begin{IEEEeqnarray}{rCl}\label{B_Khop}
\mathcal{B}_{k}&\triangleq &
	 \{(y_0^n, \ldots, y_{K}^n) \in \mathcal{T}_{\mu_n}^{(n)}(P_{Y_0\cdots Y_{K}}) \colon \quad   \hat{\mathcal{H}}_{k}=0 \} 
\end{IEEEeqnarray}
and for each subset $\mI \in \mathcal{P}(K)$ the set 
\begin{IEEEeqnarray}{rCl}\label{D_Khop}
	\lefteqn{\D_{\mI} \triangleq} \\
	&& \Big\{(y_0^n, \ldots, y_{K}^n) \in \mathcal{T}_{\mu_n}^{(n)}(P_{Y_0\cdots Y_{K}}) \colon \quad \hat{\mathcal{H}}_{k}=0 \quad   \forall k\in \mI  
	\qquad \textnormal{and }  \qquad \hat{\mathcal{H}}_{k}=1  \quad \forall k\notin \mI \}. 
\end{IEEEeqnarray}
Notice that the sets $\{\D_{\mI}\}_{\mI}$ are disjoint and 
\begin{equation}\label{eq:union}
	\bigcup_{\substack{\mI \in \mathcal{P}(K)\colon \\ k \in \mI }} \D_{\mI} = \mathcal{B}_{k}. 
\end{equation}

We continue to notice that by \cite[Remark to Lemma~2.12]{Csiszarbook} and the type-I error probability constraints in \eqref{type1constraint1_Khop}, for any  $k\in\{1,\ldots,K\}$:
\begin{IEEEeqnarray}{rCl}\label{eq:PB_KHop}
	P_{Y_0^nY_1^n\cdots Y_{K}^n}(\mathcal{B}_{k}) &\geq& 1 - \epsilon_k - {\vert{\mathcal{Y}_0}\vert \cdots\vert{\mathcal{Y}_{K}}\vert \over{4\mu_n^{2} n}}. \IEEEeqnarraynumspace
\end{IEEEeqnarray}
Defining
\begin{equation}
	\Delta_{\mI}:=	P_{Y_0^nY_1^n\cdots Y_{K}^n}(\D_{\mI}),
\end{equation}
we conclude by \eqref{eq:union}, by standard laws of probability, and the disjointness of the sets $\{\D_{\mI}\}_{\mI}$, that in the limit as $n \to \infty$, for any subset $\mathcal{S} \subseteq \{1,\ldots, K\}$:
\begin{equation}\label{eq:sum_KHop}
 \varliminf_{n \to \infty} \sum_{\substack{\mI \in \mathcal{P}(K)\colon \\ \mathcal{S} \subseteq  \mI }} \Delta_{\mI} \geq \max \left\{1-\sum_{k \in \mathcal{S}} \epsilon_k, 0\right\}. 
\end{equation}

We now apply Lemma~\ref{lem:receiverconverse} to every subset $\D_{\mathcal{I}}$, for $\mI \in \mathcal{P}(K)$ with  $\Delta_{\mI}> \delta$, for a given small $\delta$. This allows to conclude that for any such $\mI$ there exist random variables $\{U_{\mI, 1}, \ldots, U_{\mI, \ell_{\mI}^*}\}$ so that the random variables  ($\tilde{\M}_{\mI,1},\tilde{\M}_{\mI,2},\cdots,\tilde{\M}_{\mI,\ell_{\mI}^*},\tilde{Y}_{\mI,0}^n,\tilde{Y}_{\mI,1}^n,\cdots,\tilde{Y}_{\mI,K}^n$) defined in the lemma satisfy 
 for any $k \in \{1,\ldots, K\}$ the (in)equalities 
\begin{subequations}\label{eq:conditions_K_converse}
	\begin{IEEEeqnarray}{rCl}
		H(\tilde{M}_{\mI,k}) &\geq& nI(U_{\mI,k};\tilde{Y}_{k-1}) + \log \Delta_{\mathcal{I}},  \label{eq:Mji_Khop} \IEEEeqnarraynumspace\\
		I(U_{\mI,k}; \tilde{Y}_{\mI, k}|\tilde{Y}_{\mI,k-1}) & = & \o_{1,\mI,k}(n), \label{eq:MarkovK}
	\end{IEEEeqnarray}
	and for indices $k\in \mI$ moreover:
	\begin{IEEEeqnarray}{rCl}
		\lefteqn{	-{1\over n}\log\Pr[\hat{\mathcal{H}}_{k} = 0| \mathcal{H}=1,(Y_0^n,\ldots,Y_{K}^n)\in\D_{\mathcal{I}}]}  \nonumber\\
		&& \qquad \leq \sum_{\substack{\ell =1}}^k I(U_{\mI,\ell};\tilde{Y}_{\mI,\ell})+ \o_{2,\mI,k}(n), \IEEEeqnarraynumspace \hspace{1.5cm}
	\end{IEEEeqnarray}
\end{subequations}
where for each $(\mI,k)$   the functions $\o_{1,\mI,k}(n)$ and $\o_{2,\mI,k}(n)$ tend to $0$ as $n\to \infty$.


In the sequel, we assume that $\delta$ is very small all  $\Delta_\mI\geq \delta$ for all subsets $\mI$. Otherwise the proof is similar; details are omitted for brevity.

We continue with  the total law of probability to obtain:
\begin{IEEEeqnarray}{rCl}
	-\frac{1}{n} \log \beta_{k,n}  &\leq& \min_{\substack{\mI \in \mathcal{P}(K) \colon \\ k \in \mI}}  \sum_{\substack{\ell =1}}^k I(U_{\mI,\ell};\tilde{Y}_{\mI,\ell})  + \o_3(n),\label{eq:Rkthetaslasteq} \IEEEeqnarraynumspace
\end{IEEEeqnarray}
where  $\o_{3,k}(n)$ is a  function that tends to 0 as $n\to \infty$.
We further define the following random variables for $\mI \in \mathcal{P}(K)$ and $k \in\{1, \ldots, \ell_{\mI}^*\}$:
\begin{equation}
	{\tilde{L}_{\mI, k }} \triangleq \mathrm{len}({\tilde{\M}_{\mI,k}}).
\end{equation}
By the rate constraints  \eqref{eq:Ratek} and  the total law of expectations:
\begin{IEEEeqnarray}{rCl}
	nR_k &\geq&\sum_{\substack{\mI \in\mathcal{P}(K) \colon\\ \ell_{\mI}^* \geq k}}\mathbb{E}[\tilde{L}_{\mI,k}]\Delta_{\mI}, \label{ELI}
\end{IEEEeqnarray} 
and,  similarly to \eqref{M2ub'}, we obtain
\begin{IEEEeqnarray}{rCl}\label{eq:M2ubd}
	\sum_{\substack{\mI \in\mathcal{P}(K) \colon\\ \ell_{\mI}^* \geq k}} \Delta_{\mI} H(\tilde{\M}_{\mI,k}) \leq {nR_k} \left(1 +\sum_{\substack{\mI \in\mathcal{P}(K) \colon\\ \ell_{\mI}^* \geq k}}h_b\left({\Delta_{\mI} \over nR_k}\right)\right). \nonumber \\
	 \label{MKub'}
\end{IEEEeqnarray}
Then  combining  \eqref{eq:M2ubd}  with \eqref{eq:Mji_Khop} and \eqref{eq:Rkthetaslasteq}, and taking $n\to \infty$, we obtain that \begin{subequations}\label{eq:conditions_KHop}
	\begin{IEEEeqnarray}{rCl}			
		R_k &\geq  &\sum_{\substack{\mI \in\mathcal{P}(K) \colon\\ \ell_{\mI}^* \geq k}}\Delta_{\mI}^* \cdot I(U_{\mI,k}^*;\tilde{Y}_{k-1}^*)  ,\label{eq:Rhhh} \IEEEeqnarraynumspace \\
		\theta_k &\leq& \min_{\substack{\mI \in \mathcal{P}(K) \colon \\ k \in \mI}}    \sum_{\substack{\ell =1}}^k I(U_{\mI,\ell}^*;\tilde{Y}_{\mI,\ell}^*) \label{eq:thetakkkk}\IEEEeqnarraynumspace
	\end{IEEEeqnarray}
\end{subequations}
for some random variables   $(\tilde{Y}_{0}^*, \ldots, \tilde{Y}_K^*)\sim P_{Y_0Y_1\cdots Y_K}$ and $\{U^*_{\mI,1},\ldots, U^*_{\mI,\ell_{\mI}^*}\}_{\mI}$ that by \eqref{eq:MarkovK} satisfy the Markov chains 
\begin{equation}
	U^*_{\mI,k} \to Y^*_{k-1} \to Y^*_k
\end{equation}
and nonnegative numbers $\{\Delta_{\mI}^*\}_{\mI}$ that by \eqref{eq:sum_KHop} satisfy for any subset $\mathcal{S} \subseteq \{1,\ldots, K\}$:
\begin{equation}
	\sum_{\substack{\mI \in \mathcal{P}(K)\colon \\ \mathcal{S} \subseteq  \mI }} \Delta^*_{\mI} \geq \max \left\{1-\sum_{k \in \mathcal{S}} \epsilon_k, 0\right\}. 
\end{equation}

\section{Proof of Converse to Proposition~\ref{prop:K3}} \label{sec:Simp}
We start with two auxiliary lemmas.
\begin{lemma} \label{lemma_parameters}
 Let $K=3$. In Theorem~\ref{thm3} it suffices to consider values $\{\sigma_{\mI}\}_{\mI \in \mathcal{P}(3)}$ so that 
\begin{IEEEeqnarray}{rCl}\label{eq:eps1}
\sigma_{\s{1}{2}{3}}+\sigma_{\s{\pi(1)}{\pi(2)}}+\sigma_{\s{\pi(1)}{\pi(3)}}+\sigma_{\s{\pi(1)}} &=&1-\epsilon_{\pi(1)} \nonumber\\\\
 \sigma_{\s{\pi(1)}{\pi(2)}}  + \sigma_{\s{\pi(2)}} &\geq & \sigma_{\s{\pi(1)}{\pi(3)}} \nonumber \\ \label{eq:ass}
\end{IEEEeqnarray}
\end{lemma}
\begin{IEEEproof}
See  Appendix~\ref{app:proof_parameter_lemma}. \end{IEEEproof}

We thus continue with nonnegative numbers  $\{\sigma_{\mI}\}_{\mI \in \mathcal{P}(3)}$,  and $\{R_{\mI,1}, \ldots, R_{\mI, \ell^*_{\mI}}\}_{\mI\in\mathcal{P}(3)}$ satisfying 
\eqref{eq:E1_Khop} for $K=3$ as well as \eqref{eq:ass}. The proof of the desired proposition follows by the next lemma (which holds for any positive integer $K$) and by  an appropriate choice of parameters $\{c_{\mJ}\}$, see \eqref{eq:cJ} ahead.

\begin{lemma}\label{lem:above}
	Let  
	\begin{IEEEeqnarray}{C}
		\{c_{\mathcal{J}} \colon \; \mathcal{J} \in \mathcal{P}(K) \} ,\\
		\{ \delta_{\mathcal{I},\mathcal{J}} \colon\; \mathcal{I},\mathcal{J} \in \mathcal{P}(K) \; \textnormal{and} \; \mathcal{I}\cap \mathcal{J} \neq \emptyset\}
	\end{IEEEeqnarray}
	be sets of nonnegative integers   satisfying
	\begin{IEEEeqnarray}{rCl}\label{eq:assumption1}
		\sum_{\substack{\mathcal{J}\in\mathcal{P}(K)\colon\\\mathcal{I}\cap \mathcal{J} \neq \emptyset}}\delta_{\mathcal{I},\mathcal{J}} &\leq & \sigma_{\mathcal{I}},   \quad \mathcal{I}\in\mathcal{P}(K), \IEEEeqnarraynumspace
	\end{IEEEeqnarray}
	and  
	\begin{IEEEeqnarray}{rCl}\label{eq:assumption2}
		\sum_{\substack{\mathcal{I}\in \mathcal{P}(K) \colon \\k \in \mathcal{I}}}  \delta_{\mI, \mathcal{J}}\geq c_{\mathcal{J}}, \qquad \forall k \in \mathcal{J}, \; \mJ \in \mathcal{P}(K).
	\end{IEEEeqnarray}
	Then, the rates 
	\begin{IEEEeqnarray}{rCl}\label{eq:def1}
		\tilde{R}_{\mathcal{J},k} & := &\max_{\substack{j \in \mathcal{J} \colon \\ j \geq k }}  \sum_{\substack{\mathcal{I}\in \mathcal{P}(K) \colon\\ j \in \mathcal{I}}}\; \frac{\delta_{\mathcal{I},\mathcal{J}}}{c_{\mJ}} R_{\mathcal{I},k},  \nonumber\\
		 && \hspace{3cm}\quad k \leq \ell_{\mathcal{J}}^* , \mathcal{J}\in\mathcal{P}(K).\IEEEeqnarraynumspace
	\end{IEEEeqnarray}
	satisfy the following inequalities: 
	\begin{equation}\label{eq:thetai}
		\theta_k \leq    \min_{\substack{\mathcal{J} \in\mathcal{P}(K) \colon \\ k\in\mathcal{J}}}  \sum_{\ell=1}^{k}\eta_{\ell}\left(\tilde{R}_{\mathcal{J},\ell}\right), \qquad k\in\{1,\ldots, K\},
	\end{equation}
	and 
	\begin{equation}\label{eq:Rj}
		R_{k} \geq   \sum_{\substack{\mathcal{J} \in \mathcal{P}(K)\colon\\ k  \leq \ell^*_{\mathcal{J}}}} c_{\mathcal{J}} \cdot \tilde{R}_{\mathcal{J},k} , \qquad k\in \{1,\ldots, K\}.
	\end{equation}
\end{lemma}
\begin{IEEEproof}
We start by proving \eqref{eq:thetai}. By \eqref{eq:56a}, for any $k\in\{1,\ldots, K\}$ and any set $\mathcal{J}\subseteq \mathcal{P}(K)$ containing index $k$:
	\begin{IEEEeqnarray}{rCl}
		\theta_{k} &\leq& \min_{\substack{\mathcal{I} \in\mathcal{P}(K) \colon \\ k\in\mathcal{I}}} \sum_{\ell=1}^{k}\eta_{\ell}(R_{\mathcal{I},\ell}) \\
		&\stackrel{(a)}{\leq}& 
		\sum_{\substack{\mathcal{I} \in\mathcal{P}(K) \colon \\ k\in\mathcal{I}}}   \frac{ \delta_{\mathcal{I},\mathcal{J}}}{{\sum_{\substack{\mathcal{I} \in\mathcal{P}(K) \colon \\ k\in\mathcal{I}}} \delta_{\mathcal{I},\mathcal{J}}}} \ \cdot \sum_{\ell=1}^{k}\eta_{\ell}(R_{\mathcal{I},\ell}) \IEEEeqnarraynumspace\\
		&\stackrel{(b)}{\leq}&
		\sum_{\ell=1}^{k}\eta_{\ell}\left(\sum_{\substack{\mathcal{I} \in\mathcal{P}(K) \colon \\ k\in\mathcal{I}}}  \frac{\delta_{\mathcal{I},\mathcal{J}} }{\sum_{\substack{\mathcal{I} \in\mathcal{P}(K) \colon \\ k\in\mathcal{I}}} \delta_{\mathcal{I},\mathcal{J}}} \cdot R_{\mathcal{I},\ell} \right)\IEEEeqnarraynumspace\\
		& \stackrel{(c)}{\leq} & 		\sum_{\ell=1}^{k}\eta_{\ell}\left(\sum_{\substack{\mathcal{I} \in\mathcal{P}(K) \colon \\ k\in\mathcal{I}}}  \frac{\delta_{\mathcal{I},\mathcal{J}} }{c_{\mathcal{J}}} \cdot R_{\mathcal{I},\ell} \right)\IEEEeqnarraynumspace\\
		&\stackrel{(d)}{\leq}& 
		\sum_{\ell=1}^{k}\eta_{\ell}\left(\tilde{R}_{\mathcal{J},\ell}\right),\label{eq:ineqtheta}
	\end{IEEEeqnarray}
	where $(a)$ holds because the minimum of a set of numbers is never larger than any convex combination of these numbers; $(b)$ holds by the concavity of the functions $\eta_1(\cdot), \ldots, \eta_{k}(\cdot)$; $(c)$ holds by assumption \eqref{eq:assumption2} and by the monotonicity of the functions $\eta_1(\cdot), \ldots, \eta_{k}(\cdot)$; and $(d)$ holds by the definition of $\tilde{R}_{\mathcal{J},k}$ in \eqref{eq:def1} because $k\geq \ell$ and $k\in \mJ$ thus 
	$\ell \leq \ell_{\mJ}^*$.
	
	To prove \eqref{eq:Rj}, fix $k\in\{1,\ldots, K\}$ and for each subset $\mathcal{J}\subseteq \mathcal{P}(K)$ with $\ell_{\mathcal{J}}^* \geq k$  pick  an index $j_{\mathcal{J}} \in \mathcal{J}$ so that $j_{\mathcal{J}} \geq k$. Then, by \eqref{eq:Rkkk}:
	\begin{IEEEeqnarray}{rCl}
		R_k &\geq& 	\sum_{\substack{\mathcal{I} \in \mathcal{P}(K)\colon\\ k  \leq \ell^*_{\mathcal{I}}}}  \sigma_{\mathcal{I}}  \cdot R_{\mathcal{I}, k} , \IEEEeqnarraynumspace\\
		&\stackrel{(e)}{\geq}&	\sum_{\substack{\mathcal{I} \in \mathcal{P}(K)\colon\\ k  \leq \ell^*_{\mathcal{I}}}} \sum_{\substack{\mathcal{J}\in\mathcal{P}(K)\colon\\\mathcal{I}\cap \mathcal{J} \neq \emptyset}}\delta_{\mathcal{I},\mathcal{J}}  \cdot R_{\mathcal{I}, k} \\
		&=&	\sum_{\substack{\mathcal{J} \in \mathcal{P}(K)}} \; \sum_{\substack{\mathcal{I}\in\mathcal{P}(K)\colon\\k\leq \ell^*_{\mathcal{I}}\\ \mathcal{I}\cap \mathcal{J} \neq \emptyset}}\delta_{\mathcal{I},\mathcal{J}}  \cdot R_{\mathcal{I}, k} \\
		&\stackrel{(f)}{\geq}&
		\sum_{\substack{\mathcal{J} \in \mathcal{P}(K) \colon \\ 
				k \leq \ell_{\mathcal{J}}^*}}\; \sum_{\substack{\mathcal{I}\in\mathcal{P}(K)\colon\\k\leq \ell^*_{\mathcal{I}}\\ \mathcal{I}\cap \mathcal{J} \neq \emptyset}}\delta_{\mathcal{I},\mathcal{J}}  \cdot R_{\mathcal{I}, k} \\
		&\stackrel{(f)}{\geq}&
		\sum_{\substack{\mathcal{J} \in \mathcal{P}(K) \colon \\ 
				k \leq \ell_{\mathcal{J}}^*} }\sum_{\substack{\mathcal{I}\in \mathcal{P}(K) \colon\\ k \leq \ell_{\mI}^* \\j_{\mathcal{J}} \in \mathcal{I}}}  \delta_{\mathcal{I},\mathcal{J}}R_{\mathcal{I},k}\\
		&\stackrel{(g)}{=}&
		\sum_{\substack{\mathcal{J} \in \mathcal{P}(K) \colon \\ 
				k \leq \ell_{\mathcal{J}}^*} }\sum_{\substack{\mathcal{I}\in \mathcal{P}(K) \colon \\j_{\mathcal{J}} \in \mathcal{I}}}  \delta_{\mathcal{I},\mathcal{J}}R_{\mathcal{I},k}\\
		&{=}&
		\sum_{\substack{\mathcal{J} \in \mathcal{P}(K)\colon\\ k  \leq \ell^*_{\mathcal{J}}}} c_{\mathcal{J}} \cdot \frac{\sum_{\substack{\mathcal{I}\in \mathcal{P}(K) \colon \\ j_{\mathcal{J}} \in \mathcal{I}}} \delta_{\mathcal{I},\mathcal{J}}R_{\mathcal{I},k}}{c_{\mJ}},\label{eq:ineqrate}
	\end{IEEEeqnarray}
	where  $(e)$ holds by Assumption \eqref{eq:assumption1};  inequalities $(f)$ hold because we consider less summands and each summand is nonnegative (recall that $j_{\mathcal{J}}\in\mathcal{J}$); and finally  $(g)$ holds because the two conditions $j_{\mathcal{J}}\geq k$ and $j_{\mathcal{J}} \in \mathcal{I}$ imply that $\ell_{\mathcal{I}}^* \geq k$. 
	
	The proof of the lemma is concluded by recalling the definition of rate $\tilde{R}_{\mathcal{J},k}$ in \eqref{eq:def1}  and noting that Inequality \eqref{eq:ineqtheta} holds for any set $\mathcal{J}$ containing $k$ whereas Inequality \eqref{eq:ineqrate} holds for any 
	index $j_{\mathcal{J}}\in \mathcal{J}$ larger than $k$.\end{IEEEproof}

To obtain the desired simplification in Proposition~\ref{prop:K3} from Theorem~\ref{thm3}, define the subsets
\begin{IEEEeqnarray}{rCl}
	\mathcal{J}_{k}&:=&\{\pi(k), \ldots, \pi(K)\}, \quad k \in \{1,\ldots, K\},
\end{IEEEeqnarray}
and the values $\pi(0):=0$ and $\epsilon_0:=1$. Applying above Lemma~\ref{lem:above} to the choice 
\begin{equation}\label{eq:cJ}
	c_{\mathcal{J}}:= \begin{cases}
		\epsilon_{\pi(k-1)} - \epsilon_{\pi(k)}, & \quad \mathcal{J} = \mJ_k ,
		\\
		0,  &\quad  \text{otherwise},
	\end{cases}
\end{equation}
establishes the converse to   Conjecture~\ref{cor:simplification_3hop_generalcase} for general values of $K$, if one renames rates $\tilde{R}_{\mathcal{J}_k,\ell}$ as $\tilde{R}_{k,\ell}$. The proof is concluded by showing that above parameter choice is permissible, i.e., that  there exist nonnegative numbers $\{\delta_{\mathcal{I}, \mathcal{J}}\}$ satisfying conditions \eqref{eq:assumption1} and \eqref{eq:assumption2} for $\{c_{\mathcal{J}}\}$ in \eqref{eq:cJ}. For general  values of $K$ this seems cumbersome. 

For $K=3$, this can be achieved by means of the Fourier-Motzkin Elimination algorithm \cite{FME}, which shows the existence of nonnegative numbers $\{\delta_{\mathcal{I}, \mathcal{J}}\}$ satisfying conditions \eqref{eq:assumption1} and \eqref{eq:assumption2} for $\{c_{\mathcal{J}}\}$ in \eqref{eq:cJ}, whenever  (redundant conditions are omitted)
\begin{subequations}
\begin{IEEEeqnarray}{rCl}
\sigma_{\s{1}{2}{3}}+\sigma_{\s{\pi(1)}{\pi(2)}}+\sigma_{\s{\pi(1)}{\pi(3)}}+\sigma_{\s{\pi(1)}} & \geq & 1-\epsilon_{\pi(1)}\nonumber\\ \label{eq:ass1} \\
\sigma_{\s{1}{2}{3}}+\sigma_{\s{\pi(1)}{\pi(2)}}+\sigma_{\s{\pi(2)}{\pi(3)}}+\sigma_{\s{\pi(2)}} & \geq & 1-\epsilon_{\pi(2)} \nonumber\\ \label{eq:ass2}\\
\sigma_{\s{1}{2}{3}}+\sigma_{\s{\pi(1)}{\pi(3)}}+\sigma_{\s{\pi(2)}{\pi(3)}}+\sigma_{\s{\pi(3)}} & \geq & 1-\epsilon_{\pi(3)} \nonumber\\\label{eq:ass3}
\end{IEEEeqnarray}
and 
\begin{IEEEeqnarray}{rCl}
\lefteqn{ 2\sigma_{\s{1}{2}{3}}+2\sigma_{\s{\pi(1)}{\pi(2)}}+\sigma_{\s{\pi(1)}{\pi(3)}}+\sigma_{\s{\pi(2)}{\pi(3)}}}\quad \nonumber \\
 && +\sigma_{\s{\pi(1)}}+\sigma_{\s{\pi(2)}}+\sigma_{\s{\pi(3)}} \geq  1-\epsilon_{\pi(1)} +  1-\epsilon_{\pi(3)}. \nonumber \\
 \label{eq:ass4}\end{IEEEeqnarray}
\end{subequations}
Since Conditions \eqref{eq:ass1}--\eqref{eq:ass3} are satisfied by Assumption \eqref{eq:sigma_sum_constraint_lb} and Condition~\eqref{eq:ass4} is implied by \eqref{eq:ass1}, \eqref{eq:ass2}, and \eqref{eq:ass}, this concludes the proof for $K=3$ and thus establishes Proposition~\ref{prop:K3}.

	\section{Discussion and Outlook}
We derived the optimal type-II exponents region under expected-rate constraints for the $K$-hop network with $K$ decision centers (DC) for testing against independence and when the observations at the sensors respect some Markov chain. Equivalent simplified expressions were proved  for  $K=2$ and $K=3$, and  conjectured for arbitrary $K\geq 2$.  When the various DCs have different admissible type-I errors, then the derived exponents region illustrates a tradeoff  between the   error exponents that are simultaneously achievable at the various DCs. In general, an increase in exponents region is observed compared to the setup with maximum-rate constraints.  When all DCs have  equal permissible type-I error probability $\epsilon$, then the exponents region degenerates to a $K$-dimensional hypercube meaning that all DCs can simultaneously achieve their optimal error exponents. This optimal exponent  coincides with the optimal exponent under maximum-rate constraint where the rates have to be boosted by the factor $(1-\epsilon)^{-1}$.

To achieve the optimal tradeoff, a novel coding and testing scheme based on multiplexing and rate-sharing is proposed. The idea is that the transmitter chooses one of $2^K$ subschemes with appropriate probabilities and applies each subscheme with a well-chosen rate tuple.  Notice that the various rate-tuples determine the error exponents achieved at the various DCs, and thus steer the tradeoff between the error exponents at the different DCs. We multiplex schemes in a way that each of the subschemes is meant to help only a subset of the DCs in their decision; all other DCs simply raise an alarm so as not to compromise their type-II error exponents. The probabilities of the various subschemes then have to be chosen such that the probability of each DC raising an alarm does not exceed its permissible type-I error probability.  We conjecture that it suffices to multiplex only $K+1$ subschemes and that they should be chosen with probabilities determined by the type-I error probabilities. We managed to prove this conjecture for $K=2$ and $K=3$, but proofs for larger values of $K$ seem cumbersome. 

Notice that the proposed multiplexing and rate-sharing strategy is also optimal for other multi-terminal hypothesis testing setups, as we show in \cite{HWS22_BC}.

Our converse proof methods rely on applying $2^K$ change of measure arguments in parallel, and to separately bound the achievable error exponents and the required rates for each of them.  Moreover, we prove the desired Markov chains of the auxiliary random variables that arise in the typical single-letterization steps, in the asymptotic regimes of infinite blocklengths. We think that the proof technique of using asymptotic Markov chains in connection with change of measure arguments can also be used to prove strong converse results of source coding and channel coding theorems, see \cite{arxiv} for first results.

Interesting future research directions include results for other types of hypothesis testing, not necessarily testing against independence or not assuming a Markov chain under the null hypothesis. Other network structures are also of practical importance. Intriguing following-up questions exist also from an optimization perspective. For example, finding the optimal rate-distribution across the various links so as to maximize a weighted sum of the exponents. 
	 
	\section*{Acknowledgment}
	M. Wigger and M. Hamad have been supported by the European Union’s Horizon 2020 Research And Innovation Programme under grant agreement no. 715111.
	
	\vspace{-2mm}
	
	\bibliographystyle{ieeetr}
	\bibliography{references2}
	\vspace{-2mm}
	\appendices
	
	\section{Proof of Lemma~\ref{lem:concavity}: Concavity and Monotonicity of the Function $\fxy$}\label{app:concavity} 

The 	function $\fxy(R)$ is monotonically non-decreasing because larger values of $R$ imply larger optimization domains. Continuity follows simply by the continuity of mutual information. 

The concavity of $\fxy(R)$ follows by the following arguments.	
Consider rates $R$ and $\tilde{R}$, and let $U^*$ and $\tilde{U}^*$ be the corresponding solutions to the optimizations in the definition of $\fxy$. Pick any $\lambda \in [0,1]$, define
	$Q \sim \text{Bern} (\lambda)$ independent of $(Y_0,Y_1,U^*,\tilde{U}^*)$, and  set 
	\begin{equation}U_Q^* = \begin{cases} U^* & \textnormal{ if } Q=0\\ 
 \tilde{U}^* & \textnormal{ if } Q=1.
	\end{cases}
	\end{equation}
Defining the random variable  $V:= (U_Q^*,Q)$, we obtain
\begin{IEEEeqnarray}{rCl}
		\lefteqn{\lambda \cdot \fxy(R) + (1-\lambda)\cdot \fxy(\tilde{R})}\qquad\qquad \nonumber\\
		&=& \lambda I({U}^*;Y_1) + (1- \lambda) I(\tilde{U}^*;Y_1)\IEEEeqnarraynumspace\\
		&=& I(U_Q^*; Y_1|Q)\\
		&=& I(U_Q^*,Q; Y_1) \label{eq:Qindep}\\
		&=& I(V; Y_1) \label{eq:Vdef}\\
		& \leq & \fxy(I(V;Y_0)) \label{eq:deffxy}\\
		&\leq& \fxy(\lambda R + (1-\lambda)\tilde{R}) \label{eq:concavity}
	\end{IEEEeqnarray}
	where \eqref{eq:Qindep} holds because Q is independent of $Y_1$, \eqref{eq:deffxy} holds by the definition of the function $\fxy$, and \eqref{eq:concavity} holds by the monotonicity of the function $\fxy$ and the following set of (in)equalities:
\begin{IEEEeqnarray}{rCl}
		I(V;Y_0) &=& I(U_Q^*,Q; Y_0) = I(U_Q^*; Y_0|Q) \\
		&=& \lambda I(U^*;Y_0) + (1-\lambda) I(\tilde{U}^*;Y_0)\\
		&\leq&  \lambda R + (1-\lambda) \tilde{R}.
	\end{IEEEeqnarray}
	$\hfill \blacksquare$

\section{Analysis of the coding scheme in Subsection~\ref{sec:scheme_same} for $\epsilon_1=\epsilon_2=\epsilon$}\label{app1}

Consider the  two-hop scheme  employed when $Y_0^n \in \D_{\s{1}{2}}$, and let ${\hmH}_{\s{1}{2},1}$ and ${\hmH}_{\s{1}{2},2}$ denote the  guesses produced at {\Rel} and {\Rec} when employing this scheme for any $Y_0^n \in \mathcal{Y}_0^n$. Notice that by assumption the type-I error probabilities of this scheme tend to 0 as $n\to \infty$: 
\begin{equation}\label{eq:limtyp1}
\lim_{n\to \infty} \Pr[{\hmH}_{\s{1}{2},k} = 1|\mathcal{H}=0] =0, \quad k\in\{1,2\}.
\end{equation}

Noticing that when $Y_0^n \in  \D_{\emptyset}$, then  $\hat{\mH}_1=\hat{\mH}_2=1$, and applying the total law of probability, we can  write
 for $k\in\{1,2\}$:
\begin{IEEEeqnarray}{rCl}
	\alpha_{k,n} &=&\Pr[\hat{\mathcal{H}}_k=1| \mathcal{H}=0]\\
	&=& \Pr[\hat{\mathcal{H}}_k= 1, Y_0^n \in  \D_{\emptyset} |\mathcal{H} = 0]\nonumber \\
	&& + \Pr[\hat{\mathcal{H}}_k = 1, Y_0^n \in  \D_{\s{1}{2}}|\mathcal{H} = 0]\IEEEeqnarraynumspace\\
	&=& \Pr[Y_0^n \in  \D_{\emptyset}|\mathcal{H}=0] \nonumber\\
	&& +  \Pr[{\hmH}_{\s{1}{2},k}= 1, Y_0^n \in  \D_{\s{1}{2}}|\mathcal{H}=0]  \\
		&\leq & \Pr[Y_0^n \in  \D_{\emptyset}|\mathcal{H}=0] \nonumber\\
		&& +  \Pr[{\hmH}_{\s{1}{2},k}= 1|\mathcal{H}=0] 
\end{IEEEeqnarray}

Combining these inequalities with \eqref{eq:limtyp1}, and because in the limit $n\to \infty$ Inequality \eqref{eq:y0eps} turns into an equality, we conclude that   the overall scheme satisfies the type-I error constraints: 
\begin{equation}
\varlimsup_{n\to\infty}\alpha_{k,n} \leq \epsilon, \quad \quad k\in\{1,2\}.
\end{equation}

For the type-II error probabilities of the overall scheme  we observe for $k\in\{1,2\}$:
\begin{IEEEeqnarray}{rCl}
	\beta_{1,n} &=& \Pr[\hat{\mathcal{H}}_k=0|\mathcal{H}=1]\\
	&=& \Pr[\hat{\mathcal{H}}_k=0,Y_0^n\in  \D_{\emptyset}|\mathcal{H}=1] \nonumber\\
	&& + \Pr[\hat{\mathcal{H}}_k=0, Y_0^n\in  \D_{\s{1}{2}} |\mathcal{H}=1]\\
	&=&  \Pr[\hat{\mathcal{H}}_k=0, Y_0^n\in  \D_{\s{1}{2}} |\mathcal{H}=1]\\
	&=& \Pr[{\hmH}_{\s{1}{2},k}=0, Y_0^n\in  \D_{\s{1}{2}} |\mathcal{H}=1]\\
	&\leq& \Pr[{\hmH}_{\s{1}{2},k}=0|\mathcal{H}=1]. \label{BetaZLAPP1}
\end{IEEEeqnarray}

The type-II error exponents of the overall scheme are thus given by the error exponents of the  two-hop scheme employed under $Y_0^n \in\D_{\s{1}{2}}$. By \cite{Michele} and because the two-hop scheme has to have vanishing type-I error probabilities and respect the rate constraints $R_{\s{1}{2},1}$ and $R_{\s{1}{2},2}$, the exponents in \eqref{eq:E1} are proved achievable.

\section{Analysis of the coding scheme in Subsection~\ref{sec:scheme_larger} for  $\epsilon_2>\epsilon_1$}\label{app2}

Consider the  two-hop scheme  employed when $Y_0^n \in \D_{\s{1}{2}}$, and let ${\hmH}_{\s{1}{2},1}$ and ${\hmH}_{\s{1}{2},2}$ denote the  guesses produced at {\Rel} and {\Rec} when employing this scheme for any $y_0^n \in \mathcal{Y}_0^n$. Similarly, let ${\hmH}_{\s{1},1}$ and ${\hmH}_{\s{1},2}$ denote  the  guesses produced at {\Rel} and {\Rec} when employing the scheme for  $Y_0^n \in \D_{\s{1}}$, where we again extend the scheme to the entire set $\mathcal{{Y}}_0$.

By assumption, the type-I error probabilities of these schemes tend to 0 as $n\to \infty$: 
\begin{subequations}	\label{eq:limtyp1b}
	\begin{IEEEeqnarray}{rCl}
\lim_{n\to \infty} \Pr[{\hmH}_{\s{1},k} = 1|\mathcal{H}=0] &=&0, \quad k\in\{1,2\}\\
\lim_{n\to \infty} \Pr[{\hmH}_{\s{1}{2},k} = 1|\mathcal{H}=0] &=&0, \quad k\in\{1,2\}. 
\end{IEEEeqnarray}
\end{subequations}

Notice that for $Y_0^n \in  \D_{\emptyset}$ both {\Rel} and {\Rec} decide on   $\hat{\mH}_1=\hat{\mH}_2=1$. Applying the total law of probability, we can  write
\begin{IEEEeqnarray}{rCl}
	\alpha_{1,n} &=&\Pr[\hat{\mathcal{H}}_1=1| \mathcal{H}=0]\\
	&=& \Pr[\hat{\mathcal{H}}_1= 1, Y_0^n \in  \D_{\emptyset} |\mathcal{H} = 0]\nonumber \\
		&& + \Pr[\hat{\mathcal{H}}_1= 1, Y_0^n \in  \D_{\s{1}}|\mathcal{H} = 0]\nonumber\\
	&& + \Pr[\hat{\mathcal{H}}_1= 1, Y_0^n \in  \D_{\s{1}{2}}|\mathcal{H} = 0]\IEEEeqnarraynumspace\\
	&=& \Pr[Y_0^n \in  \D_{\emptyset}|\mathcal{H}=0]\nonumber\\
	&& +  \Pr[{\hmH}_{\s{1},1}= 1, Y_0^n \in  \D_{\s{1}}|\mathcal{H}=0] \nonumber\\
	&& +  \Pr[{\hmH}_{\s{1}{2},1}= 1, Y_0^n \in  \D_{\s{1}{2}}|\mathcal{H}=0]  \\
	&\leq & \Pr[Y_0^n \in  \D_{\emptyset}|\mathcal{H}=0] \nonumber\\
	&& +  \Pr[{\hmH}_{\s{1},1}= 1|\mathcal{H}=0]  \nonumber\\
	&& +  \Pr[{\hmH}_{\s{1}{2},1}= 1|\mathcal{H}=0] 
\end{IEEEeqnarray}
Combining this inequality with \eqref{eq:limtyp1b}, and because in the limit $n\to \infty$ Inequality \eqref{eq:De} turns into an equality, we conclude that   the overall scheme satisfies the type-I error constraint: 
\begin{equation}
\varlimsup_{n\to\infty}\alpha_{1,n} \leq \epsilon_1.
\end{equation}

Similarly we have: 
\begin{IEEEeqnarray}{rCl}
	\alpha_{2,n} &=&\Pr[\hat{\mathcal{H}}_2=1| \mathcal{H}=0]\\
	&=& \Pr[\hat{\mathcal{H}}_2= 1, Y_0^n \in  (\D_{\emptyset} \cup \D_{\s{1}}) |\mathcal{H} = 0]\nonumber\\
	&& + \Pr[\hat{\mathcal{H}}_2= 1, Y_0^n \in  \D_{\s{1}{2}}|\mathcal{H} = 0]\IEEEeqnarraynumspace\\
	&=& \Pr[Y_0^n \in ( \D_{\emptyset}\cup  \D_{\s{1}}) |\mathcal{H}=0] \nonumber\\
	&& +  \Pr[{\hmH}_{\s{1}{2},2}= 1, Y_0^n \in  \D_{\s{1}{2}}|\mathcal{H}=0]  \\
	&\leq &\Pr[Y_0^n \in ( \D_{\emptyset}\cup  \D_{\s{1}}) |\mathcal{H}=0]\\
	&& +  \Pr[{\hmH}_{\s{1}{2},2}= 1|\mathcal{H}=0] .
\end{IEEEeqnarray}
Combining this inequality with \eqref{eq:limtyp1b}, and because in the limit $n\to \infty$ Inequalities \eqref{eq:De0} and \eqref{eq:De} turn into  equalities, we conclude that   the overall scheme satisfies the type-I error constraint: 
\begin{equation}
\varlimsup_{n\to\infty}\alpha_{2,n} \leq \epsilon_2.
\end{equation}

For the relay's type-II error probability in  the overall scheme  we observe: 
\begin{IEEEeqnarray}{rCl}
	\beta_{1,n} &=& \Pr[\hat{\mathcal{H}}_1=0|\mathcal{H}=1]\\
	&=& \Pr[\hat{\mathcal{H}}_1=0,Y_0^n\in  \D_{\emptyset}|\mathcal{H}=1] \nonumber\\
	&& + \Pr[\hat{\mathcal{H}}_1=0, Y_0^n\in  \D_{\s{1}} |\mathcal{H}=1]\nonumber\\
	&& + \Pr[\hat{\mathcal{H}}_1=0, Y_0^n\in  \D_{\s{1}{2}} |\mathcal{H}=1]\\
	&=&  \Pr[\hat{\mathcal{H}}_1=0, Y_0^n\in  \D_{\s{1}} |\mathcal{H}=1]\nonumber \\
	& & +  \Pr[\hat{\mathcal{H}}_1=0, Y_0^n\in  \D_{\s{1}{2}} |\mathcal{H}=1]\\
	&=&  \Pr[{\hmH}_{\s{1},1}=0, Y_0^n\in  \D_{\s{1}} |\mathcal{H}=1]\nonumber \\
	& & +  \Pr[{\hmH}_{\s{1}{2},1}=0, Y_0^n\in  \D_{\s{1}{2}} |\mathcal{H}=1]\\
	&\leq& \Pr[{\hmH}_{\s{1},1}=0|\mathcal{H}=1] \nonumber\\
	& & + \Pr[{\hmH}_{\s{1}{2},1}=0|\mathcal{H}=1]. 
\end{IEEEeqnarray}
The relay's type-II error exponent of the overall scheme is  thus given by the minimum of the error exponents of the single-hop scheme employed under $Y_0^n \in \D_{\s{1}}$ and of  two-hop scheme employed under $Y_0^n \in\D_{\s{1}{2}}$. By \cite{Han} and \cite{Michele} and because these schemes  have vanishing type-I error probabilities and respect the rate constraints $R_{\s{1},1}$ and $(R_{\s{1}{2},1},R_{\s{1}{2},2})$, respectively, the exponent $\theta_1$  in \eqref{eq:E2} is proved achievable.

It remains to analyze the receiver's type-II error exponent:  
\begin{IEEEeqnarray}{rCl}
	\beta_{2,n} &=& \Pr[\hat{\mathcal{H}}_2=0|\mathcal{H}=1]\\
	&=& \Pr[\hat{\mathcal{H}}_2=0,Y_0^n\in  ( \D_{\emptyset} \cup  \D_{\s{1}} )|\mathcal{H}=1] \nonumber\\
	&& + \Pr[\hat{\mathcal{H}}_2=0, Y_0^n\in  \D_{\s{1}{2}} |\mathcal{H}=1]\\
	&=&  \Pr[\hat{\mathcal{H}}_2=0, Y_0^n\in  \D_{\s{1}{2}} |\mathcal{H}=1]\\
	&=&   \Pr[{\hmH}_{\s{1}{2},2}=0, Y_0^n\in  \D_{\s{1}{2}} |\mathcal{H}=1]\\
	&\leq& \Pr[{\hmH}_{\s{1}{2},2}=0|\mathcal{H}=1]. 
\end{IEEEeqnarray}
The receiver's type-II error exponent of the overall scheme is  thus given by the  error exponent of the two-hop scheme employed under $Y_0^n \in\D_{\s{1}{2}}$. By  \cite{Michele} and because this scheme has vanishing type-I error probabilities and respects the rate constraints  $(R_{\s{1}{2},1},R_{\s{1}{2},2})$, the exponent $\theta_2$  in \eqref{eq:E2} is proved achievable.

\section{Proof of Lemma~\ref{lem:receiverconverse}}\label{app_lemma1}

Note first that by \eqref{pmftildedoubleprime2_lemma}:
\begin{equation}\label{tildedivergencerelation2_lemma}
D(P_{\tilde{Y}_0^n\tilde{Y}_1^n}||P_{Y_0Y_1}^{n}) \leq \log{\Delta_n^{-1}},
\end{equation}
where we defined $\Delta_n \triangleq P_{Y_0^nY_1^n}(\D)$.

Further define  
\begin{IEEEeqnarray}{rCl}
\tilde{U}_{2,t} & \triangleq & (\tilde{\M}_2,\tilde{Y}_0^{t-1},\tilde{Y}_1^{t-1})\\
\tilde{U}_{1,t} &\triangleq & (\tilde{\M}_1,\tilde{Y}_0^{t-1},\tilde{Y}_1^{t-1}),
\end{IEEEeqnarray} and notice:
\begin{IEEEeqnarray}{rCl}
	H(\tilde{\M}_1)	&\geq& I(\tilde{\M}_1;\tilde{Y}_0^n\tilde{Y}_1^n) + D(P_{\tilde{Y}_0^n\tilde{Y}_1^n}||P_{Y_0Y_1}^n) + \log\Delta_{n}\IEEEeqnarraynumspace\label{m1entropylbstep1_lemma}\\
	&=& H(\tilde{Y}_0^n\tilde{Y}_1^n) + D(P_{\tilde{Y}_0^n\tilde{Y}_1^n}||P_{Y_0Y_1}^n) \nonumber \\ && -  H(\tilde{Y}_0^n\tilde{Y}_1^n|\tilde{\M}_1) + \log\Delta_{n}\IEEEeqnarraynumspace\\
	&\geq& n [H(\tilde{Y}_{0,T}\tilde{Y}_{1,T}) + D(P_{\tilde{Y}_{0,T}\tilde{Y}_{1,T}}||P_{Y_0Y_1})] \nonumber \\ && - \sum_{t=1}^{n} H(\tilde{Y}_{0,t}\tilde{Y}_{1,t}|\tilde{U}_{1,t})+ \log\Delta_{n}\IEEEeqnarraynumspace\label{m1entropylbstep4_lemma}\\
	&=& n [H(\tilde{Y}_{0,T}\tilde{Y}_{1,T}) + D(P_{\tilde{Y}_{0,T}\tilde{Y}_{1,T}}||P_{Y_0Y_1}) \nonumber \\ && - H(\tilde{Y}_{0,T}\tilde{Y}_{1,T}|\tilde{U}_{1,T},T)]+ \log\Delta_{n} \label{Tuniformdef_lemma}\\
	&\geq & n [H(\tilde{Y}_{0,T}\tilde{Y}_{1,T})  - H(\tilde{Y}_{0,T}\tilde{Y}_{1,T}|\tilde{U}_{1,T},T)] \nonumber \\
	&& + \log\Delta_{n} \IEEEeqnarraynumspace \\
	&=& n [I(\tilde{Y}_0\tilde{Y}_1;U_1)] +  \log{\Delta_{n}} \label{eq:HM1_LB_lemma_last_eq} \\
	&\geq& n \left[I(\tilde{Y}_0;U_1) + {1 \over n} \log{\Delta_{n}}\right] .\IEEEeqnarraynumspace \label{eq:HM1_LB_lemma}
\end{IEEEeqnarray}
Here, (\ref{m1entropylbstep1_lemma}) holds by (\ref{tildedivergencerelation2_lemma}); (\ref{m1entropylbstep4_lemma}) holds by the super-additivity property in \cite[Proposition 1]{tyagi2019strong}, by the chain rule, and by the definition of $\tilde{U}_{1,t}$ and  by defining $T$ uniform over $\{1,\dots,n\}$ independent of all other random variables; and \eqref{eq:HM1_LB_lemma_last_eq} holds by the definitions of $U_1,\tilde{Y}_0,\tilde{Y}_1$ in the lemma.

We can lower bound the entropy of $\tilde{\M}_2$ in a similar way to obtain:
\begin{IEEEeqnarray}{rCl}
	H(\tilde{\M}_2) &\geq& n \left[I(\tilde{Y}_1;U_2) + {1 \over n} \log{\Delta_{n}}\right]. 
\end{IEEEeqnarray}

We next  upper bound the error exponent at the receiver. To this end, fix a sequence of real numbers $\{\ell_n\}_{n=1}^\infty$ satisfying $\lim_{n \rightarrow \infty} {\ell_n/n} =0 $ and $\lim_{n \to \infty} {\ell_n/\sqrt{n}} =\infty$.
Define
\begin{equation}\label{eq:yz_acceptance}
\mathcal{A}_{Y_2}(\m_2) \triangleq \{y_2^n \colon g_2(\m_2,y_2^n) = 0\},
\end{equation}
and its Hamming neighborhood:
\begin{equation}
\hat{\mathcal{A}}_{Y_2}^{\ell_n}(\m_2) \triangleq \{\tilde{y}_2^n : \exists \, y_2^n \in \mathcal{A}_{Y_2}(\m_2) \textnormal{ s.t.} \; d_H(y_2^n,\tilde{y}_2^n)\leq\ell_n\}.
\end{equation}
Since by Condition \eqref{eq:lemma1_cond1}, 
\begin{equation}\label{blowupcond2_lemma}
P_{\tilde{Y}_2^n|\tilde{Y}_0^n\tilde{Y}_1^n}(\mathcal{A}_{Y_2}(\m_2)|y_0^n,y_1^n) \geq \eta , \quad \forall (y_0^n,y_1^n) \in \D,
\end{equation}
the blowing-up lemma \cite{MartonBU} yields
\begin{equation}\label{blowup2_lemma}
P_{\tilde{Y}_2^n|\tilde{Y}_0^n\tilde{Y}_1^n}(\hat{\mathcal{A}}_{Y_2}^{\ell_n}(\m_2)|y_0^n,y_1^n) \geq 1 - \zeta_n, \quad \forall (y_0^n,y_1^n) \in \D,
\end{equation}
for real numbers $\zeta_n > 0$ such that $\lim\limits_{n \to \infty} \zeta_n = 0$.\\
Define
\begin{equation}
{\mathcal{A}}_{Y_2} \triangleq \bigcup\limits_{\m_2 \in \mathcal{M}_2} \{\m_2\} \times {\mathcal{A}}_{Y_2}(\m_2),
\end{equation}
\begin{equation}
\hat{\mathcal{A}}_{Y_2}^{\ell_n} \triangleq \bigcup\limits_{\m_2 \in \mathcal{M}_2} \{\m_2\} \times \hat{\mathcal{A}}_{Y_2}^{\ell_n}(\m_2),
\end{equation} 
and notice that
\begin{IEEEeqnarray}{rCl}
	\lefteqn{P_{\tilde{\M}_2\tilde{Y}_2^n}\left(\hat{\mathcal{A}}_{Y_2}^{\ell_n}\right) } \quad \nonumber \\
	&= & \sum_{ (y_0^n,y_1^n)\in\D } \; P_{\tilde{Y}_0^n\tilde{Y}_1^n}(y_0^n,y_1^n) \nonumber \\	
	& & \hspace{.8cm}\cdot P_{\tilde{Y}_2^n|\tilde{Y}_0^n\tilde{Y}_1^n}(\mathcal{A}_{Y_2}(\phi_2(\phi_1(y_0^n),y_1^n)))|y_0^n,y_1^n)\IEEEeqnarraynumspace\\
	&\geq&  (1-\zeta_n).\label{eq:zetan}
\end{IEEEeqnarray}
Defining
\begin{equation}
Q_{\tilde{\M}_2}(\m_2) \triangleq \sum_{y_1^n,\m_1} P_{\tilde{\M}_1}(\m_1)P_{\tilde{Y}_1^n}(y_1^n)\cdot \mathbbm{1}\{\phi_2(\m_1,y_1^n)=\m_2\},
\end{equation}
we can write
\begin{IEEEeqnarray}{rCl}
	\lefteqn{Q_{\tilde{\M}_2}P_{\tilde{Y}_2^n}\left(\hat{\mathcal{A}}_{Y_2,n}^{\ell_n}\right)}\qquad \nonumber\\
	&\leq& Q_{\M_2}P_{Y_2}^n\left(\hat{\mathcal{A}}_{Y_2,n}^{\ell_n}\right)\Delta_n^{-3}\\
	& = & \sum_{ \m_2 \in\mathcal{M}_2 } Q_{\M_2} (\m_2)P_{{Y}_2}^n\left( \hat{\mathcal{A}}_{Y_2}^{\ell_n}(\m_2)\right)\Delta_n^{-3}\IEEEeqnarraynumspace\\
	&\leq& \sum_{ \m_2 \in\mathcal{M}_2 } Q_{\M_2}(\m_2) P_{{Y}_2}^n\left( {\mathcal{A}}_{Y_2}(\m_2)\right) \nonumber \\
	& & \hspace{1cm} \cdot e^{nh_b(\ell_n/n)}|\mathcal{Y}_2|^{\ell_n}k_n^{\ell_n}\Delta_{n}^{-3}\IEEEeqnarraynumspace\\
	&=& \beta_{2,n} e^{n \delta_n},\label{Eq:ByCsiszarKornerLemma_lemma}
\end{IEEEeqnarray}
where  $\delta_n\triangleq h_b(\ell_n/n) + \frac{\ell_n}{n} \log ( |\mathcal{Y}_2|\cdot k_n) -\frac{3}{n}\log \Delta_n $ and $k_n \triangleq \min\limits_{\substack{y_2,y_2':\\P_{Y_2}(y_2') > 0}}{P_{Y_2}(y_2) \over P_{Y_2}(y_2')}$.
Here, (\ref{Eq:ByCsiszarKornerLemma_lemma}) holds by \cite[Proof of Lemma 5.1]{Csiszarbook}.

Combining \eqref{Eq:ByCsiszarKornerLemma_lemma} with  \eqref{eq:zetan} and standard inequalities (see \cite[Lemma~1]{JSAIT}), we  then obtain:\vspace{1mm}
\begin{IEEEeqnarray}{rCl}\label{theta_ub_lemma}
	\lefteqn{-{1\over n}\log \beta_{2,n}} \qquad \nonumber\\
	& \leq & 	-{1\over n}\log \left( Q_{\tilde{\M}_2}P_{\tilde{Y}_2^n}\left(\hat{\mathcal{A}}_{Y_2}^{\ell_n}\right) \right) +  \delta_n\\
	&\leq& {1 \over n (1-\zeta_n)} D(P_{\tilde{\M}_2\tilde{Y}_2^n}||Q_{\tilde{\M}_2}P_{\tilde{Y}_2^n}) + \delta_n +\frac{1}{n}.\IEEEeqnarraynumspace
\end{IEEEeqnarray}

We continue to upper bound the divergence term as
\begin{IEEEeqnarray}{rCl}
	\lefteqn{D(P_{\tilde{\M}_2\tilde{Y}_2^n}||Q_{\tilde{\M}_2}P_{\tilde{Y}_2^n})}\qquad \nonumber\\
	&=& I(\tilde{\M}_2;\tilde{Y}_2^n) + D(P_{\tilde{\M}_2}||Q_{\tilde{\M}_2}) \\
	&\leq& I(\tilde{\M}_2;\tilde{Y}_2^n) + D(P_{\tilde{Y}_1^n\tilde{\M}_1}||P_{\tilde{Y}_1^n}P_{\tilde{\M}_1})\label{eq:dp_ineq_relative_entropy}\\
	&=& I(\tilde{\M}_2;\tilde{Y}_2^n) + I(\tilde{\M}_1;\tilde{Y}_1^n) \IEEEeqnarraynumspace\\
	&=& \sum_{t=1}^n I(\tilde{\M}_2;\tilde{Y}_{2,t}|\tilde{Y}_2^{t-1}) + I(\tilde{\M}_1;\tilde{Y}_{1,t}|\tilde{Y}_1^{t-1})\label{eq:divergence_chainrule}\\
	&\leq& \sum_{t=1}^n I(\tilde{\M}_2\tilde{Y}_0^{t-1}\tilde{Y}_1^{t-1};\tilde{Y}_{2,t}) \nonumber \\
	&& \qquad + I(\tilde{\M}_1\tilde{Y}_0^{t-1}\tilde{Y}_1^{t-1};\tilde{Y}_{1,t})\IEEEeqnarraynumspace\label{eq:divergence_markovcahins}\\
	&=& \sum_{t=1}^n I(\tilde{U}_{2,t};\tilde{Y}_{2,t}) + I(\tilde{U}_{1,t};\tilde{Y}_{1,t})\label{eq:divergence_end1}\\
	&=& n[I(\tilde{U}_{2,T};\tilde{Y}_{2,T}|T) + I(\tilde{U}_{1,T};\tilde{Y}_{1,T}|T)]\\
	&\leq& n[I(\tilde{U}_{2,T}T;\tilde{Y}_{2,T}) + I(\tilde{U}_{1,T}T;\tilde{Y}_{1,T})]\\
	&=& n [I(U_2;\tilde{Y}_2) + I(U_1;\tilde{Y}_1)]\label{theta_ub2_lemma}.
\end{IEEEeqnarray}
Here \eqref{eq:dp_ineq_relative_entropy} is obtained by the data processing inequality for KL-divergence; \eqref{eq:divergence_chainrule} by the chain rule; \eqref{eq:divergence_markovcahins} by the Markov chain $\tilde{Y}_2^{t-1} \to (\tilde{Y}_0^{t-1}\tilde{Y}_1^{t-1}) \to \tilde{Y}_{2,t}$; and \eqref{eq:divergence_end1}--\eqref{theta_ub2_lemma} by the definitions of $\tilde{U}_{1,t},\tilde{U}_{2,t},U_1,U_2,\tilde{Y}_1,\tilde{Y}_2$.

Following similar steps, we now prove the desired upper bound on the relay's error exponent. 
Define the acceptance region  at {\Rel} as
\begin{equation}
\mathcal{A}_{Y_1} \triangleq \{ (\m_1,y_1^n)\colon g_1(\m_1,y_1^n) =0 \}.
\end{equation} 
Notice that for any given  $(y_0^n, y_1^n)$ the pair $(\m_1=\phi_0(y_0^n), y_1^n)$  lies inside the acceptance region $\mathcal{A}_{Y_1}$ with probability either $0$ or $1$. Thus,   for any $\eta>0$ Condition \eqref{lem:cond2} implies that for all pairs $(y_0^n,y_1^n)\in \mathcal{D}$ the corresponding pairs $(\m_1,y_1^n)$ lie in $\mathcal{A}_{Y_1}$  with probability 1:
\begin{IEEEeqnarray}{rCl}
	P_{\tilde{\M}_1\tilde{Y}_1^n}({\mathcal{A}}_{Y_1})
	&=&  1.\label{eq:12}
\end{IEEEeqnarray}
Following similar steps as in the analysis of the receiver's error exponent:
\begin{IEEEeqnarray}{rCl}
	P_{\tilde{\M}_1}P_{\tilde{Y}_1^n}\left({\mathcal{A}}_{Y_1}\right)
	&\leq& P_{{\M}_1}P_{{Y}_1^n}\left({\mathcal{A}}_{Y_1}\right)\Delta_n^{-2}\\
	&=& \beta_{1,n} \Delta_n^{-2}. \label{eq:11}
\end{IEEEeqnarray}
Combining \eqref{eq:12} and  \eqref{eq:11}  with standard inequalities (see \cite[Lemma~1]{JSAIT}), we  further obtain 
\begin{IEEEeqnarray}{rCl}\label{theta_ub_lemma_relay}
	-{1\over n}\log \beta_{1,n} & \leq &-{1\over n}\log \left( P_{\tilde{\M}_1}P_{\tilde{Y}_1^n}\left({\mathcal{A}}_{Y_1,n}\right)  \right) - \frac{2}{n} \log \Delta_n \IEEEeqnarraynumspace\\
	&\leq& {1 \over n } D(P_{\tilde{\M}_1\tilde{Y}_1^n}||P_{\tilde{\M}_1}P_{\tilde{Y}_1^n}) + \delta_n'\IEEEeqnarraynumspace
\end{IEEEeqnarray}
where $\delta_n'\triangleq - \frac{2}{n} \log \Delta_n +\frac{1}{n}$ and tends to 0 as $n \to \infty$.

We continue to upper bound the divergence term as
\begin{IEEEeqnarray}{rCl}
	D(P_{\tilde{\M}_1\tilde{Y}_1^n}||P_{\tilde{\M}_1}P_{\tilde{Y}_1^n})	&=& I(\tilde{\M}_1;\tilde{Y}_1^n)\\
	&=& \sum_{t=1}^n I(\tilde{\M}_1;\tilde{Y}_{1,t}|\tilde{Y}_1^{t-1})\label{eq:divergence_chainrule_relay}\\
	&\leq& \sum_{t=1}^n I(\tilde{\M}_1\tilde{Y}_0^{t-1}\tilde{Y}_1^{t-1};\tilde{Y}_{1,t})\IEEEeqnarraynumspace\label{eq:divergence_markovcahins_relay}\\
	&=& \sum_{t=1}^n I(\tilde{U}_{1,t};\tilde{Y}_{1,t})\label{eq:divergence_end1_relay}\\
	&=& n[I(\tilde{U}_{1,T};\tilde{Y}_{1,T}|T)]\\
	&\leq& n[I(\tilde{U}_{1,T}T;\tilde{Y}_{1,T})]\\
	&=& n [I(U_1;\tilde{Y}_1)]\label{theta_ub2_lemma_relay}.
\end{IEEEeqnarray}
Here \eqref{eq:divergence_chainrule_relay} holds by the chain rule and \eqref{eq:divergence_end1_relay}--\eqref{theta_ub2_lemma_relay} hold by the definitions of $\tilde{U}_{1,t},U_1,\tilde{Y}_1$.

Finally, we proceed to prove the Markov chain $U_1 \to \tilde{Y}_0 \to \tilde{Y}_1$ in the limit as $n \to \infty$. To this end, notice the Markov chain $\tilde{\M}_1 \to \tilde{Y}_0^n \to \tilde{Y}_1^n$, and thus similar to the analysis in \cite[Section V.C]{HWS20}:
\begin{IEEEeqnarray}{rCl}
	0 &=& I(\tilde{\M}_1;\tilde{Y}_1^n|\tilde{Y}_0^n) \\ 
	&\geq& H(\tilde{Y}_1^n|\tilde{Y}_0^n)  - H(\tilde{Y}_1^n|\tilde{Y}_0^n\tilde{\M}_1) \nonumber \\
	&& + D(P_{\tilde{Y}_0^n\tilde{Y}_1^n}||P_{Y_0Y_1}^n) + \log{\Delta_{n}}
	\label{MC1proofstep1}\\
	&{\geq}& n[H(\tilde{Y}_{1,T}|\tilde{Y}_{0,T}) + D(P_{\tilde{Y}_{0,T}\tilde{Y}_{1,T}}||P_{Y_0Y_1})] +\log{\Delta_{n}} \nonumber\\
	&& - H(\tilde{Y}_1^n|\tilde{Y}_0^n\tilde{\M}_1) \label{MC1proofstep2}\\
	&\geq& n[H(\tilde{Y}_{1,T}|\tilde{Y}_{0,T}) + D(P_{\tilde{Y}_{0,T}\tilde{Y}_{1,T}}||P_{Y_0Y_1})] +\log{\Delta_{n}} \nonumber \\ 	&&-  \sum_{t=1}^{n}H(\tilde{Y}_{1,t}|\tilde{Y}_{0,t}\tilde{Y}_0^{t-1}\tilde{Y}_1^{t-1}\tilde{\M}_1)\label{MC1proofstep3}\\
	&=& n[H(\tilde{Y}_{1,T}|\tilde{Y}_{0,T}) + D(P_{\tilde{Y}_{0,T}\tilde{Y}_{1,T}}||P_{Y_0Y_1})] +\log{\Delta_{n}} \nonumber \\
	&&-  \sum_{t=1}^{n}H(\tilde{Y}_{1,t}|\tilde{Y}_{0,t}\tilde{U}_{1,t})\label{MC1proofstep4}\\
	&\geq& n[H(\tilde{Y}_{1,T}|\tilde{Y}_{0,T}) - H(\tilde{Y}_{1,T}|\tilde{Y}_{0,T},\tilde{U}_{1,T},T) ]+ \log{\Delta_{n}}  \IEEEeqnarraynumspace\label{MC1proofstep4b}\\
	&\geq& nI(\tilde{Y}_1;{U}_1|\tilde{Y}_0) + \log{\Delta_{n}},\label{MC1proofstep5}
\end{IEEEeqnarray}
where \eqref{MC1proofstep2} holds by the super-additivity property in \cite[Proposition 1]{tyagi2019strong}; \eqref{MC1proofstep3} by the chain rule and since conditioning reduces entropy; \eqref{MC1proofstep4} by the definition of $\tilde{U}_{1,t}$ and by recalling that $T$ is uniform over $\{1,\ldots,n\}$ independent of all other random quantities; \eqref{MC1proofstep4b} by the non-negativity of the Kullback-Leibler divergence; and finally \eqref{MC1proofstep5} holds by the definitions of ${U}_1,\tilde{Y}_0, \tilde{Y}_1$.

\section{Analysis of the coding scheme in Section~\ref{sec:schemeK}}\label{app:schemeK}

Consider the  $\ell_{\mI}^*$-hop hypothesis testing scheme  employed when $Y_0^n \in \D_{\mI}$, for $\mI \in\mathcal{P}(K)$.  For any $\mI \in\mathcal{P}(K)$,  let ${\hmH}_{\mI,1}, \ldots, \hmH_{\mI,\ell_{\mI}^*}$ denote the  guesses produced at  terminals $1,\ldots, \ell_{\mI}^*$ when employing this scheme. 

By assumption, the type-I error probabilities of these decisions tend to 0 as $n\to \infty$ for any $\mI \in\mathcal{P}(K)$:
	\begin{IEEEeqnarray}{rCl}	\label{eq:limtyp1c}
\lim_{n\to \infty} \Pr[{\hmH}_{\mI,k} = 1| \mathcal{H}=0, Y_0^n \in \D_{\mI}] &=&0, \quad  k\in\mI.\nonumber\\
\end{IEEEeqnarray}
Recalling that decision center $k$ declares $\hat{\mH}_k=1$ whenever $Y_0^n \in \D_{\emptyset}$ or $Y_0^n\in \D_{\mI}$ for  a set $\mI$ not containing $k$, and applying the total law of probability, we can  write
\begin{IEEEeqnarray}{rCl}
	\alpha_{k,n} &=&\Pr[\hat{\mathcal{H}}_k=1| \mathcal{H}=0]\\
	&=&  \sum_{\mI \in (\mathcal{P}(K) \cup \emptyset)} \Pr[\hat{\mathcal{H}}_k= 1, Y_0^n \in  \D_{\mI} |\mathcal{H} = 0] \\
	&=&  \Pr[Y_0^n \in \D_{\emptyset}|\mathcal{H}=0] +\sum_{\substack{\mI\in\mathcal{P}(K) \colon \\ {k\notin\mI} }} \Pr[Y_0^n \in \D_{\mI}|\mathcal{H}=0]\nonumber\\
		&& + \sum_{\substack{\mI\in\mathcal{P}(K) \colon \\ {k\in\mI} }}   \Pr[{\hmH}_{k}= 1,Y_0^n \in  \D_{\mI}| \mathcal{H}=0]  \\
			&\leq&  \Pr[Y_0^n \in \D_{\emptyset}|\mathcal{H}=0] +\sum_{\substack{\mI\in\mathcal{P}(K) \colon \\ {k\notin\mI} }} \Pr[Y_0^n \in \D_{\mI}|\mathcal{H}=0] \nonumber\\
		&& +  \sum_{\substack{\mI\in\mathcal{P}(K) \colon \\ {k\in\mI} }}   \Pr[{\hmH}_{\mI,k}= 1| \mathcal{H}=0,Y_0^n \in  \D_{\mI}] .
\end{IEEEeqnarray}
Combining this inequality with \eqref{eq:limtyp1c}, and by Inequalities \eqref{eq:sigma_eps}, we conclude that   the overall scheme satisfies the type-I error constraints: 
\begin{equation}
\varlimsup_{n\to\infty}\alpha_{k,n} \leq \epsilon_k, \quad k\in\{1,\ldots, K\}.
\end{equation}

For the  type-II error exponent at a decision center $k$ we observe: 
\begin{IEEEeqnarray}{rCl}
	\beta_{k,n} &=& \Pr[\hat{\mathcal{H}}_k=0|\mathcal{H}=1]\\
	&=&  \sum_{\mI \in( \mathcal{P}(K) \cup \emptyset)}   \Pr[\hat{\mathcal{H}}_k=0, Y_0^n\in  \D_{\mI} |\mathcal{H}=1]\\
	&=&    \sum_{\substack{\mI \in \mathcal{P}(K) \colon \\ k\in\mI}}   \Pr[\hat{\mathcal{H}}_{\mI,k}=0, Y_0^n\in  \D_{\mI} |\mathcal{H}=1]\\
		&\leq&   \sum_{\substack{\mI \in \mathcal{P}(K) \colon \\ k\in\mI}}   \Pr[\hat{\mathcal{H}}_{\mI,k}=0|\mathcal{H}=1, Y_0^n\in  \D_{\mI}]. \label{eq:KB}
\end{IEEEeqnarray}
Defining
\begin{IEEEeqnarray}{rCl}
\theta_{k,\mI}:= \varliminf_{n\to \infty} - \frac{1}{n} \log  \Pr[\hat{\mathcal{H}}_{\mI,k}=0|\mathcal{H}=1, Y_0^n\in  \D_{\mI}], \IEEEeqnarraynumspace
\end{IEEEeqnarray}
we  conclude by \eqref{eq:KB}  that the exponent 
\begin{IEEEeqnarray}{rCl}
\min_{\substack{\mI \in \mathcal{P}(K) \colon \\ k\in\mI}} \theta_{k,\mI} 
\end{IEEEeqnarray}
is achievable at decision center $k$.  This proves in particular that when applying an instance of  the multi-hop scheme in \cite{salehkalaibar2020hypothesisv1} for each set $\mathcal{I}\in\mathcal{P}(K)$,    the exponents $\theta_1,\ldots, \theta_K$  in \eqref{thm3} are proved achievable.

\section{Proof of Lemma~\ref{lem:receiverconverseKhop}}\label{app_lemmaK}
Note first that by \eqref{pmftildedoubleprime2_lemma_Khop}:
\begin{equation}\label{tildedivergencerelation2_lemmaKHop}
	D(P_{\tilde{Y}_0^n\cdots\tilde{Y}_{K}^n}||P_{Y_0\cdots Y_{K}}^{n}) \leq \log{\Delta_n^{-1}},
\end{equation}
where we defined $\Delta_n \triangleq P_{Y_0^n\cdots Y_{K}^n}(\D)$.

Further define  $\tilde{U}_{i,t}\triangleq(\tilde{\M}_i,\tilde{Y}_0^{t-1},\ldots,\tilde{Y}_{K}^{t-1})$ for $i \in \{1,\ldots,K\}$ and notice:
\begin{IEEEeqnarray}{rCl}
	H(\tilde{\M}_i)	&\geq& I(\tilde{\M}_i;\tilde{Y}_0^n\cdots \tilde{Y}_{K}^n)  \nonumber \\
	&& + D(P_{\tilde{Y}_0^n\cdots \tilde{Y}_{K}^n}||P_{Y_0\cdots Y_{K}}^n)  + \log\Delta_{n}\IEEEeqnarraynumspace\label{m1entropylbstep1_lemmaKHop}\\
	&=& H(\tilde{Y}_0^n\cdots\tilde{Y}_{K}^n) + D(P_{\tilde{Y}_0^n\cdots \tilde{Y}_{K}^n}||P_{Y_0\cdots Y_{K}}^n) \nonumber \\ 
	&& -  H(\tilde{Y}_0^n\cdots \tilde{Y}_{K}^n|\tilde{\M}_i) + \log\Delta_{n}\IEEEeqnarraynumspace\\
	&\geq& n [H(\tilde{Y}_{0,T}\cdots\tilde{Y}_{K,T}) \nonumber \\
	&& \quad  + D(P_{\tilde{Y}_{0,T}\cdots \tilde{Y}_{K,T}}||P_{Y_0\cdots Y_{K}})] \nonumber \\ && - \sum_{t=1}^{n} H(\tilde{Y}_{0,t}\cdots \tilde{Y}_{K,t}|\tilde{U}_{i,t})+ \log\Delta_{n}\IEEEeqnarraynumspace\label{m1entropylbstep4_lemmaKHop}\\
	&=& n [H(\tilde{Y}_{0,T}\cdots \tilde{Y}_{K,T}) \nonumber \\
	&& \quad+ D(P_{\tilde{Y}_{0,T}\cdots \tilde{Y}_{K,T}}||P_{Y_0\cdots Y_{K}}) \nonumber \\ 
	&& \quad- H(\tilde{Y}_{0,T}\cdots \tilde{Y}_{K,T}|\tilde{U}_{i,T},T)]+ \log\Delta_{n} \label{Tuniformdef_lemmaKHop}\\
	&\geq & n [H(\tilde{Y}_{0,T}\cdots \tilde{Y}_{K,T})  \nonumber \\
	&& \quad - H(\tilde{Y}_{0,T}\cdots\tilde{Y}_{K,T}|\tilde{U}_{i,T},T)]+ \log\Delta_{n} \IEEEeqnarraynumspace\\
	&=& n [I(\tilde{Y}_0\cdots\tilde{Y}_{K};U_i)] +  \log{\Delta_{n}} \label{eq:HM1_LB_lemma_last_eqKHop} \\
	&\geq& n \left[I(\tilde{Y}_{i-1};U_i) + {1 \over n} \log{\Delta_{n}}\right] .\IEEEeqnarraynumspace \label{eq:HM1_LB_lemmaKHop}
\end{IEEEeqnarray}
Here, (\ref{m1entropylbstep1_lemmaKHop}) holds by (\ref{tildedivergencerelation2_lemmaKHop}); (\ref{m1entropylbstep4_lemmaKHop}) holds by the super-additivity property in \cite[Proposition 1]{tyagi2019strong}, by the chain rule,  by the definition of $\tilde{U}_{i,t}$ and  by defining $T$ uniform over $\{1,\dots,n\}$ independent of the previously defined random variables; and \eqref{eq:HM1_LB_lemma_last_eqKHop} by the definitions of $U_i,\tilde{Y}_i,\tilde{Y}_{i-1}$ in the lemma.
This proves Inequality \eqref{eq:r} in the lemma.

We next upper  bound the type-II error exponent at R$_{k}$, for $k\in\{1,\ldots, K\}$. 
To this end, 
define R$_{k}$'s acceptance region 
\begin{equation}
	\mathcal{A}_{Y_k} \triangleq \{ (\m_k,y_k^n)\colon g_k(\m_k,y_k^n) =0 \}.
\end{equation}
 Since for any  tuple of sequences $(y_0^n,\cdots,y_{K}^n)$, the corresponding $(\m_k,y_k^n)$ either lie inside or outside the acceptance region $\mathcal{A}_{Y_k}$, for any $k\in\{1,\ldots, K\}$, 
Condition \eqref{eq:lemma_cond1Khop} 
implies 
\begin{IEEEeqnarray}{rCl}
	P_{\tilde{\M}_k\tilde{Y}_k^n}({\mathcal{A}}_{Y_k})
	&=&  1.\label{eq:12KHop}
\end{IEEEeqnarray}
Define for any $k\in\{1,\ldots, K\}$:
\begin{IEEEeqnarray}{rCl}
	\lefteqn{Q_{\tilde{\M}_k}(\m_k) } \nonumber\\
	&\triangleq& \sum_{y_0^n,y_1^n,\ldots, y_{k-1}^n}   P_{\tilde{Y}_{0}^n}(y_0^n)\cdots P_{\tilde{Y}_{k-1}^n}(y_{k-1}^n)  \nonumber\\
	&&  \quad \cdot \mathbbm{1}\{\m_k=\phi_k(\phi_{k-1}(\cdots(\phi_1(y_0^n)\cdots)), y_{k-1}^n)\}, \IEEEeqnarraynumspace
\end{IEEEeqnarray}
and 
\begin{IEEEeqnarray}{rCl}
	\lefteqn{Q_{{\M}_k}(\m_K) } \nonumber\\
	&\triangleq& \sum_{y_0^n,y_1^n,\ldots, y_{k-1}^n}   P_{{Y}_{0}^n}(y_0^n)\cdots P_{{Y}_{k-1}^n}(y_{k-1}^n)  \nonumber\\
	&&  \quad \cdot \mathbbm{1}\{\m_k=\phi_k(\phi_{k-1}(\cdots(\phi_1(y_0^n)\cdots)), y_{k-1}^n)\}. \IEEEeqnarraynumspace
\end{IEEEeqnarray}
and notice that
\begin{IEEEeqnarray}{rCl}
	Q_{\tilde{\M}_k}P_{\tilde{Y}_k^n}\left({\mathcal{A}}_{Y_k}\right)
	&\leq& Q_{{\M}_k}P_{{Y}_k^n}\left({\mathcal{A}}_{Y_k}\right)\Delta_n^{-(k+1)}\\
	&=& \beta_{k,n} \Delta_n^{-(k+1)}. \label{eq:11KHop}
\end{IEEEeqnarray}
By \eqref{eq:12KHop}, \eqref{eq:11KHop},  and standard inequalities (see \cite[Lemma~1]{JSAIT}), we  further obtain 
\begin{IEEEeqnarray}{rCl}\label{theta_ub_lemma_relayKHop}
	-{1\over n}\log \beta_{k,n} & \leq &-{1\over n}\log \left( Q_{\tilde{\M}_k}P_{\tilde{Y}_k^n}\left({\mathcal{A}}_{Y_k}\right)  \right) - \frac{(k+1)}{n} \log \Delta_n \nonumber\\\\
	&\leq& {1 \over n } D(P_{\tilde{\M}_k\tilde{Y}_k^n}||Q_{\tilde{\M}_k}P_{\tilde{Y}_k^n}) + \delta_n'\IEEEeqnarraynumspace
\end{IEEEeqnarray}
where $\delta_n'\triangleq - \frac{(k+1)}{n} \log \Delta_n +\frac{1}{n}$ and tends to 0 as $n \to \infty$.

We continue to upper bound the divergence term as
\begin{IEEEeqnarray}{rCl}
	\lefteqn{D(P_{\tilde{\M}_k\tilde{Y}_k^n}||Q_{\tilde{\M}_k}P_{\tilde{Y}_k^n})}\qquad \nonumber\\
	&=& I(\tilde{\M}_k;\tilde{Y}_k^n) + D(P_{\tilde{\M}_k}||Q_{\tilde{\M}_k}) \\
	&\leq& I(\tilde{\M}_k;\tilde{Y}_k^n) + D(P_{\tilde{Y}_{k-1}^n\tilde{\M}_{k-1}}||P_{\tilde{Y}_{k-1}^n}Q_{\tilde{\M}_{k-1}})\label{eq:dp_ineq_relative_entropyKHop}\\
	&\leq& I(\tilde{\M}_k;\tilde{Y}_{k}^n) + I(\tilde{\M}_{k-1};\tilde{Y}_{k-1}^n) \nonumber \\
	&& \qquad \qquad \qquad + D(P_{\tilde{Y}_{k-2}^n\tilde{\M}_{k-2}}||P_{\tilde{Y}_{k-2}^n}Q_{\tilde{\M}_{k-2}})\IEEEeqnarraynumspace\\
		&\vdots& \nonumber\\
	&\leq& \sum_{i=1}^{k} I(\tilde{\M}_i;\tilde{Y}_{i}^n)\\
	&=& \sum_{i=1}^k \sum_{t=1}^n I(\tilde{\M}_i;\tilde{Y}_{i,t}|\tilde{Y}_i^{t-1}) \label{eq:divergence_chainruleKHop}\\
	&\leq& \sum_{i=1}^k \sum_{t=1}^n I(\tilde{\M}_i\tilde{Y}_0^{t-1}\cdots \tilde{Y}_{K}^{t-1};\tilde{Y}_{i,t})\IEEEeqnarraynumspace\label{eq:divergence_markovcahinsKHop}\\
	&=& \sum_{i=1}^k \sum_{t=1}^n I(\tilde{U}_{i,t};\tilde{Y}_{i,t})\label{eq:divergence_end1KHop}\\
	&=& \sum_{i=1}^k nI(\tilde{U}_{i,T};\tilde{Y}_{i,T}|T)\\
	&\leq&  n \sum_{i=1}^k I(U_i;\tilde{Y}_i) \label{theta_ub2_lemmaKHop}.
\end{IEEEeqnarray}
Here \eqref{eq:dp_ineq_relative_entropyKHop} is obtained by the data processing inequality for KL-divergence; \eqref{eq:divergence_chainruleKHop} by the chain rule; 
 and \eqref{eq:divergence_end1KHop}--\eqref{theta_ub2_lemmaKHop} by the definitions of $\tilde{U}_{i,t},U_i,\tilde{Y}_i$ and $T$.

Combined with \eqref{theta_ub_lemma_relayKHop} this establishes Inequality~\ref{eq:t} for $k\in\{1,\ldots, K-1\}$.

Finally, we proceed to prove that for any $k\in\{1,\ldots, K\}$  the Markov chain $U_k \to \tilde{Y}_{k-1} \to \tilde{Y}_{k}$ holds in the limit as $n \to \infty$. We start by noticing the Markov chain $\tilde{\M}_1 \to \tilde{Y}_0^n \to (\tilde{Y}_1^n,\cdots,\tilde{Y}_K^n)$, and thus similar to the analysis in \cite[Section V.C]{HWS20}:
\begin{IEEEeqnarray}{rCl}
	0 &=& I(\tilde{\M}_1;\tilde{Y}_1^n\cdots \tilde{Y}_{K}^n|\tilde{Y}_0^n) \label{MC1proofstep0KHop}\\ 
	&\geq& H(\tilde{Y}_1^n\cdots \tilde{Y}_{K}^n|\tilde{Y}_0^n)  - H(\tilde{Y}_1^n\cdots \tilde{Y}_{K}^n|\tilde{Y}_0^n\tilde{\M}_1) \nonumber \\
	&&\quad  + D(P_{\tilde{Y}_0^n\cdots \tilde{Y}_{K}^n}||P_{Y_0\cdots Y_{K}}^n) + \log{\Delta_{n}}
	\label{MC1proofstep1KHop}\\
	&{\geq}& n[H(\tilde{Y}_{1,T}\cdots \tilde{Y}_{K,T}|\tilde{Y}_{0,T}) \nonumber \\
	&& \quad + D(P_{\tilde{Y}_{0,T}\cdots \tilde{Y}_{K,T}}||P_{Y_0\cdots Y_{K}})] +\log{\Delta_{n}} \nonumber\\
	&& - H(\tilde{Y}_1^n \cdots \tilde{Y}_{K}^n|\tilde{Y}_0^n\tilde{\M}_1) \label{MC1proofstep2KHop}\\
	&\geq& n[H(\tilde{Y}_{1,T}\cdots \tilde{Y}_{K,T}|\tilde{Y}_{0,T}) \nonumber \\
	&&\quad  + D(P_{\tilde{Y}_{0,T}\cdots \tilde{Y}_{K,T}}||P_{Y_0\cdots Y_{K}})] +\log{\Delta_{n}} \nonumber \\ 	&&-  \sum_{t=1}^{n}H(\tilde{Y}_{1,t}\cdots \tilde{Y}_{K,t} |\tilde{Y}_{0,t}\tilde{Y}_0^{t-1}\cdots \tilde{Y}_{K}^{t-1}\tilde{Y}_{0,t+1}^n\tilde{\M}_1)\nonumber\\\label{MC1proofstep3KHop}\\
	&=& n[H(\tilde{Y}_{1,T}\cdots\tilde{Y}_{K,T}|\tilde{Y}_{0,T}) \nonumber \\
	&& \quad + D(P_{\tilde{Y}_{0,T}\cdots \tilde{Y}_{K,T}}||P_{Y_0\cdots Y_{K}})] +\log{\Delta_{n}} \nonumber \\
	&&-  n H(\tilde{Y}_{1,T}\cdots \tilde{Y}_{K,T} |\tilde{Y}_{0,T}\tilde{Y}_0^{T-1}\cdots \tilde{Y}_{K}^{T-1}\tilde{Y}_{0,T+1}^n\tilde{\M}_1T)\label{MC1proofstep4KHop}\nonumber\\\\
	&\geq& nI(\tilde{Y}_{1,T}\cdots \tilde{Y}_{K,T};\tilde{Y}_0^{T-1}\cdots \tilde{Y}_{K}^{T-1}\tilde{Y}_{0,T+1}^n\tilde{\M}_1T|\tilde{Y}_{0,T}) \nonumber \\
	&& \qquad  + \log{\Delta_{n}}\label{MC1proofstep5KHop}\\
	&\geq& nI(\tilde{Y}_{1}\cdots \tilde{Y}_{K};\tilde{U}_1|Y_0)  + \log{\Delta_{n}},
\end{IEEEeqnarray}
where \eqref{MC1proofstep2KHop} holds by the super-additivity property in \cite[Proposition 1]{tyagi2019strong}; \eqref{MC1proofstep3KHop} by the chain rule; \eqref{MC1proofstep5KHop} by the non-negativity of the Kullback-Leibler divergence. 

Since $\Delta_n$ is bounded,  $\frac{1}{n}\log \Delta$ tends to 0 as $n\to \infty$, and we can conclude that
\begin{equation}
\lim_{n\to \infty}I(\tilde{Y}_{1}\cdots \tilde{Y}_{K};\tilde{U}_1|Y_0)=0,
\end{equation}
thus proving \eqref{eq:s} for $k=1$. 

Notice next that for any $k\in\{2,\ldots, K\}$:
\begin{IEEEeqnarray}{rCl}
I(U_k;\tilde{Y}_{k}| \tilde{Y}_{k-1}) &\leq& I(U_k\tilde{Y}_0\cdots \tilde{Y}_{k-2};\tilde{Y}_{k}| \tilde{Y}_{k-1})\\
&=& I(U_k;\tilde{Y}_{k}| \tilde{Y}_0 \cdots \tilde{Y}_{k-1}) \nonumber \\
&& + I(\tilde{Y}_0\cdots \tilde{Y}_{k-2};\tilde{Y}_{k}| \tilde{Y}_{k-1}) . \label{eq:sum}\IEEEeqnarraynumspace
\end{IEEEeqnarray}
In the following we show that both quantities $I(U_k;\tilde{Y}_{k}| \tilde{Y}_0 \cdots \tilde{Y}_{k-1})$ and $I(\tilde{Y}_0\cdots \tilde{Y}_{k-2};\tilde{Y}_{k}| \tilde{Y}_{k-1})$ tend to 0 as $n \to \infty$, which  establishes  \eqref{eq:s} for $k\in\{2,\ldots, K\}$. 

To prove that $I(\tilde{Y}_0\cdots \tilde{Y}_{k-2};\tilde{Y}_{k}| \tilde{Y}_{k-1})$ tends to 0 as $n \to \infty$, we notice that for any $k\in\{1,\ldots, K-1\}$:
\begin{IEEEeqnarray}{rCl}
\lefteqn{D(P_{\tilde{Y}_{0}\cdots \tilde{Y}_{K}}||P_{Y_0\cdots Y_{K}})} \qquad \qquad \\ &\geq&  D(P_{\tilde{Y}_{0}\cdots \tilde{Y}_{k}}||P_{Y_0\cdots Y_{k}}) \\
&=& D(P_{\tilde{Y}_{0}\cdots \tilde{Y}_{k}}||P_{Y_0\cdots Y_{k-1}}P_{Y_{k}|Y_{k-1}}) \\
&= &  D(P_{\tilde{Y}_{0}\cdots \tilde{Y}_{k-1}}|| P_{\tilde{Y}_0\cdots\tilde{Y}_{k-1} }P_{\tilde{Y}_{k}|\tilde{Y}_{k-1}} ) \nonumber \\ 
& & +\mathbb{E}_{P_{\tilde{Y}_{k-1}} }\left[ D(P_{\tilde{Y}_{k}|\tilde{Y}_{k-1}}||P_{{Y}_{k}| {Y}_{k-1}}) \right] \nonumber \\
&& \qquad + D(P_{\tilde{Y}_0\cdots\tilde{Y}_{k-1}} ||P_{Y_0\cdots Y_{k-1}})    \\
&\geq & D(P_{\tilde{Y}_{0}\cdots \tilde{Y}_{k}}|| P_{\tilde{Y}_0\cdots\tilde{Y}_{k-1}} P_{\tilde{Y}_{k}|\tilde{Y}_{k-1}} ) \\
&\geq& I(\tilde{Y}_0\cdots \tilde{Y}_{k-2};\tilde{Y}_{k}| \tilde{Y}_{k-1}) \label{eq:la}.
\end{IEEEeqnarray}
 Since $(\tilde{Y}_0\cdots \tilde{Y}_{K})$ lie in the jointly typical set $\mathcal{T}_{\mu_n}^{(n)}(P_{Y_0\cdots Y_{K}})$:
 \begin{equation}|P_{\tilde{Y}_0\cdots \tilde{Y}_{K}} - P_{Y_0\cdots Y_{K}}| \leq \mu_n.
 \end{equation}
 Recalling that $\mu_n \downarrow 0$ as $n \to \infty$,  and by the continuity of the KL-divergence, we conclude that  $D(P_{\tilde{Y}_{0}\cdots \tilde{Y}_{K}}||P_{Y_0\cdots Y_{K}})$ tends to 0 as $n\to \infty$, and thus by \eqref{eq:la} and the nonnegativity of mutual information:
 \begin{IEEEeqnarray}{rCl}\label{eq:summand1}
 \lim_{n\to \infty} I(\tilde{Y}_0\cdots \tilde{Y}_{k-2};\tilde{Y}_{k}| \tilde{Y}_{k-1}) =0.
 \end{IEEEeqnarray}


Following similar steps to \eqref{MC1proofstep0KHop}--\eqref{MC1proofstep5KHop}, we further obtain: 
\begin{IEEEeqnarray}{rCl}
	0 &=& I(\tilde{\M}_k;\tilde{Y}_k^n\cdots \tilde{Y}_{K}^n|\tilde{Y}_0^n\cdots\tilde{Y}_{k-1}^n ) \nonumber \\
	& \geq & H(\tilde{Y}_k^n\cdots \tilde{Y}_{K}^n|\tilde{Y}_0^n\cdots\tilde{Y}_{k-1}^n ) \nonumber\\
	&&  -H(\tilde{Y}_k^n\cdots \tilde{Y}_{K}^n|\tilde{Y}_0^n\cdots\tilde{Y}_{k-1}^n \tilde{M}_k)\\
	&& + D(P_{\tilde{Y}_0^n\cdots \tilde{Y}_{K}^n}||P_{Y_0\cdots Y_{K}}^n) + \log{\Delta_{n}}\\ 
	&\geq& n[H(\tilde{Y}_{k,T}\cdots \tilde{Y}_{K,T}|\tilde{Y}_{0,T}\cdots \tilde{Y}_{k-1,T}) \nonumber \\
	&& \quad + D(P_{\tilde{Y}_{0,T}\cdots \tilde{Y}_{K,T}}||P_{Y_0\cdots Y_{K}})] +\log{\Delta_{n}} \nonumber\\
	&& - H(\tilde{Y}_k^n\cdots \tilde{Y}_{K}^n|\tilde{Y}_0^n\cdots \tilde{Y}_{k-1}^n \tilde{\M}_k) \label{MC1proofstep2KHop_MC2}\\
	&\geq& n[H(\tilde{Y}_{k,T} \cdots \tilde{Y}_{K,T} |\tilde{Y}_{0,T} \cdots \tilde{Y}_{k-1,T}) \nonumber \\
	&& + D(P_{\tilde{Y}_{0,T}\cdots \tilde{Y}_{K,T}}||P_{Y_0\cdots Y_{K}})] +\log{\Delta_{n}} \nonumber \\ 	
	&&-  \sum_{t=1}^{n}H(\tilde{Y}_{k,t} \cdots \tilde{Y}_{K,t} |\tilde{Y}_{0,t}\cdots \tilde{Y}_{k-1,t} \nonumber \\
	&& \qquad \qquad \tilde{Y}_0^{t-1}\cdots \tilde{Y}_{K}^{t-1} \tilde{Y}_{0,t+1}^n \cdots \tilde{Y}_{k-1,t+1}^n \tilde{\M}_k)\label{MC1proofstep3KHop_MC2}\IEEEeqnarraynumspace\\
	&\geq& nH(\tilde{Y}_{k,T} \cdots \tilde{Y}_{K,T} |\tilde{Y}_{0,T} \cdots \tilde{Y}_{k-1,T})  + \log{\Delta_{n}}\nonumber \\
	&&- n H(\tilde{Y}_{k,T} \cdots \tilde{Y}_{K,T} |\tilde{Y}_{0,T}\cdots \tilde{Y}_{k-1,T} \nonumber \\
	&& \qquad \qquad \tilde{Y}_0^{T-1}\cdots \tilde{Y}_{K}^{T-1} \tilde{Y}_{0,T+1}^n \cdots \tilde{Y}_{k-1,T+1}^n \tilde{\M}_kT)\label{MC1proofstep5KHop_MC2}\\
	&\geq& nI(\tilde{Y}_{k,T}\cdots \tilde{Y}_{K,T};\tilde{Y}_0^{T-1}\cdots \tilde{Y}_{K}^{T-1}\tilde{\M_{k}}T|\tilde{Y}_{0,T}\cdots \tilde{Y}_{k-1,T} ) \nonumber \\
	&& \qquad \quad + \log{\Delta_{n}}\\
	&=& nI(\tilde{Y}_{k}\cdots \tilde{Y}_{K};U_k|\tilde{Y}_{0}\cdots \tilde{Y}_{k-1} ) + \log{\Delta_{n}}.
	\IEEEeqnarraynumspace\label{MC1proofstep6KHop_MC2}
\end{IEEEeqnarray}
Since $\Delta_n$ is bounded,  $\frac{1}{n}\log \Delta$ tends to 0 as $n\to \infty$, and we can conclude that
\begin{equation}
\lim_{n\to \infty}I(\tilde{Y}_{k};U_k|\tilde{Y}_{0}\cdots \tilde{Y}_{k-1} )=0,
\end{equation}
Combined with \eqref{eq:sum}, \eqref{eq:summand1}, and the nonnegativity of mutual information, this proves  \eqref{eq:s} for $k\in\{2,\ldots, K\}$.

\section{Proof of Lemma~\ref{lemma_parameters}}\label{app:proof_parameter_lemma}
To show sufficiency of \eqref{eq:eps1}, start by fixing any set of nonnegative numbers  $\{\sigma_{\mI}\}_{\mI \in \mathcal{P}(3)}$,  and $\{R_{\mI,1}, \ldots, R_{\mI, \ell^*_{\mI}}\}_{\mI\in\mathcal{P}(3)}$ satisfying 
\eqref{eq:E1_Khop} for $K=3$, (and possibly violating \eqref{eq:eps1}). Choose  new nonnegative numbers $\tilde{\sigma}_{\s{1}{2}{3}}, \tilde\sigma_{\s{\pi(1)}{\pi(2)}}, \tilde\sigma_{\s{\pi(1)}{\pi(3)}}, \tilde\sigma_{\s{\pi(1)}}$ satisfying
\begin{IEEEeqnarray}{rCl}
\tilde{\sigma}_{\mathcal{I}} &\leq &{\sigma}_{\mathcal{I}}, \quad \forall \mI \colon \pi(1) \in \mI,\\
\tilde{\sigma}_{\s{1}{2}{3}}+\tilde{\sigma}_{\s{\pi(1)}{\pi(2)}} & \geq &1- \epsilon_{\pi(1)}- \epsilon_{\pi(2)}\\
\tilde{\sigma}_{\s{1}{2}{3}}+\tilde{\sigma}_{\s{\pi(1)}{\pi(3)}}  &\geq &1- \epsilon_{\pi(1)}- \epsilon_{\pi(3)}
\end{IEEEeqnarray} 
and 
\begin{IEEEeqnarray}{rCl}
\tilde{\sigma}_{\s{1}{2}{3}}+\tilde{\sigma}_{\s{\pi(1)}{\pi(2)}}+\tilde{\sigma}_{\s{\pi(1)}{\pi(3)}}+\tilde{\sigma}_{\s{\pi(1)}}  &= &1-\epsilon_{\pi(1)}. \nonumber\\
\end{IEEEeqnarray} 
The existence of the desired numbers can be checked by applying the Fourier-Motzkin Elimination algorithm \cite{FME} and by noting  Constraints \eqref{eq:E1_Khop}. 
Further choose for any set $\mI$ containing $\pi(1)$ and $\ell \in \{1,2,3\}$ the rate: 
\begin{equation}\label{eq:I_1}
\tilde{R}_{\mathcal{I},\ell} := {R}_{\mathcal{I},\ell} ,
\end{equation}
and for any set $\mI$ not containing $\pi(1)$ and $\ell \in\{1,2,3\}$: 
\begin{IEEEeqnarray}{rCl}
	\tilde{\sigma}_{\mathcal{I}} &:=& \sigma_{\mathcal{I}} + \sigma_{\mI_{\pi(1)}} - \tilde{\sigma}_{\mI_{\pi(1)}} \\
	\tilde{R}_{\mathcal{I},\ell} &:=& \frac{\sigma_{\mathcal{I}}}{\tilde{\sigma}_{\mathcal{I}}}{R}_{\mathcal{I},\ell} + \frac{\sigma_{\mI_{\pi(1)}}-\tilde{\sigma}_{\mI_{\pi(1)}}}{\tilde{\sigma}_{\mathcal{I}}}{R}_{\mI_{\pi(1)},\ell},  \IEEEeqnarraynumspace \label{eq:new_sgima_tilde_cond5}
\end{IEEEeqnarray}
where we defined  $\mathcal{I}_{\pi(1)} := \mI \cup \{\pi(1)\}$. 

By Lemma~\ref{lem:moving}, the new set of numbers  $\{\tilde \sigma_{\mI}\}_{\mI \in \mathcal{P}(3)}$,  and $\{\tilde{R}_{\mI,1}, \ldots, \tilde R_{\mI, \ell^*_{\mI}}\}_{\mI\in\mathcal{P}(3)}$ also satisfies Constraints \eqref{eq:E1_Khop},
 which proves that one can restrict to  numbers $\{\sigma_{\mI}\}_{\mI \in \mathcal{P}(3)}$ satisfying \eqref{eq:eps1}. Since
$\epsilon_{\pi(1)}\geq \epsilon_{\pi(2)}$ and
\begin{equation}
\sigma_{\s{1}{2}{3}}+\sigma_{\s{\pi(1)}{\pi(2)}}+\sigma_{\s{\pi(2)}{\pi(3)}}+\sigma_{\s{\pi(2)}} \geq 1-\epsilon_{\pi(2)},
\end{equation}
this further implies that one can restrict to numbers $\{\sigma_{\mI}\}_{\mI \in \mathcal{P}(3)}$ satisfying
\begin{IEEEeqnarray}{rCl}
\sigma_{\s{\pi(2)}{\pi(3)}} & \geq  &\sigma_{\s{\pi(1)}{\pi(3)}} +\sigma_{\s{\pi(1)}}  - \sigma_{\s{\pi(2)}}\\
& \geq& \sigma_{\s{\pi(1)}{\pi(3)}} - \sigma_{\s{\pi(2)}} - \sigma_{\s{\pi(1)}{\pi(2)}}. \label{eq:ass13_first}\IEEEeqnarraynumspace
\end{IEEEeqnarray}

 We next show that one can further restrict to nonnegative numbers satisfying also \eqref{eq:ass}. 
 To this end, assume that \eqref{eq:ass} is violated and define
  \begin{equation}\label{eq:a}
  a:=\tilde \sigma_{\s{\pi(1)}{\pi(3)}} - \tilde \sigma_{\s{\pi(2)}} - \tilde\sigma_{\s{\pi(1)}{\pi(2)}} >0. 
 \end{equation}
Define also the new parameters
 \begin{IEEEeqnarray}{rCl}
 \sigma_{\s{1}{2}{3}}' & := &\tilde \sigma_{\s{1}{2}{3}} +a \\
 \sigma_{\s{\pi{(3)}}}' & := & \tilde\sigma_{\s{\pi{(3)}}} +a \\
   \sigma_{\s{\pi(1)}{\pi(3)}}' & :=&     \tilde\sigma_{\s{\pi(1)}{\pi(3)}} -a\\
   \sigma_{\s{\pi(2)}{\pi(3)}}'& :=&   \tilde  \sigma_{\s{\pi(2)}{\pi(3)}} -a \\
   \sigma_{\mI}' & :=&\tilde \sigma_{\mI}, \quad \pi(3) \notin \mI,
 \end{IEEEeqnarray} 
 and  the new rates 
 \begin{subequations}\label{eq:sub:new_rates}
  \begin{IEEEeqnarray}{rCl}
 {R}_{\s{1}{2}{3},\ell} '&=&\frac{  a \left( \lambda_{\ell} \tilde{R}_{ \s{\pi(1)}{\pi(3)},\ell} + (1-\lambda_{\ell})\tilde {R}_{ \s{\pi(2)}{\pi(3)},\ell} \right)}{\sigma_{\s{1}{2}{3}}'} \nonumber \\
 && + \frac{\tilde\sigma_{\s{1}{2}{3}} \tilde{R}_{\s{1}{2}{3},\ell} }{\sigma_{\s{1}{2}{3}}'}  , \qquad \ell \in \{1,2,3\},\nonumber\\\\
{R}_{\s{\pi(3)},\ell}'&=& \frac{  a \left( (1-\lambda_{\ell}) \tilde{R}_{ \s{\pi(1)}{\pi(3)},\ell} +\lambda_{\ell} \tilde{R}_{ \s{\pi(2)}{\pi(3)},\ell} \right)}{\sigma_{\s{\pi(3)}}'}  \nonumber\\
  && +
   \frac{\tilde\sigma_{\s{\pi(3)}} \tilde{R}_{\s{\pi(3)},\ell } }{\sigma_{\s{\pi(3)}}'},  \qquad \ell\in \{1,\ldots,\pi(3)\}\nonumber \\ \\
   {R}_{\mI,\ell}'& = &   \tilde{R}_{\mI,\ell}, \quad \mI \in \mathcal{P}(3)\backslash \{ \s{1}{2}{3},       \s{\pi(3)}\}.
  \end{IEEEeqnarray} 
  \end{subequations}
 Notice that by the definition of $a$ and by \eqref{eq:ass13_first}, the parameters $\{\sigma_{\mI}'\}$ are all nonnegative, and it is easily verified that they continue to satisfy \eqref{eq:E1_Khop} for any choice of $\lambda_1, \lambda_2, \lambda_3 \in [0,1]$.
  
  We next choose the parameters $\lambda_1, \lambda_2, \lambda_3 \in [0,1]$ in function of the  rates $\{\tilde{R}_{\mI,\ell}\}$ and the ordering $\pi(\cdot)$, and show that for the proposed choice of rates in \eqref{eq:sub:new_rates}, the exponents $\theta_1,\theta_2, \theta_3$ are only increased.
 We distinguish three cases. 

For notational simplicity we assume $\pi(1)<\pi(2)$. (The proof for $\pi(1)>\pi(2)$ is analogous.)
 This implies  that 
 \begin{equation}
 1=\pi(1) < \pi(3) \qquad \textnormal{or} \qquad 1=\pi(3)<\pi(2)=2
 \end{equation}
 and 
  \begin{equation}
  2=\pi(2) < \pi(3)=3 \qquad \textnormal{or} \qquad \pi(3)<\pi(2)=3.
  \end{equation}
  
\textit{Case 1:} If 
\begin{equation}\label{eq:assumption_case1}
 \eta_{1}\left( \tilde R_{\{\pi(1),\pi(3)\},1} \right)\leq  \eta_1\left(\tilde R_{\{\pi(2),\pi(3)\},1}\right),
\end{equation}
 choose 
\begin{IEEEeqnarray}{rCl}
\lambda_{\ell} &=& 0, \qquad \qquad \qquad \qquad  \ell \in \{1,\ldots,\pi(3)\},\\
\lambda_{\ell} &=&\mathbbm{1}\left\{\tilde{R}_{ \s{\pi(1)}{\pi(3)},\ell} \geq \tilde{R}_{ \s{\pi(2)}{\pi(3)},\ell}\right\}, \nonumber \\
&& \qquad \qquad \qquad \qquad \quad \ell \in \{\pi(3)+1,\ldots,3\}. \IEEEeqnarraynumspace \label{eq:lambda_ell_2}
\end{IEEEeqnarray}
Using the same proof steps as in Lemma~\ref{lem:moving}, it can be shown that for this choice of the $\lambda$s the new rates  in  \eqref{eq:sub:new_rates} still satisfy Constraint \eqref{eq:56a} for  $\theta_{\pi(3)}$ because $\lambda_1=\cdots=\lambda_{\pi(3)}$. 

To see that they  satisfy \eqref{eq:56a} also for $\theta_{\pi(1)}$, notice that:
\begin{IEEEeqnarray}{rCl}
 \lefteqn{\min \left\{ \sum_{\ell=1}^{\pi(1)} \eta_\ell\left( \tilde{R}_{\{\pi(1),\pi(3)\}, \ell}\right) , \; \sum_{\ell=1}^{\pi(1)} \eta_\ell\left( \tilde{R}_{\{1,2,3\}, \ell}\right) \right\}} \nonumber\\
& \leq	& \min \left\{\sum_{\ell=1}^{\pi(1)} \eta_\ell\left( \tilde{R}_{\{\pi(1),\pi(3)\}, \ell}\right) ,  \right.  \nonumber \\
& & \hspace{1cm} \;  \frac{a}{\sigma_{\{1,2,3\}}'}  \sum_{\ell=1}^{\pi(1)} \eta_\ell\left( \tilde{R}_{\{\pi(1),\pi(3)\}, \ell}\right)  \nonumber \\
 &&\hspace{1.2cm} \; \ \left.+ \frac{\tilde{\sigma}_{\{1,2,3\}}}{\sigma_{\{1,2,3\}}'}   \sum_{\ell=1}^{\pi(1)} \eta_\ell\left( \tilde{R}_{\{1,2,3\}, \ell}\right) \right\}\\
&  \leq	& \min \left\{\sum_{\ell=1}^{\pi(1)} \eta_\ell\left( \tilde{R}_{\{\pi(1),\pi(3)\}, \ell}\right) ,  \right.  \nonumber \\
& & \hspace{1cm} \; \frac{a}{\sigma_{\{1,2,3\}}'}   \eta_1\left( \tilde{R}_{\{\pi(2),\pi(3)\}, 1}\right)  \nonumber \\
& & \hspace{1cm} \; + \frac{a}{\sigma_{\{1,2,3\}}'}  \mathbbm{1}\left\{ \pi(1)=2\right\}  \nonumber\\
&& \hspace{1.2cm} \; \cdot\max\left\{\eta_2\left( \tilde{R}_{\{\pi(1),\pi(3)\}, 2}\right),  \eta_2\left( \tilde{R}_{\{\pi(2),\pi(3)\}, 2}\right)\right\}  \nonumber \\
&&\hspace{3.6cm} \; \ \left.+ \frac{\tilde{\sigma}_{\{1,2,3\}}}{\sigma_{\{1,2,3\}}'}   \sum_{\ell=1}^{\pi(1)} \eta_\ell\left( \tilde{R}_{\{1,2,3\}, \ell}\right) \right\} \nonumber \\ \\
& \leq & \min \left\{  \sum_{\ell=1}^{\pi(1)} \eta_\ell\left( {R}_{\{\pi(1),\pi(3)\}, \ell}'\right) , \;  \sum_{\ell=1}^{\pi(1)}  \eta_\ell \left( {R}_{\{1,2,3\}, \ell}'\right)   \right\}. \hspace{.6cm}\nonumber \\ 
 \end{IEEEeqnarray}
 where the second inequality holds by Assumption \eqref{eq:assumption_case1} and the third inequality holds by  the definitions of the rates $\{R'_{\{1,2,3\},\ell}\}$  and by the concavity and monotonicity of the functions $\{\eta_\ell(\cdot)\}$. 
 
Similarly, we notice for $\theta_{\pi(2)}$:
 \begin{IEEEeqnarray}{rCl}
		\lefteqn{\min \left\{ \sum_{\ell=1}^{\pi(2)} \eta_\ell\left( \tilde{R}_{\{\pi(2),\pi(3)\}, \ell}\right) , \; \sum_{\ell=1}^{\pi(2)} \eta_\ell\left( \tilde{R}_{\{1,2,3\}, \ell}\right) \right\}} \nonumber\\
		& \leq	& \min \left\{\sum_{\ell=1}^{\pi(2)} \eta_\ell\left( \tilde{R}_{\{\pi(2),\pi(3)\}, \ell}\right) ,  \right.  \nonumber \\
		& & \hspace{1cm} \;  \frac{a}{\sigma_{\{1,2,3\}}'}  \sum_{\ell=1}^{\pi(2)} \eta_\ell\left( \tilde{R}_{\{\pi(2),\pi(3)\}, \ell}\right)  \nonumber \\
		&&\hspace{1.2cm} \; \ \left.+ \frac{\tilde{\sigma}_{\{1,2,3\}}}{\sigma_{\{1,2,3\}}'}   \sum_{\ell=1}^{\pi(2)} \eta_\ell\left( \tilde{R}_{\{1,2,3\}, \ell}\right) \right\}\\
		&  \leq	& \min \left\{\sum_{\ell=1}^{\pi(2)} \eta_\ell\left( \tilde{R}_{\{\pi(2),\pi(3)\}, \ell}\right) ,  \right.  \nonumber \\
		& & \hspace{1cm} \; \frac{a}{\sigma_{\{1,2,3\}}'}  \sum_{\ell=1}^{\min\{\pi(2),\pi(3)\}} \eta_\ell\left( \tilde{R}_{\{\pi(2),\pi(3)\}, \ell}\right)  \nonumber \\
		& & \hspace{1.2cm} \; + \frac{a}{\sigma_{\{1,2,3\}}'}  \sum_{\ell=\pi(3)+1}^{\pi(2)} \max\left\{\eta_\ell\left( \tilde{R}_{\{\pi(1),\pi(3)\}, \ell}\right), \right. \nonumber \\
		& & \left. \hspace{4.6cm} \; \eta_\ell\left( \tilde{R}_{\{\pi(2),\pi(3)\}, \ell}\right)\right\}  \nonumber \\
		&&\hspace{1.6cm} \; \ \left.+ \frac{\tilde{\sigma}_{\{1,2,3\}}}{\sigma_{\{1,2,3\}}'}   \sum_{\ell=1}^{\pi(2)} \eta_\ell\left( \tilde{R}_{\{1,2,3\}, \ell}\right) \right\}\IEEEeqnarraynumspace \label{eq:second_line}\\
		& \leq & \min \left\{  \sum_{\ell=1}^{\pi(2)} \eta_\ell\left( {R}_{\{\pi(2),\pi(3)\}, \ell}'\right) , \;  \sum_{\ell=1}^{\pi(2)}  \eta_\ell \left( {R}_{\{1,2,3\}, \ell}'\right)   \right\}, \nonumber \\ \IEEEeqnarraynumspace
\end{IEEEeqnarray} where notice that the sum in the second line of \eqref{eq:second_line} is empty when $\pi(2) \leq \pi(3)$. Here, the last inequality holds by the definitions of the rates $\{R'_{\{1,2,3\},\ell}\}$ and by the choice of the $\lambda$s  and  the concavity and monotonicity of the functions $\{\eta_\ell(\cdot)\}_{\ell}$. 
 
 \medskip 
 
\textit{Case 2:} If 
\begin{equation}\label{eq:case2_assumption}
\sum_{\ell=1}^{2}\eta_\ell\left( \tilde R_{\{\pi(2),\pi(3)\},\ell}\right)\leq \sum_{\ell=1}^{2}\eta_\ell\left(\tilde R_{\{\pi(1),\pi(3)\},\ell}\right),
\end{equation}  choose 
\begin{IEEEeqnarray}{rCl}
	\lambda_{\ell} &=& 1, \qquad \qquad \qquad   \ell \in \{1,\ldots,\max\{2,\pi(3)\}\}, \label{eq:lambda_ell_case2_1}\\
	\lambda_{\ell} &=&\mathbbm{1}\left\{\tilde{R}_{ \s{\pi(1)}{\pi(3)},\ell} \geq \tilde{R}_{ \s{\pi(2)}{\pi(3)},\ell}\right\}, \nonumber \\
	&& \qquad \qquad \qquad \quad \ell \in \{\max\{2,\pi(3)\}+1,\ldots,3\}. \IEEEeqnarraynumspace
\end{IEEEeqnarray}
Using similar arguments as in the previous case, one can conclude that the new rates in \eqref{eq:sub:new_rates} still satisfy \eqref{eq:56a}. More specifically, since $\lambda_1 = \cdots = \lambda_{\pi(3)} = 1$ by \eqref{eq:lambda_ell_case2_1}, similar  
 proof steps as in Lemma~\ref{lem:moving} can be used to show that \eqref{eq:56a} holds for $\theta_{\pi(3)}$.
 
 To see that \eqref{eq:56a} holds  for $\theta_{\pi(2)}$, recall that $\pi(2)\geq 2$ and notice:
	 \begin{IEEEeqnarray}{rCl}
		 \lefteqn{\min \left\{ \sum_{\ell=1}^{\pi(2)} \eta_\ell\left( \tilde{R}_{\{\pi(2),\pi(3)\}, \ell}\right) , \; \sum_{\ell=1}^{\pi(2)} \eta_\ell\left( \tilde{R}_{\{1,2,3\}, \ell}\right) \right\}} \nonumber\\
			& \leq	& \min \left\{\sum_{\ell=1}^{\pi(2)} \eta_\ell\left( \tilde{R}_{\{\pi(2),\pi(3)\}, \ell}\right) ,  \right.  \nonumber \\
			& & \hspace{1cm} \;  \frac{a}{\sigma_{\{1,2,3\}}'}  \sum_{\ell=1}^{\pi(2)} \eta_\ell\left( \tilde{R}_{\{\pi(2),\pi(3)\}, \ell}\right)  \nonumber \\
			&&\hspace{1.2cm} \; \ \left.+ \frac{\tilde{\sigma}_{\{1,2,3\}}}{\sigma_{\{1,2,3\}}'}   \sum_{\ell=1}^{\pi(2)} \eta_\ell\left( \tilde{R}_{\{1,2,3\}, \ell}\right) \right\}\\
			&  \leq	& \min \left\{\sum_{\ell=1}^{\pi(2)} \eta_\ell\left( \tilde{R}_{\{\pi(2),\pi(3)\}, \ell}\right) ,  \right.  \nonumber \\
			& & \hspace{1cm} \; \frac{a}{\sigma_{\{1,2,3\}}'}  \sum_{\ell=1}^{2} \eta_\ell\left( \tilde{R}_{\{\pi(1),\pi(3)\}, \ell}\right)  \nonumber \\
			& & \hspace{1cm} \; + \frac{a}{\sigma_{\{1,2,3\}}'}  \mathbbm{1}\{\pi(2)=3\}  \nonumber \\
			&& \hspace{1.2cm}  \cdot \max\left\{\eta_{{3}}\left( \tilde{R}_{\{\pi(1),\pi(3)\}, {3}}\right), \; \eta_{{3}}\left( \tilde{R}_{\{\pi(2),\pi(3)\}, {3}}\right)\right\}  \nonumber \\
			&&\hspace{1.6cm} \; \ \left.+ \frac{\tilde{\sigma}_{\{1,2,3\}}}{\sigma_{\{1,2,3\}}'}   \sum_{\ell=1}^{\pi(1)} \eta_\ell\left( \tilde{R}_{\{1,2,3\}, \ell}\right) \right\} \\
			& \leq & \min \left\{  \sum_{\ell=1}^{\pi(2)} \eta_\ell\left( {R}_{\{\pi(2),\pi(3)\}, \ell}'\right) , \;  \sum_{\ell=1}^{\pi(2)}  \eta_\ell \left( {R}_{\{1,2,3\}, \ell}'\right)   \right\}, \nonumber \\ 
	\end{IEEEeqnarray}
where the second inequality holds by our assumption \eqref{eq:case2_assumption} and since $\pi(2)\geq2$. 

 Finally, \eqref{eq:56a} holds for $\theta_{\pi(1)}$, because:
 \begin{IEEEeqnarray}{rCl}
		\lefteqn{\min \left\{ \sum_{\ell=1}^{\pi(1)} \eta_\ell\left( \tilde{R}_{\{\pi(1),\pi(3)\}, \ell}\right) , \; \sum_{\ell=1}^{\pi(1)} \eta_\ell\left( \tilde{R}_{\{1,2,3\}, \ell}\right) \right\}} \nonumber\\
		& \leq	& \min \left\{\sum_{\ell=1}^{\pi(1)} \eta_\ell\left( \tilde{R}_{\{\pi(1),\pi(3)\}, \ell}\right) ,  \right.  \nonumber \\
		& & \hspace{1cm} \;  \frac{a}{\sigma_{\{1,2,3\}}'}  \sum_{\ell=1}^{\pi(1)} \eta_\ell\left( \tilde{R}_{\{\pi(1),\pi(3)\}, \ell}\right)  \nonumber \\
		&&\hspace{1.2cm} \; \ \left.+ \frac{\tilde{\sigma}_{\{1,2,3\}}}{\sigma_{\{1,2,3\}}'}   \sum_{\ell=1}^{\pi(1)} \eta_\ell\left( \tilde{R}_{\{1,2,3\}, \ell}\right) \right\}\\
		& \leq & \min \left\{  \sum_{\ell=1}^{\pi(1)} \eta_\ell\left( {R}_{\{\pi(1),\pi(3)\}, \ell}'\right) , \;  \sum_{\ell=1}^{\pi(1)}  \eta_\ell \left( {R}_{\{1,2,3\}, \ell}'\right)   \right\}. \nonumber \\ \IEEEeqnarraynumspace
\end{IEEEeqnarray}
where the second inequality holds by the assumption $\pi(1)<\pi(2)$ and thus $\pi(1)\leq 2$. 
\medskip

\textit{Case 3:} Else, i.e., if
\begin{IEEEeqnarray}{rCl} 
\eta_1\left( \tilde R_{\{\pi(1),\pi(3)\},1}\right)> \eta_1\left(\tilde R_{\{\pi(2),\pi(3)\},1}\right) \label{eq:case3_assumption1}
 \end{IEEEeqnarray}
 and 
 \begin{IEEEeqnarray}{rCl} 
 	\sum_{\ell=1}^{2}\eta_\ell\left( \tilde R_{\{\pi(1),\pi(3)\},\ell}\right)< \sum_{\ell=1}^{2}\eta_\ell\left(\tilde R_{\{\pi(2),\pi(3)\},\ell}\right). \label{eq:case3_assumption2}
 \end{IEEEeqnarray}
  Choose 
 \begin{equation}\label{eq:lambda_case3}
 \lambda_1=1, \quad \lambda_2=\lambda, \quad \textnormal{and} \quad \lambda_3=0,
 \end{equation}
for a value of  $\lambda\in[0,1]$  so that the auxiliary rates 
\begin{IEEEeqnarray}{rCl}
\bar R_{\{\pi(1),\pi(3)\},2} & : = & \lambda R_{\{\pi(1),\pi(3)\},2} + (1-\lambda) R_{\{\pi(2),\pi(3)\},2}\IEEEeqnarraynumspace  \\
\bar R_{\{\pi(2),\pi(3)\},2} & := &(1- \lambda)R_{\{\pi(1),\pi(3)\},2} + \lambda R_{\{\pi(2),\pi(3)\},2} 
\end{IEEEeqnarray}
satisfy
\begin{IEEEeqnarray}{rCl}
\lefteqn{\eta_1\left(\tilde R_{\{\pi(1),\pi(3)\},1}\right) + \eta_2\left( \bar{R}_{\{\pi(1),\pi(3)\},2}\right)} \qquad \nonumber\\  &= &\sum_{\ell=1}^{2} \eta_\ell\left(\tilde R_{\{\pi(2),\pi(3)\},\ell}\right)\IEEEeqnarraynumspace \label{eq:R_13_R_23_improved}\\
\lefteqn{\eta_1\left(\tilde R_{\{\pi(2),\pi(3)\},1}\right) + \eta_2\left( \bar{R}_{\{\pi(2),\pi(3)\},2}\right)} \qquad \nonumber\\  &\geq&\sum_{\ell=1}^{2} \eta_\ell\left(\tilde R_{\{\pi(1),\pi(3)\},\ell}\right) .\label{eq:inc}
\end{IEEEeqnarray}
Existence of the desired choice of $\lambda$ can be seen as follows. Notice first that for $\lambda=0$,  relation~\eqref{eq:R_13_R_23_improved} holds with a $>$ sign because of Assumption~\eqref{eq:case3_assumption1}. For $\lambda=1$, relation~\eqref{eq:R_13_R_23_improved} holds with a $<$ sign because of Assumption~\eqref{eq:case3_assumption2}.
By the continuity of the functions $\{\eta_\ell(\cdot)\}$  and the intermediate value theorem, there is thus a value $\lambda \in (0,1)$ such that \eqref{eq:R_13_R_23_improved} holds with equality. Let $\lambda$ be this  value and notice that  by the concavity of the functions  $\{\eta_\ell(\cdot)\}$:
\begin{IEEEeqnarray}{rCl}
\lefteqn{\eta_1\left(\tilde R_{\{\pi(1),\pi(3)\},1}\right) + \eta_2\left( \bar R_{\{\pi(1),\pi(3)\},2}\right)} \quad  \nonumber \\
	&& + \eta_1\left(\tilde R_{\{\pi(2),\pi(3)\},1}\right) + \eta_2\left( \bar R_{\{\pi(2),\pi(3)\},2}\right)  \nonumber\\
& \geq &\sum_{\ell=1}^{2} \eta_\ell\left(\tilde R_{\{\pi(1),\pi(3)\},\ell}\right)+\sum_{\ell=1}^{2} \eta_\ell\left(\tilde R_{\{\pi(2),\pi(3)\},\ell}\right),\IEEEeqnarraynumspace
\end{IEEEeqnarray}
which combined with~\eqref{eq:R_13_R_23_improved} implies \eqref{eq:inc}.

Now that we established the existence of the desired value  $\lambda$, we continue to show that for the choice in \eqref{eq:lambda_case3}, Constraints \eqref{eq:56a} remain valid. For $\theta_{\pi(1)}$ this can be verified through the following steps, where recall that $\pi(1)\leq 2$:
 \begin{IEEEeqnarray}{rCl}
 \lefteqn{\min \left\{ \sum_{\ell=1}^{\pi(1)} \eta_\ell\left( \tilde{R}_{\{\pi(1),\pi(3)\}, \ell}\right) , \; \sum_{\ell=1}^{\pi(1)} \eta_\ell\left( \tilde{R}_{\{1,2,3\}, \ell}\right) \right\}} \nonumber\\
& \leq	& \min \left\{\sum_{\ell=1}^{\pi(1)} \eta_\ell\left( \tilde{R}_{\{\pi(1),\pi(3)\}, \ell}\right) ,  \right.  \nonumber \\
& & \hspace{1cm} \;  \frac{a}{\sigma_{\{1,2,3\}}'}  \sum_{\ell=1}^{\pi(1)} \eta_\ell\left( \tilde{R}_{\{\pi(1),\pi(3)\}, \ell}\right)  \nonumber \\
 &&\hspace{1.2cm} \; \ \left.+ \frac{\tilde{\sigma}_{\{1,2,3\}}}{\sigma_{\{1,2,3\}}'}   \sum_{\ell=1}^{\pi(1)} \eta_\ell\left( \tilde{R}_{\{1,2,3\}, \ell}\right) \right\}\\
 & =	& \min \left\{\sum_{\ell=1}^{\pi(1)} \eta_\ell\left( \tilde{R}_{\{\pi(1),\pi(3)\}, \ell}\right) ,  \right.  \nonumber \\
 & & \hspace{1cm} \;  \frac{a}{\sigma_{\{1,2,3\}}'}   \Big(\eta_1\left( \tilde{R}_{\{\pi(1),\pi(3)\}, 1}\right)   \nonumber \\
  && \hspace{2.3cm}+\left. \mathbbm{1}\left\{\pi(1)=2\right\} \cdot \eta_2\left( \bar{R}_{\{\pi(1),\pi(3)\}, 2}\right) \right) \nonumber \\
 &&\hspace{1.2cm} \; \ \left.+ \frac{\tilde{\sigma}_{\{1,2,3\}}}{\sigma_{\{1,2,3\}}'}   \sum_{\ell=1}^{\pi(1)} \eta_\ell\left( \tilde{R}_{\{1,2,3\}, \ell}\right) \right\}\\
& \leq & \min \left\{  \sum_{\ell=1}^{\pi(1)} \eta_\ell\left( {R}_{\{\pi(1),\pi(3)\}, \ell}'\right) , \;  \sum_{\ell=1}^{\pi(1)}  \eta_\ell \left( {R}_{\{1,2,3\}, \ell}'\right)   \right\},\nonumber\\
 \end{IEEEeqnarray}
where the second inequality holds since $\pi(1)\leq 2$, and by \eqref{eq:case3_assumption2} and \eqref{eq:R_13_R_23_improved}, and the last inequality holds by the definitions of the rates  $\{R'_{\{1,2,3\},\ell}\}$, the choice of the $\lambda$s, and the concavity and monotonicity of the functions $\{\eta_\ell(\cdot)\}$.

To verify that Constraint \eqref{eq:56a} remains valid for $\theta_{\pi(2)}$, recall that $\pi(2)\geq 2$ and notice:
 \begin{IEEEeqnarray}{rCl}
 \lefteqn{\min \left\{ \sum_{\ell=1}^{\pi(2)} \eta_\ell\left( \tilde{R}_{\{\pi(2),\pi(3)\}, \ell}\right) , \; \sum_{\ell=1}^{\pi(2)} \eta_\ell\left( \tilde{R}_{\{1,2,3\}, \ell}\right) \right\}} \nonumber\\
& \leq	& \min \left\{\sum_{\ell=1}^{\pi(2)} \eta_\ell\left( \tilde{R}_{\{\pi(2),\pi(3)\}, \ell}\right) ,  \right.  \nonumber \\
& & \hspace{1cm} \;  \frac{a}{\sigma_{\{1,2,3\}}'}  \sum_{\ell=1}^{\pi(2)} \eta_\ell\left( \tilde{R}_{\{\pi(2),\pi(3)\}, \ell}\right)  \nonumber \\
 &&\hspace{1.2cm} \; \ \left.+ \frac{\tilde{\sigma}_{\{1,2,3\}}}{\sigma_{\{1,2,3\}}'}   \sum_{\ell=1}^{\pi(2)} \eta_\ell\left( \tilde{R}_{\{1,2,3\}, \ell}\right) \right\}\\
&  =	& \min \left\{\sum_{\ell=1}^{\pi(2)} \eta_\ell\left( \tilde{R}_{\{\pi(2),\pi(3)\}, \ell}\right) ,  \right.  \nonumber \\
& & \hspace{1cm} \; \frac{a}{\sigma_{\{1,2,3\}}'}  \sum_{\ell=1}^{2} \eta_\ell\left( \tilde{R}_{\{\pi(2),\pi(3)\}, \ell}\right)  \nonumber \\
& & \hspace{1.2cm} +  \frac{a}{\sigma_{\{1,2,3\}}'} \mathbbm{1}\left\{\pi(2)=3\right\} \cdot \eta_3\left( \tilde{R}_{\{\pi(2),\pi(3)\} ,3} \right)\nonumber \\
&&\hspace{1.4cm} \; \ \left.+ \frac{\tilde{\sigma}_{\{1,2,3\}}}{\sigma_{\{1,2,3\}}'}   \sum_{\ell=1}^{\pi(2)} \eta_\ell\left( \tilde{R}_{\{1,2,3\}, \ell}\right) \right\}\\
&  =	& \min \left\{\sum_{\ell=1}^{\pi(2)} \eta_\ell\left( \tilde{R}_{\{\pi(2),\pi(3)\}, \ell}\right) ,  \right.  \nonumber \\
& & \hspace{1cm} \;  \frac{a}{\sigma_{\{1,2,3\}}'}  \left( \eta_1\left( \tilde{R}_{\{\pi(1),\pi(3)\}, 1}\right) + \eta_2\left( \bar{R}_{\{\pi(1),\pi(3)\}, 2}\right) \right) \nonumber \\
& & \hspace{1.2cm} +  \frac{a}{\sigma_{\{1,2,3\}}'} \mathbbm{1}\left\{\pi(2)=3\right\} \cdot \eta_3\left( \tilde{R}_{\{\pi(2),\pi(3)\} ,3} \right)\nonumber \\
&&\hspace{1.4cm} \; \ \left. + \frac{\tilde{\sigma}_{\{1,2,3\}}}{\sigma_{\{1,2,3\}}'}   \sum_{\ell=1}^{\pi(2)} \eta_\ell\left( \tilde{R}_{\{1,2,3\}, \ell}\right) \right\}\\
& \leq & \min \left\{  \sum_{\ell=1}^{\pi(2)} \eta_\ell\left( {R}_{\{\pi(2),\pi(3)\}, \ell}'\right) , \;  \sum_{\ell=1}^{\pi(2)}  \eta_\ell \left( {R}_{\{1,2,3\}, \ell}'\right)   \right\}. \IEEEeqnarraynumspace
 \end{IEEEeqnarray}
Here,  the second equality holds by \eqref{eq:R_13_R_23_improved}.

Finally, to see that Constraint \eqref{eq:56a} is also satisfied for $\theta_{\pi(3)}$, we distinguish two cases. If $\pi(3)=1$, the proof is similar to the proof of Lemma~\ref{lem:moving} because $\lambda_{1}=1$. For the proof in the case $\pi(3)\geq 2$, notice first:
\begin{IEEEeqnarray}{rCl}
	\lefteqn{\min \left\{ \sum_{\ell=1}^{\pi(3)} \eta_\ell\left( \tilde{R}_{\{\pi(2),\pi(3)\}, \ell}\right) , \; \sum_{\ell=1}^{\pi(3)} \eta_\ell\left( \tilde{R}_{\{1,2,3\}, \ell}\right) \right\}} \nonumber\\
	& \leq	& \min \left\{\sum_{\ell=1}^{\pi(3)} \eta_\ell\left( \tilde{R}_{\{\pi(2),\pi(3)\}, \ell}\right) ,  \right.  \nonumber \\
	& & \hspace{1cm} \;  \frac{a}{\sigma_{\{1,2,3\}}'}  \sum_{\ell=1}^{\pi(3)} \eta_\ell\left( \tilde{R}_{\{\pi(2),\pi(3)\}, \ell}\right)  \nonumber \\
	&&\hspace{1.2cm} \; \ \left.+ \frac{\tilde{\sigma}_{\{1,2,3\}}}{\sigma_{\{1,2,3\}}'}   \sum_{\ell=1}^{\pi(3)} \eta_\ell\left( \tilde{R}_{\{1,2,3\}, \ell}\right) \right\}\\
	&  {=}	& \min \left\{\sum_{\ell=1}^{\pi(3)} \eta_\ell\left( \tilde{R}_{\{\pi(2),\pi(3)\}, \ell}\right) ,  \right.  \nonumber \\
	& & \hspace{1cm} \; \frac{a}{\sigma_{\{1,2,3\}}'} \eta_1\left( \tilde{R}_{\{\pi(1),\pi(3)\}, 1}\right) \nonumber \\
	& & \hspace{1.2cm} \; + \frac{a}{\sigma_{\{1,2,3\}}'} \eta_{2}\left( \bar{R}_{\{\pi(1),\pi(3)\}, 2}\right)  \nonumber \\
	& & \hspace{1.2cm} \; +  \frac{a}{\sigma_{\{1,2,3\}}'} \mathbbm{1}\{\pi(3)=3\}\ . \ \eta_\ell\left( \tilde{R}_{\{\pi(2),\pi(3)\} ,\ell} \right) \nonumber \\
	&&\hspace{3cm} \; \ \left. + \frac{\tilde{\sigma}_{\{1,2,3\}}}{\sigma_{\{1,2,3\}}'}   \sum_{\ell=1}^{\pi(3)} \eta_\ell\left( \tilde{R}_{\{1,2,3\}, \ell}\right) \right\}\\
	& \leq & \min \left\{  \sum_{\ell=1}^{\pi(3)} \eta_\ell\left( {R}_{\{\pi(2),\pi(3)\}, \ell}'\right) , \;  \sum_{\ell=1}^{\pi(3)}  \eta_\ell \left( {R}_{\{1,2,3\}, \ell}'\right)   \right\}, \IEEEeqnarraynumspace
\end{IEEEeqnarray}
where the equality holds by Assumption  \eqref{eq:R_13_R_23_improved}. 

Notice further that 
\begin{IEEEeqnarray}{rCl}
	\lefteqn{\min \left\{ \sum_{\ell=1}^{\pi(3)} \eta_\ell\left( \tilde{R}_{\{\pi(1),\pi(3)\}, \ell}\right) , \; \sum_{\ell=1}^{\pi(3)} \eta_\ell\left( \tilde{R}_{\{\pi(3)\}, \ell}\right) \right\}} \nonumber\\
	& \leq	& \min \left\{\sum_{\ell=1}^{\pi(3)} \eta_\ell\left( \tilde{R}_{\{\pi(1),\pi(3)\}, \ell}\right) ,  \right.  \nonumber \\
	& & \hspace{1cm} \;  \frac{a}{\sigma_{\{\pi(3)\}}'}  \sum_{\ell=1}^{\pi(3)} \eta_\ell\left( \tilde{R}_{\{\pi(1),\pi(3)\}, \ell}\right)  \nonumber \\
	&&\hspace{1.2cm} \; \ \left.+ \frac{\tilde{\sigma}_{\{\pi(3)\}}}{\sigma_{\{\pi(3)\}}'}   \sum_{\ell=1}^{\pi(3)} \eta_\ell\left( \tilde{R}_{\{\pi(3)\}, \ell}\right) \right\}\\
	&  \leq	& \min \left\{\sum_{\ell=1}^{\pi(3)} \eta_\ell\left( \tilde{R}_{\{\pi(1),\pi(3)\}, \ell}\right) ,  \right.  \nonumber \\
	& & \hspace{1cm} \; \frac{a}{\sigma_{\{\pi(3)\}}'} \eta_1\left( \tilde{R}_{\{\pi(2),\pi(3)\}, 1}\right) \nonumber \\
	& & \hspace{1.2cm} \; + \frac{a}{{\sigma_{\{\pi(3)\}}'}}   \eta_{2}\left( \bar{R}_{\{\pi(2),\pi(3)\}, 2}\right)  \nonumber \\
	& & \hspace{1.2cm} \; +  \frac{a}{{\sigma_{\{\pi(3)\}}'}} \mathbbm{1}\{\pi(3)=3\}\ . \ \eta_{{3}}\left( \tilde{R}_{\{\pi({1}),\pi(3)\} ,{3}} \right) \nonumber \\
	&&\hspace{3cm} \; \ \left. + {\frac{\tilde{\sigma}_{\{\pi(3)\}}}{\sigma_{\{\pi(3)\}}'}  }   \sum_{\ell=1}^{\pi(3)} \eta_\ell\left( \tilde{R}_{\{1,2,3\}, \ell}\right) \right\}\\
	& \leq & \min \left\{  \sum_{\ell=1}^{\pi(3)} \eta_\ell\left( {R}_{\{\pi({1}),\pi(3)\}, \ell}'\right) , \;  \sum_{\ell=1}^{\pi(3)}  \eta_\ell \left( {R}_{\{{\pi(3)}\}, \ell}'\right)   \right\}, \IEEEeqnarraynumspace
\end{IEEEeqnarray}
where the second inequality holds by \eqref{eq:inc}.

\end{document}